\pdfoutput=1

\documentclass{article}
\usepackage{fullpage}
\usepackage{orcidlink}

\usepackage{amsmath}
\usepackage{amsthm}
\usepackage{amsfonts}

\usepackage{graphicx}
\usepackage{xcolor}
\usepackage{tikz}

\usepackage[T1]{fontenc}
\usepackage{microtype}
\usepackage{mathpazo}
\usepackage{xspace}

\usepackage{float}

\usepackage{enumitem}

\usepackage{cryptocode}
\usepackage{mathtools}

\usepackage[nocompress]{cite}
\usepackage{xurl}
\usepackage{hyperref}
\AtBeginDocument{\pcfixhyperref}
\hypersetup{pdflang=en,
    urlcolor=blue,colorlinks=true,
    citecolor=black!70!green,
    linkcolor=black!70!red,
}

\usepackage{dsfont}
\usepackage{pifont}

\DeclarePairedDelimiter\floor{\lfloor}{\rfloor}
\DeclarePairedDelimiter\ceil{\lceil}{\rceil}

\def\secpar{\lambda}

\newcommand{\poly}{\mathrm{poly}}

\newcommand{\negl}{\mathsf{negl}}

\newcommand{\bra}[1]{\left\langle #1\right|}
\newcommand{\ket}[1]{\left|#1\right\rangle}

\newcommand{\bits}{\{0,1\}}

\newcommand{\getsr}{\overset{{\scriptscriptstyle\$}}{\leftarrow}}

\newcommand{\calJ}{\mathcal{J}}
\newcommand{\calO}{\mathcal{O}}

\newcommand{\calP}{\mathcal{P}}
\newcommand{\FF}{\mathbb{F}_q}

\newcommand{\Ex}{\mathbb{E}}

\newcommand{\heading}[1]{\medskip\noindent{\textbf{#1}}}

\newcommand{\thmref}[1]{Theorem~\ref{#1}}

\newtheorem{thm}{Theorem}[section]
\newtheorem{obs}[thm]{Observation}
\newtheorem{assm}[thm]{Assumption}
\newtheorem{defn}[thm]{Definition}
\newtheorem{lem}[thm]{Lemma}
\newtheorem{cor}[thm]{Corollary}

\newtheorem{rem}[thm]{Remark}

\newenvironment{remark}{\begin{rem}}{\end{rem}}
\newenvironment{lemma}{\begin{lem}}{\end{lem}}
\newenvironment{definition}{\begin{defn}}{\end{defn}}
\newenvironment{theorem}{\begin{thm}}{\end{thm}}
\newenvironment{corollary}{\begin{cor}}{\end{cor}}

\newcommand{\init}{\mathsf{Init}}
\newcommand{\query}{\mathsf{Query}}
\newcommand{\ans}{\mathsf{ans}}
\newcommand{\bans}{\mathbf{ans}}
\newcommand{\rs}{r_s}

\newcommand{\brc}{\mathbf{r_c}}
\newcommand{\brs}{\mathbf{r_s}}
\newcommand{\tx}{T}
\newcommand{\btx}{\mathbf{T}}
\newcommand{\calA}{\mathcal{A}}
\newcommand{\calE}{\mathcal{E}}
\newcommand{\G}{\mathbf{G}}
\newcommand{\bD}{\mathbf{D}}
\newcommand{\Sim}{\mathsf{Sim}}
\newcommand{\calF}{\mathcal{F}}
\newcommand{\Fsum}{\mathcal{F}_{\mathrm{sum}}}
\newcommand{\Prep}{\Pi_{\mathrm{rep}}}
\newcommand{\phase}[1]{\textsc{#1}}
\newcommand{\send}{\text{send }}
\newcommand{\sto}{\text{ to }}
\newcommand{\View}{\mathsf{View}}
\newcommand{\Ideal}{\mathsf{Ideal}}

\title{On the (Quantum) Communication Complexity of\\ Secure MPC from a Random Oracle}
\author{
    Alexander Hoover \orcidlink{0009-0003-9818-1419}\footnote{
    Stevens Institute of Technology, Email: \texttt{\href{mailto:ahoover@stevens.edu}{ahoover@stevens.edu}}}
    \and
    Giuseppe Persiano \orcidlink{0000-0001-6579-4807}\footnote{Universit\`a di Salerno, Email: \texttt{\href{mailto:giuper@gmail.com}{giuper@gmail.com}}}
    \and
    Kevin Yeo \orcidlink{0009-0009-4997-6307}\footnote{Google, Email: \texttt{\href{mailto:kwlyeo@google.com}{kwlyeo@google.com}}}
}
\date{}

\begin{document}

{
\hypersetup{urlcolor=black}
\maketitle
}
\begin{abstract}
    Information-theoretic MPC among $n$ parties with $t$ corruptions can securely evaluate a circuit $C$ with $O(|C| \cdot t/(n-t))$ online communication in the preprocessing model and $O(|C| \cdot t/(n-2t))$ communication in the honest majority setting. Damg{\aa}rd, Larsen and Nielsen (CRYPTO 2019) showed that both communication bounds are tight for inner products. Fully homomorphic encryption and other assumptions known to imply public-key cryptography beat both communication bounds, but the communication complexity with symmetric-key cryptography alone is less understood. For MPC with preprocessing, we prove that the same $\Omega(|C| \cdot t/(n-t))$ lower bound holds in the random oracle model, so a random oracle cannot reduce the online communication. For honest majority MPC relative to a random oracle, we present an $\Omega(|C|)$ communication lower bound that is matched by a new construction for degree-2 polynomials (including inner products) when $n = O(\log \secpar)$, beating the information-theoretic bound.

We also study quantum MPC. First, we prove that information-theoretic protocols with quantum communication (and preprocessing) cannot reduce communication compared to their classical counterparts. Furthermore, we extend our lower bound for MPC with preprocessing to protocols with a quantum-accessible random oracle, but use only classical communication. In contrast, with both quantum communication and a post-quantum one-way function, we present a simple two-party protocol that evaluates the two-party instance of these circuits with $\poly(\secpar, \log |C|)$ communication, beating these lower bounds. Finally, such communication savings do not come for free. We prove that there exist circuits $C$ such that every blackbox protocol with $o(|C|)$ quantum communication must perform $\Omega(|C|)$ cryptographic operations.

    \vfill
\end{abstract}

\pagebreak
{
\hypersetup{linkcolor=black}
\pagenumbering{gobble}
\tableofcontents
}
\pagebreak

\pagenumbering{arabic}
\setcounter{page}{2}

\section{Introduction}\label{sec:intro}

Secure multi-party computation (MPC) is a well-studied problem in cryptography that considers $n \ge 2$ parties whose goal is to securely evaluate some function that is represented as a circuit $C$. For security, the goal is to guarantee any subset of $t < n$ corrupted parties cannot learn any more information than may already be learned from the inputs and outputs of the corrupted parties. MPC is a core primitive for cryptography as it lies at the heart of practically-motivated privacy-preserving systems (including ad conversion~\cite{ads_conversion}, federated learning~\cite{CCS:BIKMMP17}, inventory matching~\cite{USENIX:PADBBH23}, password leak check~\cite{USENIX:TPYRKI19} and telemetry gathering~\cite{prio}) as well as being closely connected to other theoretical areas (such as circuit complexity~\cite{FOCS:AppIshKus04}, coding theory~\cite{EC:IshKus04}, communication complexity~\cite{JC:AppVas21} and proof systems~\cite{STOC:IKOS07}).

A central problem for MPC is minimizing the communication complexity necessary to securely evaluate any circuit $C$. In particular, one can consider the amount of additional communication cost necessary for secure evaluation of $C$ using MPC compared to plaintext solutions.
Without security, a trivial protocol is for one party to send their input to another party that can compute the circuit locally. As a simple example, if we consider two-party MPC where one party has a much smaller input, then the plaintext solution has communication sublinear in the circuit size.
Going back to MPC, a natural question is what are the settings and cryptographic assumptions where sublinear circuit communication (similar to the plaintext solution) is possible?
Furthermore, one can also wonder in what MPC settings must the communication cost be linear (or even super-linear) in the circuit size. In our work, we will exclusively focus on circuits whose size is linear in the input size to avoid the more challenging question of whether super-linear circuits are required for specific problems (this remains a long-standing open problem in theoretical computer science).

\heading{Classical MPC.} The communication complexity of MPC has been well-studied in a long line of works considering a wide range of settings and security properties (see~\cite{FOCS:Yao82,STOC:GolMicWig87,STOC:BenGolWig88,STOC:ChaCreDam88,STOC:FraYun92} and references therein as some examples).
For information-theoretically secure MPC (without any cryptographic assumptions)
in the honest majority setting, there have been many constructions that achieve $\tilde{O}(|C| \cdot t/(n-2t))$ communication for any $n > 2t$ (including~\cite{C:DamNie07,EC:DamIshKro10,C:GenIshPol15,C:GIOZ17,C:CCXY18,EC:BGJK21,C:GoyPolSon21}).
It is well known that dishonest majority MPC requires computational assumptions~\cite{STOC:BenGolWig88,STOC:ChoKus89}. 
Nevertheless, one can consider the preprocessing model where parties may communicate to generate correlated randomness prior to receiving their inputs. Using the preprocessing, one can construct MPC protocols that achieve information-theoretically secure online phases 
with $\tilde{O}(|C| \cdot t/(n-t))$ online communication for any $n > t$ (see~\cite{C:DPSZ12,EC:BDOZ11,TCC:IKMOP13,CCS:KelOrsSch16,EC:KelPasRot18,C:GoyPolSon22,EC:EGPSW23,EC:BieYeo25} as examples). 
However, all these dishonest majority MPC protocols with information-theoretically secure online phases require a secure preprocessing phase that must be instantiated with computational assumptions (or even stronger requirements such as trusted setup).
Damgård, Larsen and Nielsen~\cite{C:DamLarNie19} proved communication lower bounds for information-theoretic MPC showing that the above communication costs are tight. In particular, they show that information-theoretically secure evaluation of the inner product problem requires communication $\Omega(|C| \cdot t/(n-2t))$ in the honest majority setting 
(i.e., $n > 2t$) and online communication $\Omega(|C| \cdot t/(n-t))$ in the preprocessing model with $n>t$.

To contrast the above results, one can also consider building MPC using stronger cryptographic assumptions with the goal of reducing communication costs.
For example, if one assumes fully homomorphic encryption~\cite{STOC:Gentry09}, it is trivial to construct an MPC protocol whose communication cost is nearly identical to the plaintext solution (ignoring security parameter factors incurred by encryption) by having one party encrypt their input and send it to the other party for local circuit evaluation.
However, all such homomorphic encryption schemes require computational assumptions that are known to imply public-key cryptography.
Finally, several prior works (such as~\cite{TCC:IKMOP13,C:BoyGilIsh16,EC:Couteau19,EC:CouMey21,C:DIJL23,EC:AbrRoySch24}) obtain $o(|C|)$ communication for circuits that are super-linear in the input size, but all of them have communication linear in the input size and so do not help for the linear-size circuits we consider (see Section~\ref{sec:related}).

Given the above state of affairs for MPC, it leaves open the question of whether one can improve the communication costs of MPC using symmetric-key cryptography. Song and Ye~\cite{EC:SonYe25} show that, with security parameter $\secpar$, one can construct honest majority MPC with $n = 2t + O(1)$ communicating $O(|C| \cdot (\log q + \lambda))$ field elements for fields of order $q \ge 2^{\lambda}$ using a random oracle.
For Boolean circuits, this was further improved to a constant round protocol with $O(|C| \lambda)$ communication in~\cite{C:GLOS25}.
Both these protocols using random oracles achieve communication independent of $n$ and $t$. So, they obtain communication asymptotically impossible for information-theoretic MPC with $n = 2t + O(1)$ for large enough $t = \omega(\lambda)$. However, these communication improvements using random oracles have been limited to the honest majority setting. For dishonest majority MPC, 1-out-of-2 oblivious transfer is complete~\cite{STOC:Kilian88,C:IshPraSah08,EC:BenLin18,EC:GarSri18} and cannot be built from only a random oracle~\cite{STOC:ImpRud89}, so we consider the online communication of MPC with preprocessing, where the online phase may use only the random oracle but the preprocessing may use other computational assumptions. This leaves the following question.
\begin{center}
\emph{What is the communication complexity of honest majority MPC with a random oracle,\\ and can a random oracle reduce the online communication of MPC with preprocessing?}
\end{center}

\heading{MPC with Quantum Resources.} Next, we can consider the MPC problem of classical functionalities where parties have access to quantum computation (and communication). It is already well known that (post-quantum) one-way functions substantially improve MPC with quantum communication. In particular, quantum oblivious transfer (OT) may be built from post-quantum one-way functions~\cite{EC:GLSV21,C:BCKM21b}. In turn, quantum OT can be used to build MPC with quantum communication for classical functionalities using prior works that build protocols from blackbox OT~\cite{STOC:GolMicWig87,STOC:Kilian88}. As an immediate result, it is possible to construct dishonest majority MPC with quantum communication using only one-way functions~\cite{EC:GLSV21,C:BCKM21b}. 
In fact, this can also be extended to enable MPC for quantum functionalities using MPC for classical functionalities~\cite{EC:DGJMS20}.
In contrast, classical dishonest majority MPC requires public-key cryptography, as discussed above.

While quantum communication is known to reduce the computational assumptions required for dishonest majority MPC, the communication complexity remains the same. In particular, these MPC protocols with quantum communication still follow the same template as used for building classical MPC from OT~\cite{STOC:GolMicWig87,STOC:Kilian88}. Thus, they incur the same communication costs (modulo the different communication costs of the underlying quantum and classical OT). This leaves the following question.
\begin{center}
\emph{Can quantum communication or quantum access to a one-way function reduce\\
the communication of MPC below the information-theoretic lower bounds?}
\end{center}

\subsection{Our Contributions}

In this work, we present results for both classical MPC as well as MPC with quantum computation (and communication). Throughout our work, we will consider semi-honest security. For our lower bounds, we will consider weaker adversaries with non-adaptive corruptions and protocols with arbitrary preprocessing algorithms that may use strong computational assumptions (or even unbounded computational power). 
We only restrict the protocol such that the only cryptography used in the online phase is the random oracle.
Furthermore, our lower bounds only assume the adversary observes the online phase. These restrictions only make our lower bounds stronger. For our constructions, we prove them adaptively secure.

\heading{Classical MPC.} Our main goal is to understand the impact of symmetric-key cryptography on the communication complexity of MPC. We start by studying MPC with preprocessing with $t < n,$ including the dishonest majority setting. As a reminder, dishonest majority MPC implies OT that cannot be built in a blackbox way from one-way functions~\cite{STOC:ImpRud89}. However, one can consider MPC where OT is used only in the preprocessing stage. Then, is it possible to use one-way functions to reduce the online communication of MPC compared to MPC protocols with information-theoretically secure online phases? We show that this is impossible in general with the following lower bound, which matches the known lower bound for MPC with preprocessing that obtain information-theoretically secure online phases.

\begin{theorem}[Informal]
\label{thm:informal_mpc_lb}
For every field $\mathbb{F}_q$ of size $q$,
    there exists a circuit $C$ over $\mathbb{F}_q$
    such that every $n$-party MPC with preprocessing using a random oracle that securely evaluates $C$ against $t < n$ corruptions must have $\Omega(|C|\log q \cdot t/(n-t))$ online communication.
\end{theorem}

As a technical lemma for the above lower bound, we also prove a linear lower bound on the communication of single-server PIR with correlated randomness in the random oracle model. In particular, this shows that correlated randomness cannot help reduce communication in single-server PIR.
Without it, any single-server PIR whose total communication is below the database size yields oblivious transfer by a blackbox reduction~\cite{EC:diCMalOst00}, so none exists relative to a random oracle~\cite{STOC:ImpRud89,C:BarMah09}; blackbox constructions from trapdoor permutations likewise need the server to send $\Omega(N)$ bits~\cite{TCC:HaiHocSeg08}. To our knowledge, the bound with correlated randomness was not previously known, as the reduction to OT fails there: there exist constructions for OT with correlated randomness (such as OT extension) that only use one-way functions during online evaluation~\cite{STOC:Beaver96a,C:IKNP03}. As a note, the correlated randomness shared between the client and server must be independent of the database (corresponding to the preprocessing model of MPC). If the correlated randomness may depend on the server's database, there are known single-server PIR constructions with sublinear communication 
using one-way functions~\cite{SP:ZPZS24,CCS:RenMugSun24,EC:HPPY25,EC:WanRen25}.

\begin{theorem}[Informal]
There does not exist a single-server PIR with correlated randomness in the random oracle model for a database with $N$ $b$-bit entries using $o(N \cdot b)$ communication.
\end{theorem}

In contrast, as discussed above, random oracles already reduce the communication of honest majority MPC with $n = 2t + O(1)$ to $O(|C| \cdot \lambda)$ bits for Boolean circuits~\cite{C:GLOS25} and $O(|C| (\log q + \lambda))$ field elements for $q \ge 2^\lambda$~\cite{EC:SonYe25}, independent of $n$ and $t$ and, thus, below the information-theoretic $\Omega(|C| \cdot t/(n-2t))$ bound~\cite{C:DamLarNie19} when $t = \omega(\lambda)$.
Note, our lower bound in Theorem~\ref{thm:informal_mpc_lb} for the setting of $n = 2t + O(1)$ states that
$\Omega(|C| \log q)$ communication is required for circuits
over $\mathbb{F}_q$.
We actually present a slightly stronger lower bound that holds for all values of $n > 2t$.

\begin{theorem}[Informal]
\label{thm:informal_honest_maj_lb}
For every field $\mathbb{F}_q$ of size $q$,
there exists a circuit $C$ over $\mathbb{F}_q$
    such that every $n$-party honest majority MPC with preprocessing using a random oracle that securely evaluates $C$ against $t < n/2$ non-adaptive corruptions must have $\Omega(|C|\log q)$ online communication.
\end{theorem}

As a result, there remains a multiplicative gap of $\Omega(\lambda)$ between the best constructions and our lower bound. Prior works have shown that this $O(\lambda)$ factor per gate is not required, but only in orthogonal settings.
Recent work~\cite{EPRINT:LiSon26} for garbled circuits shows that $O(|C| \log\lambda)$ communication is possible for two-party computation using OT correlations and a random oracle against unbounded adversaries that make $\poly(\lambda)$ random oracle queries.
However, generalizing to $n$-party protocols using standard frameworks~\cite{STOC:BeaMicRog90,C:DamIsh05}
would incur an additional multiplicative $O(n)$ factor.
It is also possible to obtain $O(|C|)$ communication for Boolean circuits with suboptimal honest majority corruption thresholds of $t = (1/2-\gamma)n$ for constant $0 < \gamma < 1/2$ as shown in~\cite{EPRINT:ZYFZ26}. However, this simply matches the prior information-theoretic $\Omega(|C| \cdot t/(n-2t)) = \Omega(|C|)$ lower bound in~\cite{C:DamLarNie19}.
Using preprocessing, one can construct protocols for Boolean circuits with $O(|C|)$ online communication assuming exponentially-sized preprocessing~\cite{TCC:IKMOP13} or the class of layered circuits~\cite{EC:Couteau19}.\footnote{In fact, the protocols in~\cite{TCC:IKMOP13,EC:Couteau19} with preprocessing as well as~\cite{C:BarIsh05} for constant-depth circuits also enable $o(|C|)$ communication when circuits are asymptotically larger than the input size. As a reminder, we mainly focus on circuits whose size is linear in the input size to avoid the more challenging problem of whether a super-linear circuit is required for a specific functionality.}
Finally, $O(|C|)$ communication is possible if we assume circuits with no multiplications as linearly homomorphic secret sharing schemes enable additions for free~\cite{STOC:BenGolWig88} whereas protocols for constant-depth Boolean circuits achieve $O(|C| \cdot \poly\log(|C|))$ communication~\cite{C:BarIsh05}.

To our knowledge, it remains open whether one can obtain $O(|C|\log q)$ communication using random oracles for circuits with any multiplications in the honest majority setting $n = 2t + O(1)$ without preprocessing. We show that there exists an MPC protocol that securely evaluates degree-2 functions over $\mathbb{F}_q$ (equivalently, circuits with multiplicative depth 1) with $O(|C|\log q)$ communication matching our lower bound.

\begin{theorem}[Informal]
\label{thm:informal_deg2}
For $n = O(\log\lambda)$ and any field $\mathbb{F}_q$ of size $q$,
there exists an $n$-party protocol using a random oracle that securely evaluates any circuit $C$ over $\mathbb{F}_q$ for a degree-2 function in which each multiplication gate reads the inputs of $O(1)$ parties, secure against $t < n/2$ adaptive corruptions, with $O(|C|\log q)$ communication.
\end{theorem}

The above shows that our lower bound from Theorem~\ref{thm:informal_honest_maj_lb} in the honest majority setting is asymptotically tight for the class of degree-2 functions (or circuits with multiplicative depth 1).
Our construction using random oracles obtains a provable communication gap compared to its information-theoretic counterpart with $n = 2t + O(1)$ for any super-constant number of parties $t = \omega(1)$ whereas prior works~\cite{EC:SonYe25,C:GLOS25} required larger $t = \omega(\lambda)$.
As a caveat, our construction's communication grows exponentially in $n$ and, thus, only obtains $O(|C|\log q)$ communication when $n = O(\log\lambda)$.

Furthermore, we note that the inner product functionality used in our lower bounds as well as the prior information-theoretic ones in~\cite{C:DamLarNie19} may be computed using degree-2 functions.
In fact, our construction is inspired by constructing a low-communication protocol for a three-party function considered in prior lower bounds~\cite{C:DatPraPra14,C:DamLarNie19}.
We leave it open whether one can extend our $O(|C|\log q)$ communication protocol to arbitrary circuits or obtain a higher $\omega(|C|\log q)$ communication lower bound when assuming no preprocessing.
If we wish to obtain higher $\omega(|C|\log q)$ communication lower bounds for honest majority MPC with $n = O(\log \lambda)$, then we need to consider polynomials of degree 3 or more. However, the required proof techniques would need to understand the circuit structure that leads to much harder long-standing problems in theoretical computer science about circuit depths.

\heading{Quantum MPC.} We also study MPC for classical functionalities where parties have access to quantum computation. First, we consider whether quantum communication alone may be useful to reduce the communication costs of information-theoretic MPC. In particular, we show that this is impossible:

\begin{theorem}[Informal]
For every field $\mathbb{F}_q$ of size $q$, there exists a circuit $C$ over $\mathbb{F}_q$
such that every statistically secure $n$-party MPC with quantum communication that securely evaluates $C$ against $t$ corruptions must have:
\begin{enumerate}
    \item $\Omega(|C|\log q \cdot t /(n-t))$ online communication in the preprocessing model, even with quantum preprocessing, for any $n > t$.
    \item $\Omega(|C|\log q \cdot t/(n-2t))$ communication without preprocessing in the honest majority setting for any $n > 2t$.\footnote{The formal statement (Theorem~\ref{thm:statistical_nopre}) assumes that $s = n - 2t$ is even and divides $n$.}
\end{enumerate}
\end{theorem}

The above lower bounds match the known communication complexity of information-theoretic MPC with classical communication in~\cite{C:DamLarNie19}.
In other words, quantum communication (and computation) does not immediately help reduce communication costs for MPC.

Next, we consider whether 
a quantum-accessible random oracle could help reduce communication costs.
As our next result, we show this is not the case for MPC with quantum computation and classical communication (QCCC).
Similar to our above classical lower bound, we show that a quantum-accessible random oracle does not help reduce the communication costs in the MPC with preprocessing model.

\begin{theorem}[Informal]
Assuming there does not exist a secure two-party key agreement with classical communication using a quantum-accessible random oracle, for every field $\mathbb{F}_q$ of size $q$,
there exists a circuit $C$ over $\mathbb{F}_q$ such that every $n$-party MPC with classical communication and preprocessing using a quantum-accessible random oracle that securely evaluates $C$ against $t < n$ corruptions must have $\Omega(|C|\log q \cdot t/(n-t))$ online communication.
\end{theorem}

As a note, our above lower bound only applies to MPC protocols where the preprocessing step produces classical outputs for each party. This is required as we are focusing on MPC with classical online communication as quantum teleportation enables simulating quantum online communication using quantum preprocessing and classical online communication~\cite{PRL:BBCJPW93}.
Furthermore, we only consider lower bounds for MPC with preprocessing and not the honest majority setting as our $O(|C|)$ classical communication construction in Theorem~\ref{thm:informal_deg2} with $n = 2t + O(1)$ and $n = O(\log \lambda)$ can also be made secure against adversaries with a quantum-accessible random oracle.

We note that there have been several recent works proving impossibilities for key agreement using classical communication with a quantum-accessible random oracle that make our lower bound unconditional in some settings. For MPC protocols with perfect correctness, our lower bound follows from the impossibility of perfectly correct key agreement~\cite{ARXIV:LLLL26}. In the setting of MPC with non-zero error, our lower bounds also follow if we make certain restrictions on the quantum queries to the random oracle as done for prior impossibility results for key agreement proved in~\cite{C:LLLL24,ARXIV:KNVY26}. Using the contrapositive of our theorem, any QCCC MPC protocol with preprocessing that beats this bound yields a QCCC key agreement in the quantum-accessible random oracle model, so reducing the communication of QCCC MPC would require building such key agreement.

Next, we consider MPC with quantum communication (QCQC) where all parties have quantum access to a random oracle. In this case, we show that it is indeed possible to reduce the communication costs
using a post-quantum one-way function. We present a simple two-party MPC protocol built on the QPIR protocol in~\cite{EPRINT:AnaBhaGul26} whose communication is smaller than all the prior lower bounds in certain settings:

\begin{theorem}[Informal]
Consider any circuit $C$ for two parties with input bit sizes of $s_1 \le s_2$ and $m$ output bits. Assuming the existence of a post-quantum one-way function, there exists a two-party MPC protocol with quantum communication implementing $C$ with communication $O(\secpar \cdot m \cdot s_1) + \poly(\secpar)$, independent of $|C|$ and $s_2$,
and computation $2^{s_1} \cdot \poly(\secpar, |C|)$.
\end{theorem}

We note that the communication cost of the above two-party protocol essentially matches the trivial FHE solution where one party sends their input encrypted to the other party. For the setting where one party's input is smaller $s_1 \ll s_2$, we immediately obtain an $o(|C|)$ communication protocol as $|C| = \Omega(s_1 + s_2)$. As a caveat, our construction requires computation exponential in the smaller party's input size. However, we obtain polynomial time protocols if the smaller party's input size is $s_1 = O(\log\secpar)$. Moreover, the two-party instance of the circuit used in our lower bounds, in which both inputs are long, also has a protocol with $\poly(\secpar, \log |C|)$ qubits of communication, as each party only needs to retrieve one bit of the other's input.
Note, prior classical MPC protocols that could obtain $o(|C|)$ communication~\cite{EC:Couteau19} including those with exponential time~\cite{C:BFKR90,STOC:NaoNis01} or preprocessing~\cite{TCC:IKMOP13} have communication costs linear in the total input size. Thus, they use $\Omega(|C|)$ communication for linear size circuits. In contrast, our quantum protocol obtains communication sublinear in the input size using a one-way function (without any preprocessing).
Finally, our above two-party protocol can be easily adapted to the setting of quantum preprocessing and classical online communication using quantum teleportation, showing that the restriction to classical preprocessing in our QCCC lower bound is necessary.

Lastly, we explore whether it is possible to also reduce the computational costs of MPC protocols using quantum communication. Even though our above protocol obtains $o(|C|)$ communication, we still require at least $\Omega(|C|)$ computation across both parties (and larger in many settings). In fact, our protocol requires a large number of one-way function invocations. We show that this is inherent by showing that any small communication protocol requires performing a large number of cryptographic operations. 

\begin{theorem}[Informal]
For every field $\mathbb{F}_q$ of size $q$,
there exists a circuit $C$ over $\mathbb{F}_q$
such that every $n$-party MPC with quantum communication and preprocessing using a quantum-accessible crypto oracle that securely evaluates $C$ against $t = 1$ non-adaptive corruption with $o(|C|\log q)$ communication must have a total of $\Omega(|C|\log q)$ cryptographic qubit operations across all parties.
\end{theorem}

Our lower bound is formally modeled using a crypto oracle~\cite{EC:DujHaj24,EC:LinMooWic25,FOCS:HooPerYeo26} that encompasses any cryptographic computational assumptions that may be simulated by a stateless polynomial time algorithm with access to a private random function. In particular, crypto oracles are able to implement a wide range of powerful cryptographic primitives including random oracles, virtual blackbox obfuscation and fully homomorphic encryption. Even if one uses powerful public-key cryptography to reduce computation, blackbox constructions still require a large number of cryptographic operations.

\section{Technical Overview}

We present a technical overview of all our results. For simplicity, we present everything with respect to Boolean circuits, but our results naturally generalize to circuits over finite fields $\mathbb{F}_q$.

\subsection{Classical MPC}

\noindent{\bf Lower Bounds for MPC with Preprocessing.}
We first overview the techniques introduced by Damg{\aa}rd, Larsen and Nielsen~\cite{C:DamLarNie19} for proving 
lower bounds for MPC with information-theoretically secure online phases. 
We will also highlight the changes
that must be done to adapt the proof for
protocols that use the random oracle model during
the online phase. For simplicity, we consider
the setting of $n$-party MPC with preprocessing
where the adversary may corrupt $t = n - 1$ parties.

At a high level, 
we consider a functionality in which 
each party $P_i$ receives three inputs: a database $x_i \in \{0,1\}^\ell$, an index $z_i \in [n \cdot \ell]$ and an output control bit $b_i \in \{0,1\}$.
To define the output, we consider the concatenated database $D = x_1 \| \cdots \| x_n$ and index $z = z_1 \oplus \cdots \oplus z_n$.
The output of party $P_i$ is defined as $b_i \cdot (D \cdot \mathbb{1}_z)$ where $\mathbb{1}_z$ has a one in the $z$-th entry and zero everywhere else. Note, if $b_i = 0$, the output of $P_i$ is $0$ and the party learns no information about any other inputs. With this function, it turns out one can setup the inputs to make the functionality be
identical to PIR. Suppose party $P_n$ is the PIR client and the remaining $t = n - 1$ parties are the PIR server. We can set the inputs to be $b_j = 0$ and $z_j = 0$ for all $j \ne n$. Then, we set $b_n = 1$, $x_n = 0^{\ell}$ and $z_n \in [(n-1)\ell]$ for party $P_n$. As a result, party $P_n$ obtains entry $b_n \cdot D \cdot \mathbb{1}_z = (x_1 \| \cdots \| x_{n-1})[z_n]$ that is the $z_n$-th entry from an $(n-1)\cdot \ell$ bit database
whereas all other parties receive $0$.
This functionality is identical to PIR
with party $P_n$ as the client and the remaining $n - 1$ parties acting as the server.
Suppose we had a communication lower bound of $\Omega(n \cdot \ell)$ for PIR, this immediately implies that the communication for party $P_n$ in the protocol must be $\Omega(n \cdot \ell)$. We can amplify this by considering $n$ different instances of PIR where each party $P_i$ is the client obtaining an $\Omega(n^2 \cdot \ell)$ communication lower bound
across all $n$ parties. In particular, the communication pattern must hide if any of the $n$ instances are solving PIR as well.
Note, as the circuit $C$ for computing the above inner product functionality has size $|C| = O(n \cdot \ell)$ meaning we obtain a communication lower bound
of $\Omega(|C| \cdot n)$ for any protocol securely evaluating $C$.

To adapt the above to our setting, we must prove a communication lower bound for single-server PIR
with correlated randomness in the random oracle model.
As a note, we only consider the case where the correlated randomness shared between the client and the server is independent of the inputs including the server's database, which corresponds directly to the MPC with preprocessing model. This is necessary if we wish to prove a communication lower bound linear in the database size as there exist PIR with database-dependent preprocessing schemes that obtain sublinear communication using one-way functions~\cite{SP:ZPZS24,CCS:RenMugSun24,EC:HPPY25,EC:WanRen25} or no computational assumptions at all~\cite{C:IshShiWic24}.

It is well known that there already exist blackbox separations between single-server PIR with sublinear communication and random oracles as proven by Di Crescenzo, Malkin and Ostrovsky~\cite{EC:diCMalOst00}. In particular, 
their proof shows that sublinear single-server PIR enables building oblivious transfer (OT) that was already shown to be blackbox separated from random oracles by Impagliazzo and Rudich~\cite{STOC:ImpRud89}. 
However, their work considered single-server PIR without any correlated randomness prior to receiving inputs.
With correlated randomness, there are known constructions of oblivious transfer that use only one-way functions during the online phase (such as~\cite{STOC:Beaver96a,C:IKNP03}).

Therefore, we take a different approach to prove a linear communication lower bound for single-server PIR with correlated randomness. Rather than building OT (or other public-key primitives already known to be impossible with a random oracle), we show how to build a PIR adversary that may break privacy for all
sublinear communication protocols. As a reminder, the PIR adversary observes all inputs to the server including the server's correlated randomness and the input $N$-bit database $D$.

We will build a sequence of adversarial algorithms to obtain our final privacy adversary. First, we consider the setting where the adversary observes the query transcript from a single query. However, we will consider a slightly different situation where we suppose the adversary does not observe the database $D$, but the adversary is given the client's query index $i$. The goal of this adversary is to compute the client's output $D[i]$.
However, we note that the adversary has no trivial way to compute $D[i]$.
First, the adversary cannot observe the client's correlated randomness or internal coin tosses and, thus, cannot directly compute $D[i]$ using the client's reconstruction algorithm. Additionally, the adversary does not have the database $D$ to directly retrieve $D[i]$. Nevertheless, we show that there exists an adversarial algorithm that efficiently computes $D[i]$ by adapting the key exchange attacks of Barak and Mahmoody~\cite{C:BarMah09}. In particular, we prove that if there exists no such adversary that can compute $D[i]$ with high probability given this information, then we can successfully build a secure key exchange protocol in the random oracle model. At a high level, for each fixed index $i \in [N]$, we consider the input-free two-party protocol in which $i$ is hardwired into both parties, the party acting as the client generates and shares the server's correlated randomness, and the party acting as the server locally samples a random database. The two parties execute the PIR protocol, after which the client outputs its answer and the server outputs $D[i]$. Note, an eavesdropper in this key exchange protocol observes exactly the query transcript and the server's correlated randomness, and knows the index $i$, corresponding to the
setting of this adversarial algorithm.

Next, we go back to the original setup of the privacy adversary for PIR that observes the server's correlated randomness, input database $D$ and the query transcript. In particular, the adversary no longer observes the client's queried index $i$.
To break the privacy of a sublinear communication PIR, our adversary will execute the first adversarial algorithm on all indices $i \in [N]$. For each output $a_i$ on input index $i$, the adversary may check its correctness with the database $D[i]$. For a uniformly random $N$-bit database $D$ and a PIR protocol with $o(N)$ online communication, we prove that it is impossible for the first adversarial algorithm to be correct on all $N$ indices as the transcript is too small to encode a random $N$-bit string.
However, by the correctness guarantees of PIR, the client must successfully reconstruct $D[i]$ for the correct index $i$ with high probability. As a result, we obtain an adversary that successfully contradicts privacy as only the indices for which the first adversarial algorithm is correct are likely to be the client's input index.

\heading{$O(|C|)$ Honest Majority MPC.}
Our above result immediately implies an $\Omega(|C|)$ communication lower bound
for any honest majority (even with preprocessing) when $n = 2t + O(1)$.
Prior works such as~\cite{EC:SonYe25,C:GLOS25} show that $O(|C|\lambda)$ communication is sufficient
for Boolean circuits in the same setting.
We begin to close this gap by presenting a protocol with $O(|C|)$ communication matching our lower bound for the restricted circuit class of degree-2 polynomials (including inner products used in our lower bounds and prior works~\cite{C:DamLarNie19}).

To obtain our new construction, we use inspiration from the prior statistical lower bounds.
We start by presenting an improved protocol using a random oracle for a three-party inner product functionality at the core of the proof in~\cite{C:DamLarNie19}. Afterwards, we show how our three-party protocol can be extended to both $n > 3$ parties as well as degree-2 polynomials.
In this three-party functionality, two input parties $P_1$ and $P_2$ hold
inputs $x_1, x_2 \in \bits^\ell$, and the third party $P_3$ must output
$\langle x_1, x_2 \rangle$.\footnote{This can be made symmetric as we did
for the previous bound, but for the technical overview we consider this
asymmetric version.} The lower bound in~\cite{C:DamLarNie19} shows
that any statistically secure protocol for this functionality requires
$P_3$ to communicate $\Omega(\ell)$ bits.

With a random oracle $\calO_\ell \colon \bits^* \to \bits^\ell$, however,
this bound no longer holds. The rough idea is that $P_3$ can generate and
communicate pads for the inputs succinctly. Specifically, $P_3$ samples
random seeds $k_1, k_2 \getsr \bits^\secpar$ and computes pads $r_1 \gets
\calO_\ell(k_1 \| 1)$ and $r_2 \gets \calO_\ell(k_2 \| 2)$. Then, $P_3$
samples a bit $c_1 \getsr \bits$ and computes $c_2 \gets c_1 \oplus
\langle r_1, r_2 \rangle$, so that $c_1, c_2$ are shares of the inner
product of the pads. Finally, $P_3$ sends $(k_i, c_i)$ to party $P_i$,
using only $\secpar + 1$ bits of communication per party.

Then each party $P_i$, for $i \in \{1, 2\}$, recomputes its pad $r_i
\gets \calO_\ell(k_i \| i)$ and sends the masked input $u_i \gets x_i
\oplus r_i$ to the other input party $P_{3-i}$. Now $P_1$ and $P_2$ can
each compute a one-bit share of the inner product as follows:
\begin{align*}
    s_1 &= \langle r_1, u_2 \rangle \oplus c_1
         = \langle r_1, x_2 \rangle \oplus \langle r_1, r_2 \rangle \oplus c_1, \\
    s_2 &= \langle u_1, x_2 \rangle \oplus c_2
         = \langle x_1, x_2 \rangle \oplus \langle r_1, x_2 \rangle \oplus c_2.
\end{align*}
Adding the two shares gives $s_1 \oplus s_2 = \langle x_1, x_2 \rangle
\oplus \langle r_1, r_2 \rangle \oplus (c_1 \oplus c_2) = \langle x_1, x_2
\rangle$, where the last equality uses $c_1 \oplus c_2 = \langle r_1, r_2
\rangle$. As $P_3$ knows both pads and could learn $\langle r_1, x_2
\rangle$ from $s_1$, the shares are not sent to $P_3$ but added by a short
honest majority protocol~\cite{STOC:BenGolWig88} in which $P_3$ learns
only their sum, at a cost independent of $\ell$. Privacy against a corrupt
$P_1$ holds since $u_2$ is masked by $r_2 = \calO_\ell(k_2 \| 2)$, which is
uniform unless $P_1$ queries the oracle at the seed $k_2$, an event of
probability at most $Q \cdot 2^{-\secpar}$ for an adversary making $Q$
oracle queries; privacy against $P_2$ is symmetric, and $P_3$ receives
nothing that depends on the inputs.

Note that the protocol as a whole still communicates $\Omega(\ell)$ bits,
as $P_1$ and $P_2$ exchange their masked inputs; however, the proof in~\cite{C:DamLarNie19}
requires that $P_3$ specifically communicates $\Omega(\ell)$ bits, whereas
here $P_3$ communicates only $O(\secpar)$ bits. This already rules out the
proof technique used in prior work. However, we go further to show that a
stronger barrier exists, since the $n$-party functionality used to prove
the bound has protocols which beat the statistical lower bound.

For the $n$-party version of this functionality,
the $n$ parties are split into two halves.
Party $P_{j,i}$ (for $j \in \{1, 2\}$ and $i \in [n/2]$) holds a block
$x_{j,i} \in \bits^\ell$ and an output control bit $b_{j,i}$, and receives
$b_{j,i} \cdot \langle x_1, x_2 \rangle$, where $x_j$ is the concatenation
of the blocks held by the $j$-th half.
Our $n$-party protocol follows the same paradigm as the three-party one
used in tandem with pseudorandom secret sharing~\cite{TCC:CraDamIsh05}.
The circuit for computing this
functionality (without any privacy) has size $|C| = O(n \cdot \ell)$ and, in particular, is a sum of $n/2$ inner products (one per pair of
parties). Each pair runs the protocol above for its own inner product, and
the one-bit shares of all pairs are added by the same cheap honest
majority protocol, whose cost is independent of $\ell$. The difficult part
is that a single party can no longer play the dealer $P_3$, since we must
tolerate $t$ corruptions, and a corrupt dealer together with $P_1$ would
know the pad of $P_2$.

In our protocol $\Prep$, the other $n - 2$ parties, the \emph{helpers},
jointly play the dealer using the replication idea of pseudorandom secret
sharing~\cite{TCC:CraDamIsh05}. Every set $J$ of $n - t - 1$ helpers holds
a seed for each pad, each pad is the XOR of the pads $r_J$ expanded from
its seeds, and the inner product of the two pads is the XOR of the cross
terms $\langle r_J, r'_{J'} \rangle$. Any two such sets intersect, so some
helper holds both seeds of each cross term and can split it into two
shares for the pair, as $P_3$ did above; whereas any $t - 1$ helpers miss
some set, so one pad of an honest party stays hidden from the adversary,
as $r_2$ did above. The $\binom{n-2}{t-1}$ seeds per pad are
$\poly(\secpar)$ when $n = O(\log \secpar)$, giving $n\ell + \poly(\secpar)$
bits of communication in total, whereas every statistically secure
protocol communicates $\Omega(n\ell \cdot t/(n-2t))$ bits~\cite{C:DamLarNie19}.

Finally, we extend our protocol to degree-2 polynomials.
Nothing in $\Prep$ is specific to inner products or to bits. Over any
field $\FF$, with additive shares in place of XOR, the terms of a
degree-2 polynomial that involve two parties form an inner product of
two vectors that the parties compute locally, so $\Prep$ evaluates any
degree-2 polynomial by running the pair protocol once per pair of
parties whose inputs are multiplied, with $O(|C|)$ elements of $\FF$ plus
$\poly(\secpar)$ bits for every circuit $C$ in which each multiplication gate
reads the inputs of $O(1)$ parties (Section~\ref{sec:counter_example}). It is
private even against adaptive corruptions, as the simulator can program
the oracle at a seed that only honest parties hold. In another regime,
$t/(n-2t) = \omega(\secpar)$, OT extension~\cite{C:IKNP03} can replace the seeds at
$O(\secpar)$ bits per pad bit, which is how~\cite{EC:SonYe25,C:GLOS25}
beat the bound of~\cite{C:DamLarNie19} in that regime; for $n =
\omega(\log\secpar)$ with $t/(n-2t) = O(\secpar)$ the question remains open.

\subsection{Quantum MPC}

Complementing our classical MPC results, we also prove lower bounds for
MPC for classical functionalities built using quantum resources. In these results, we distinguish between:
\begin{itemize}[nosep]
\item quantum \emph{computation}, which enables parties to query their oracle
$\calO$ in superposition and perform other quantum operations,
\item quantum \emph{communication}, which allows parties to send quantum states
between each other, and
\item quantum \emph{preprocessing}, which allows parties to share a joint
(entangled) quantum state $\rho_{\mathrm{pre}}$ before the inputs are known.
\end{itemize}
Each of these resources gives
some additional degree of freedom in designing quantum MPC protocols. In the quantum setting, we use the register model
of~\cite{EC:ABCGLS19} where each party holds an input, output and preprocessing (if needed) register (see Section~\ref{sec:prelim_models} for more details). We consider \emph{specious} security~\cite{C:DupNieSal10}, the quantum analog of semi-honest security, where the adversary
follows the protocol but defers every measurement and keeps every
register it would discard, so its state after any round is that of the
\emph{purified} protocol $\bar\Pi$.
Privacy requires that this state,
jointly with a reference system $R$ that may be entangled with the
inputs, should be simulatable from the corrupted parties'
inputs, preprocessing, and outputs
(see Section~\ref{sec:prelim_quantum} for more details).

\heading{Statistical security lower bounds.}
For our first result, we study whether quantum communication can help improve costs of statistically-secure MPC. We show that this
is impossible by showing
the same communication lower bounds of Damg{\aa}rd, Larsen and
Nielsen~\cite{C:DamLarNie19} extend for this broad class of quantum MPC.
The two statistical facts that are used in their lower bounds are: (1)
a statistically secure PIR with correlated randomness requires $\Omega(N)$
communication for an $N$-bit database and (2) the output party of the inner
product protocol requires communicating $\Omega(\ell)$ bits. However, neither
of these facts obviously extends to the quantum setting, as quantum
information has different properties than its classical counterpart.
In fact, quantum preprocessing often allows one to do things that are not
classically possible, such as teleportation~\cite{PRL:BBCJPW93} and
superdense coding~\cite{PRL:BenWie92}. For example, in a relaxed threat model
where the server's database register is classical, quantum PIR (QPIR) does
exist using only $O(\log N)$ qubits~\cite{QIC:KLLR16,EC:ABCGLS19}.

Despite the existence of QPIR in this relaxed model, there are lower bounds
showing that statistically secure QPIR against specious servers requires
$\Omega(N)$ qubits of
communication~\cite{FOCS:Nayak99,FOCS:JaiRadSen02,JC:BauBro15}. Most
recently, this has been strengthened to QPIR even with any amount of quantum
preprocessing between the client and server~\cite{EC:ABCGLS19}
(Theorem~\ref{thm:abcgls}). Therefore, we directly appeal to this known
result together with the reduction of the dishonest majority setting to PIR
used above. Since this reduction works without issue in the quantum setting,
we get as an immediate corollary that any $n$-party protocol with quantum
preprocessing and quantum communication for the circuit $C$ of
Theorem~\ref{thm:main_mpc} that is statistically secure against $t$
corruptions must communicate $\Omega(|C| \cdot t/(n-t))$ qubits
(Theorem~\ref{thm:statistical_mpc}).

The second statistical fact relies on Shannon entropy in the original
classical proof, which we recall here.
Consider the protocol where parties $P_1$ and
$P_2$ hold $x_1, x_2 \in \bits^{\ell}$ and $P_3$ must output
$\langle x_1, x_2 \rangle$. Let $\tx_{a,b}$ be the transcript of the messages
exchanged between parties $P_a$ and $P_b$. Then, for uniformly random
$\mathbf{x}_1$ and any fixed $x_2$, privacy against $P_2$ gives
$H(\mathbf{x}_1\mid \btx_{1,2}, \btx_{2,3}) \ge
    \ell - \negl(\secpar)$.
We combine this with the fact that, by privacy against $P_1$, any transcript
is consistent with $x_2 = e_j$ for every $j \in [\ell]$, and by correctness
the completed execution outputs $\langle \mathbf{x}_1, e_j \rangle =
\mathbf{x}_1[j]$, so the full transcript determines $\mathbf{x}_1$. This gives
$H(\mathbf{x}_1 \mid \btx_{1,2}, \btx_{1,3}, \btx_{2,3}) \le \negl(\secpar)$.
Subtracting these inequalities, it follows that
$H(\btx_{1,3}) \ge \ell - \negl(\secpar)$, which lower bounds the
communication of $P_3$.

In the quantum setting, there are no transcripts to condition on, as message
registers move between parties, and one must work instead with von Neumann
entropy, which has different conditioning properties. Fortunately, however,
we can obtain an analog of the two inequalities above. We run the purified
protocol with $X_1$ maximally entangled with a reference register $R$, so
that measuring $R$ yields a uniformly random $\mathbf{x}_1$, and let $A_a$ be
all registers held by $P_a$ at the end. Privacy against $P_2$ then gives
$H(\mathbf{x}_1 \mid A_2) \ge \ell - 2 - \negl(\secpar)$.
For the second inequality, instead of considering setting $x_2 = e_j$ to extract
$\mathbf{x}_1$, we can use Uhlmann's theorem. Privacy against $P_1$ says its
view is nearly the same for $x_2 = 0^\ell$ and $x_2 = e_j$, which implies
that there is a
unitary on $A_2 A_3$ mapping one execution to the other, after which $P_3$
outputs $\mathbf{x}_1[j]$ by correctness. This gives $H(\mathbf{x}_1 \mid A_2 A_3) \le \negl(\secpar)$.
The difference of the two is $I(\mathbf{x}_1 ; A_3 \mid A_2)$, which is
bounded by $O(H(A_3))$. This shows then that $H(A_3) = \Omega(\ell)$, which
lower bounds the number of qubits sent or received by
$P_3$, as $P_3$ starts in a pure state (Lemmas~\ref{lem:comm_entropy}
and~\ref{lem:three_party}).
The honest majority bound then follows exactly as in~\cite{C:DamLarNie19}.
For $n = 2t + s$ parties, each group of $s$ parties is cast as $P_3$ and the
remaining $t$ parties of each half as $P_1$ and $P_2$ with $t\ell$-bit inputs,
so each group sends or receives $\Omega(t\ell)$ qubits. Summing over the
$n/s$ groups (for $s$ even and $s \mid n$) gives $\Omega(n \cdot t\ell/s) = \Omega(|C| \cdot t/(n-2t))$
qubits (Theorem~\ref{thm:statistical_nopre}).

\heading{Quantum computation with classical communication.}
Our next quantum lower bound extends our own lower bound on the
blackbox usage of a random oracle (Theorem~\ref{thm:main_mpc}) to protocols
which may query the oracle in superposition, i.e., the QCCC setting.
Here, we are still concerned
with protocols which use classical preprocessing and communication (and
therefore the classical semi-honest security definition, as the inputs,
preprocessing and messages of the corrupted parties are all classical
strings). However, both the protocol and the adversary are allowed to query
$\calO$ in superposition. Unfortunately, the well-known key agreement lower
bounds~\cite{STOC:ImpRud89,C:BarMah09} do not apply to protocols with this
quantum access, since these bounds use rewinding heavily.

Instead, our result relies on more recent works proving that quantum access
to a random oracle cannot be used to build key agreement with classical
communication, under some restrictions. This is the subject of a line of
work extending these famous impossibility results to the quantum
setting~\cite{C:ACCFLM22,C:LLLL24,EPRINT:LLLL25,ARXIV:KNVY26,ARXIV:LLLL26}.
In particular, we rely on the most recent bound of Li, Li, Li and
Liu~\cite{ARXIV:LLLL26}, which shows that any perfectly correct key agreement
protocol with classical communication in the quantum random oracle model can
be broken by an eavesdropper making polynomially many classical queries to
$\calO$ (Theorem~\ref{thm:qccc_ke_attack}). With this impossibility, we can
construct a key agreement protocol from a too efficient secure MPC protocol,
as done in our classical bound. Fortunately, this reduction is entirely
compatible with quantum computation, as every other step of our proof only
runs the eavesdropper on the classical transcript and correlated randomness,
without any rewinding.

This gives the immediate corollary that, relative to
a quantum-accessible random oracle, any perfectly correct QCCC protocol with
classical preprocessing that securely computes the circuit $C$ of
Theorem~\ref{thm:main_mpc} against $t$ corruptions must have
$\Omega(|C| \cdot t/(n-t))$ online communication
(Corollary~\ref{cor:qccc}). The restriction to perfect correctness is
inherited from~\cite{ARXIV:LLLL26}, and any attack on QCCC key agreement
with imperfect correctness, such as those of~\cite{C:LLLL24,ARXIV:KNVY26}
for restricted query patterns, carries over to our bound
(Remark~\ref{rem:qccc_imperfect}). Conversely, any QCCC MPC protocol with
preprocessing that beats this bound yields a QCCC key agreement protocol in
the quantum random oracle model, so improving the communication of such
protocols would require building QCCC key agreement.

\heading{Quantum communication or preprocessing.}
Given the above extension, it is natural to ask whether we can extend our
result to allow either quantum communication or preprocessing. Unfortunately,
there is no hope for the same proof pathway, as key agreement can be built
unconditionally using quantum communication~\cite{IEEE:BenBra84}. The same
holds one level down: classically, sublinear PIR implies oblivious
transfer~\cite{EC:diCMalOst00}, whereas with quantum communication oblivious
transfer exists from one-way functions~\cite{EC:GLSV21,C:BCKM21b}.

In fact, we go further and give a two-party protocol with no preprocessing
and quantum communication\footnote{Note that this also rules out such a
bound for quantum preprocessing with classical communication, as the quantum
communication can be replaced by classical communication using quantum
teleportation over preprocessed EPR pairs~\cite{PRL:BBCJPW93}.} that beats
our bound. Our starting point is the recent single-server quantum PIR of
Ananth, Bhardwaj and Gulati~\cite{EPRINT:AnaBhaGul26}, which is built from
one-way functions and uses only $O(\secpar \log N)$ qubits of communication
(Theorem~\ref{thm:qmpc_cpir}). The client sends the two keys of a
distributed point function~\cite{EC:GilIsh14,EC:BoyGilIsh15} in
superposition, the server applies a phase given by the inner product of $D$
with its evaluated share, and the client recovers $D[i]$ from the relative
phase. Given such a PIR, a two-party protocol for any circuit $C$ with one
short input is almost immediate. Suppose Alice holds $x \in \bits^a$ and Bob
holds $y$. Bob samples a mask $r$, computes the truth table $T[w] = C(w, y)
\oplus r$ for all $w \in \bits^a$, and lets Alice retrieve $T[x]$ as the
client of the PIR; the parties then exchange $r$ or $T[x]$ depending on who
should learn the output. Bob's view is independent of $x$ by the privacy of
the PIR, and Alice's final state depends only on $x$, her coins and $T[x]$.
The communication is $O(\secpar \cdot a)$ qubits per output bit, independent
of $|C|$ and $|y|$, and the computation is $2^a$ evaluations of $C$, so the
protocol is efficient whenever $a = O(\log \secpar)$
(Theorem~\ref{thm:qmpc_2pc}).
Next, we upgrade the
PIR to a symmetric PIR~\cite{STOC:GIKM98}
to ensure symmetric privacy and Alice only learns the output. To do this, we use the techniques
of~\cite{STOC:NaoPin99,TCC:FIPR05} where Bob masks each row $w$ with
$\bigoplus_j \calO(\kappa_{j, w_j} \| j \| w)$ for keys $\kappa_{j, 0},
\kappa_{j, 1}$, and Alice receives the $\log N$ keys $\kappa_{j, i_j}$ by
oblivious transfer, in effect an oblivious PRF evaluation, so that she can
unmask row $x$ and no other. This gives a symmetric PIR with $\poly(\secpar,
\log N)$ qubits of communication and no preprocessing
(Corollary~\ref{cor:qmpc_spir}).

This protocol does not only beat our bound for lopsided circuits. The
two-party instance of the very circuit used in Theorem~\ref{thm:main_mpc}, where
both parties hold $\ell$-bit inputs, has a protocol with $\poly(\secpar,
\log \ell)$ qubits of communication, as each party only needs to retrieve
one bit of the other's database by PIR before a constant-size two-party
computation produces the outputs (Corollary~\ref{cor:qmpc_mpc}). Together
with the previous two results, this shows that for this circuit neither
quantum preprocessing nor a quantum-accessible random oracle reduces the
communication of MPC on its own, but the two together do. Whether every
circuit admits such a protocol with polynomial computation in the input size remains an
intriguing open question.

\heading{Blackbox computation lower bound.}
Although the protocol above beats our communication lower bound, it still
requires \emph{computation} that is linear in the circuit size. But even
more than that, the PIR of~\cite{EPRINT:AnaBhaGul26} evaluates its
distributed point function on every database index, which requires
$\Theta(N)$ (superposition) queries to its random oracle. Our final quantum
result proves that this is in fact optimal by extending techniques
from~\cite{EC:DujHaj24,EC:LinMooWic25,FOCS:HooPerYeo26}.

We consider the broad class of \emph{crypto oracles}, following the syntax
of~\cite{EC:LinMooWic25,FOCS:HooPerYeo26}. A crypto oracle is a stateless,
efficient algorithm $\mathcal{B}^R$ with access to a private random function
$R$ (Definition~\ref{def:crypto_oracle}). This captures the blackbox usage
of all sorts of cryptography, from ``weak'' cryptography like random oracles
to idealized fully homomorphic encryption, virtual blackbox obfuscation and
generic multilinear groups. In previous works, it was shown that this sort
of blackbox cryptography has limited use for cryptographic objects like
doubly-efficient PIR~\cite{EC:LinMooWic25}, pseudorandom error-correcting
codes~\cite{TCC:GarGunWan25}, and PIR with client-side
preprocessing~\cite{FOCS:HooPerYeo26}.

To our knowledge, we are the first to consider \emph{quantum access} to
these crypto oracles. In particular, we extend the result
of~\cite{EC:DujHaj24}, which showed that any classical single-server PIR
with sublinear communication requires $\Omega(N)$ calls to its oracle, which
they refer to as ``public-key'' operations, regardless of the number of
symmetric-key operations. Our bound similarly shows that any PIR with
quantum communication and $o(N)$ communication, even with quantum
preprocessing, must have the server query its crypto oracle on $\Omega(N)$
qubits in total. Of course, the server must read the whole database in any
PIR, so the content of the bound is that a linear number of these
operations must be cryptographic. Interestingly, in the classical setting,
random oracle calls can be ignored because they cannot give sublinear PIR
on their own~\cite{EC:diCMalOst00}, whereas in the quantum setting they are
exactly what enables sublinear communication. Our bound shows that the
$\Theta(N)$ calls of~\cite{EPRINT:AnaBhaGul26} are optimal up to
$\poly(\secpar)$ factors, even though these are not traditional
``public-key'' operations.

To prove this result, we use the compressed oracle
technique~\cite{C:Zhandry19}. This lets us simulate the private random
function $R$ of the crypto oracle perfectly using limited space. Then, since
$\mathcal{B}$ itself is deterministic and efficient, we can simulate the
entire crypto oracle efficiently by running $\mathcal{B}^R$ coherently and
uncomputing its intermediate state. This allows one to ``compile out'' all
the oracle calls made by the server by having the client play the role of
the crypto oracle: for each query, the server sends its input and output
registers to the client, who applies the simulated oracle to them and sends
them back. As the simulation is perfect, the resulting PIR has the same
correctness and the same privacy, but no oracle. Here, it is crucial that,
as in all our lower bounds, privacy is required against computationally
unbounded adversaries that are only bounded in their number of oracle
queries, so the compiled PIR is statistically private. Its server
communication is that of the original PIR plus the qubits of the server's
queries, and appealing to the statistical lower bound
of~\cite{EC:ABCGLS19}, which allows the prior entanglement, we get that one
of the two must be $\Omega(N)$.

The bound for MPC (Theorem~\ref{thm:qmpc_comp_lb}) follows by the
reduction of Theorem~\ref{thm:main_mpc} with $t = 1$. For the circuit of
that theorem, every party $P_i$ can be cast as the server of a PIR with
its input $d_i$ as the database and the other $n - 1$ parties as the
client: the inputs can be set so that one of them receives any entry of
$d_i$ of its choice while $P_i$ receives no output, so the privacy of the
protocol against the corruption of $P_i$ alone is the privacy of the PIR
against its server. By the bound above, the qubits sent by $P_i$ plus its
cryptographic qubit operations are $\Omega(\ell)$. Both quantities are
fixed by the protocol, as the message registers and the oracle calls of a
party do not depend on the inputs, so the bounds for the $n$ parties hold
at once and sum, each message counted once by its sender: the total
communication plus the total number of cryptographic qubit operations is
$\Omega(n \ell) = \Omega(|C|)$. As a result, at least one of the two must also be $\Omega(|C|)$.

\section{Related Work}
\label{sec:related}

In this section, we provide a more detailed overview of relevant prior works
on secure multi-party computation (MPC).
Specifically, we survey MPC built with information-theoretic security, from symmetric-key assumptions,
lower bounds for MPC, and quantum MPC. 

\heading{Information-Theoretic MPC.}
For the honest majority setting, there is a long line of work that builds information-theoretic MPC~\cite{STOC:BenGolWig88,STOC:ChaCreDam88}
including recent works~\cite{C:DamNie07,EC:DamIshKro10,C:GenIshPol15,C:GIOZ17,C:CCXY18,EC:BGJK21,C:GoyPolSon21,CCS:EGPS22} that achieve optimal communication (up to log and security parameter factors) matching~\cite{C:DamLarNie19}. It is known that dishonest majority MPC requires computational assumptions~\cite{STOC:BenGolWig88,STOC:ChoKus89}. However, it is possible to obtain information-theoretically secure online phases assuming preprocessing~\cite{C:Beaver91b}. A long line of work has culminated in protocols that obtain optimal (online) communication in this model including~\cite{C:DPSZ12,EC:BDOZ11,TCC:IKMOP13,CCS:KelOrsSch16,EC:KelPasRot18,C:GoyPolSon22,EC:EGPSW23,EC:BieYeo25} matching the lower bounds in~\cite{C:DamLarNie19} up to log and security parameter factors.
Our lower bound in Theorem~\ref{thm:main_mpc} shows that the above preprocessing protocols stay
tight when the online phase may also use a random oracle.

\heading{Honest majority MPC from symmetric-key assumptions.}
We summarize the communication of honest majority MPC
for $n = 2t + s$ parties.
Without computational assumptions, $O(|C| \cdot
t/s)$ field elements suffice~\cite{C:DamNie07,C:GoyPolSon21} and are
necessary~\cite{C:DamLarNie19}, so $O(|C|)$ is reached only for $s =
\Omega(n)$, where Zeng et al.~\cite{EPRINT:ZYFZ26} achieve $O(1)$ bits per
gate of a Boolean circuit. For $s = O(1)$, the protocols below this bound
use a random oracle. Goyal et al.~\cite{C:GLOS25} evaluate Boolean circuits
with $O(|C| \secpar)$ bits in constant rounds, and Song and
Ye~\cite{EC:SonYe25} evaluate arithmetic circuits over a field with $\log q
\ge \secpar$ with $O(\log q + \secpar)$ elements per gate, both with
malicious security; from one-way functions alone, Li and
Song~\cite{TCC:LiSon25} need $O(|C| n \secpar)$ bits. The two random oracle
protocols produce their correlated randomness at a cost per gate that does
not depend on $n$, Song and Ye by OT extension, and so beat the
information-theoretic bound once $t/s$ exceeds that cost. Our lower bound
shows that the factor $|C|$ in all of them is necessary, and our protocol
shows that for degree-2 polynomials and $n = O(\log \secpar)$ nothing
else is, as $O(|C|)$ elements suffice\footnote{This requires that each multiplication gate reads the inputs of $O(1)$ parties, as in an inner product (Corollary~\ref{cor:rom_quad_circuit}).} and the information-theoretic bound is
beaten by a factor of $t/s$. Whether general circuits need the remaining factor of $\secpar$ from a
random oracle alone is open. The protocol combines the pseudorandom secret
sharing of Cramer, Damg{\aa}rd and Ishai~\cite{TCC:CraDamIsh05}, who already
evaluate degree-2 polynomials from predistributed keys but send every
masked input to every party, with the circuit randomization of
Beaver~\cite{C:Beaver91b}.
As a note, MPC for degree-2 functionalities is known to be round-complete for general MPC~\cite{EC:AppBraTsa19}, but the reduction incurs a $\poly(|C|)$ communication blowup (even with one-way functions). Furthermore, the notion of pseudorandom secret sharing has been extended to pseudorandom correlated generators in later works enabling various correlations (see~\cite{CCS:BCGI18,C:BCGIKS19,FOCS:BCGIKS20} to list as some examples).

\heading{Lower Bounds for MPC.} Several prior works have considered communication lower bounds for MPC including~\cite{Kushelvitz92,STOC:FraYun92,ChoKus93,C:DatPraPra14} as some examples. In~\cite{EC:DNOR16}, a lower bound was shown of the form $\Omega(t \cdot n)$ assuming $t$ corrupted parties for information-theoretic MPC. A lower bound growing linearly in the circuit size was presented in~\cite{C:DNPR16} for the restricted set of protocols that perform gate-by-gate computation. Tight lower bounds for both MPC with preprocessing and honest majority MPC (without preprocessing) in the information-theoretic settings were presented in~\cite{C:DamLarNie19}. These were strengthened in the perfect security setting in~\cite{ITC:DamLiSch21}.

\heading{Sublinear Communication MPC.}
A long line of work builds MPC with communication sublinear in the circuit size when the circuit is super-linear in the input size, so that the communication depends on the input size but does not grow with the circuit size. Fully homomorphic encryption~\cite{STOC:Gentry09} achieves this trivially, and MPC can also be built from somewhat homomorphic encryption~\cite{C:DPSZ12} for circuits of limited multiplicative depth, such as~\cite{TCC:BonGohNis05,C:AguGabHer10,EC:CHKV25}. Other such protocols use exponential computation~\cite{C:BFKR90,STOC:NaoNis01}, constant-depth circuits~\cite{C:BarIsh05}, preprocessing~\cite{TCC:IKMOP13,EC:Couteau19} or homomorphic secret sharing from public-key assumptions~\cite{C:BoyGilIsh16,EC:CouMey21,EC:BoyCouMey23,C:DIJL23,EC:AbrRoySch24,C:CouKum24,TCC:CCMN24,TCC:CouKumYe25}. All of these require communication at least linear in the input size, so for the linear-size circuits we consider they require $\Omega(|C|)$ communication.

\heading{Quantum MPC and PIR.}
It is known that information-theoretic dishonest majority MPC (without preprocessing) still does not exist even with quantum communication~\cite{Lo97}.
Quantum communication changes the assumptions behind
MPC but, so far, not its communication. Oblivious transfer, and with it
dishonest majority MPC~\cite{STOC:GolMicWig87,STOC:Kilian88}, exists with
quantum communication from post-quantum one-way
functions~\cite{EC:GLSV21,C:BCKM21b}, in contrast to the classical
setting~\cite{STOC:ImpRud89}.
However, from quantum computation alone,
with classical communication, there is a line of work extending this
impossibility of key
agreement~\cite{C:ACCFLM22,C:LLLL24,ARXIV:KNVY26,ARXIV:LLLL26}, currently
under restrictions on the oracle queries or on correctness.
Prior works have proved communication lower bounds for private simultaneous quantum messaging setting (i.e., non-interactive information-theoretic quantum MPC) including~\cite{ITC:KawNis21,EPRINT:GMPY26}.

We also make use of recent quantum PIR results centrally in our quantum
MPC lower bounds and constructions.
Quantum PIR with statistical privacy against a
specious server needs $\Omega(N)$ qubits of
communication, even with prior entanglement~\cite{FOCS:Nayak99,JC:BauBro15,EC:ABCGLS19}.
In the other direction, the quantum PIR of Ananth, Bhardwaj and
Gulati~\cite{EPRINT:AnaBhaGul26} from one-way functions (using the 2-to-1 server quantum compilation in~\cite{STOC:KerdeWol03} applied to~\cite{EC:GilIsh14}) is what makes our
two-party protocol possible.

Finally, our computational lower bound relies on an extension of 
the crypto oracle model~\cite{EC:DujHaj24,EC:LinMooWic25,FOCS:HooPerYeo26}.
We are the first to consider this model in the quantum setting, and we obtain
computational lower bounds for quantum MPC. As another corollary, this shows
the quantum PIR of Ananth, Bhardwaj and Gulati~\cite{EPRINT:AnaBhaGul26} is
computationally optimal.

\section{Preliminaries}\label{sec:prelim}

We write $\secpar$ for the security parameter and $[n] = \{1, \ldots, n\}$.
For a string $D$, we denote its $i$-th entry by $D[i]$, and $\|$ denotes
concatenation. Bold symbols such as $\bD$ denote random variables, and $x
\getsr X$ samples $x$ uniformly from a finite set $X$. Background on
classical and quantum information and the entropy facts used in our proofs
are deferred to Appendix~\ref{app:prelim}.

\subsection{Oracles and Computational Models}
\label{sec:prelim_models}

\heading{Random oracles.} We consider cryptographic protocols that are built
in the random oracle model, where every party is given access to an oracle
$\calO(\cdot)$ such that every output $\calO(x)$ is a uniformly random
string. Lower bounds in the random oracle model rule out every protocol that
uses a one-way function or a pseudorandom function in a blackbox way.

\begin{definition}[Random Oracle Model]
    A random oracle is a function $\calO\colon \bits^* \rightarrow \bits^\infty$ such that, for every $x \in \bits^*$, $\calO(x)$ is an independent sequence of uniformly random bits. For $\ell \ge 1$ we write $\calO_\ell(x) \in \bits^\ell$ for the first $\ell$ bits of $\calO(x)$.
\end{definition}

This is the infinite-output oracle of prior works such
as~\cite{CCS:BelRog93,C:BarMah09}, where a protocol truncates each output to
the length it needs. As we only consider efficient
cryptographic protocols, the largest length of inputs and outputs to the
random oracle will always be bounded by $\poly(\secpar)$ bits. So, in the
quantum random oracle model (QROM)~\cite{AC:BDFLSZ11}, we identify $\calO$
with the uniformly random function $\calO_m: \bits^{\le p} \to \bits^m$ with
$p, m = \poly(\secpar)$, accessed through the unitary $\ket{x}\ket{y} \mapsto
\ket{x}\ket{y \oplus \calO_m(x)}$. We call an oracle accessed in this way
\emph{quantum-accessible}. We will typically refer to $\ket{x}$ as the input register and $\ket{y}$ as the output register. Note, both must be supplied to the oracle by the caller. Parties are quantum polynomial
time and make at most $\poly(\secpar)$ superposition queries. Adversaries are
computationally unbounded and are restricted only in their number of
superposition queries, which is $\poly(\secpar)$
(Sections~\ref{sec:prelim_classical} and~\ref{sec:prelim_quantum}).

\heading{Crypto oracles.} We also consider the stronger crypto oracle model,
which encompasses blackbox cryptography that can be efficiently computed by
a stateless algorithm with access to a private random function
(see~\cite{EC:DujHaj24,EC:LinMooWic25,FOCS:HooPerYeo26}).

\begin{definition}[Crypto Oracle]
\label{def:crypto_oracle}
A crypto oracle is a function $\mathcal{B}^R$
where $\mathcal{B}$ is a stateless, polynomial time, deterministic algorithm
with oracle access to a secret random function $R: \bits^a \rightarrow \bits^b$, where $a, b = \poly(\secpar)$ for each fixed security parameter $\secpar$.
\end{definition}

As for a random oracle, a \emph{quantum-accessible} crypto oracle is accessed
through the unitary $\ket{x}\ket{y} \mapsto \ket{x}\ket{y \oplus
\mathcal{B}^R(x)}$, and no party has direct access to $R$
where $\ket{x}$ and $\ket{y}$ are the input and output registers.

\heading{Classical protocols.} We consider classical $n$-party protocols for
a function $f = (f_1, \ldots, f_n)$. The inputs and outputs of $f$ are
vectors over a finite field $\FF$, and $f$ is computed by an
arithmetic circuit $C$ over $\FF$ of fan-in $2$, whose size $|C|$
is its number of gates. So $C$ has at most $2|C|$ inputs, and Boolean
circuits are the case $q = 2$. We count communication in bits and also
state it in field elements, one element being $\lceil \log q \rceil$
bits, so that a bound of $\Theta(|C|)$ elements is one of $\Theta(|C| \log
q)$ bits. Without preprocessing, these are as
in~\cite[Section~2.2]{C:DamLarNie19}: the parties have independent random
tapes and are connected by private point-to-point channels. In addition,
every party has access to $\calO$. In a protocol \emph{with preprocessing},
a trusted dealer making $\poly(\secpar)$ queries to $\calO$ samples correlated randomness $(r_1,
\ldots, r_n)$ before the inputs are chosen and gives $r_a$ to party $P_a$.
We assume the correlated randomness is distributed securely, so the
adversary observes no information about its sampling or distribution beyond
the randomness of the corrupted parties. As we are proving lower bounds,
this only makes our results stronger. The communication is the total number
of bits sent after the preprocessing, that is, in the online phase; it may
depend on the random tapes, and the protocol has \emph{expected
communication} $c$ if its expectation is at most $c$ for every input. A
protocol has \emph{error probability} $\epsilon$ if, for every input $x$ and
every $a \in [n]$, party $P_a$ outputs $f_a(x)$ with probability at least $1
- \epsilon$ over $\calO$ and the random tapes, and it is \emph{perfectly
correct} if, for every input $x$ and every $\calO$, each party $P_a$ outputs
$f_a(x)$.

\heading{QCCC protocols.} In the quantum-computation classical-communication
(QCCC) model of~\cite{C:ACCFLM22}, the parties are quantum algorithms that may query $\calO$ in superposition, while
the correlated randomness, every message and the outputs remain classical.

\heading{Quantum preprocessing and communication.} An $n$-party protocol
$\Pi$ with quantum preprocessing for $f = (f_1, \ldots, f_n)$ is the
$n$-party analogue of a two-party protocol in the sense
of~\cite[Definition~2.1]{EC:ABCGLS19}. Party $P_a$ holds an input register
$X_a$, a preprocessing register $W_a$ and an output register $Y_a$. A trusted
dealer prepares an arbitrary state $\rho_{\mathrm{pre}}$ on $W_1 \otimes
\cdots \otimes W_n$ before the inputs are chosen. When the parties have
access to an oracle $\calO$, so does the dealer, and $\rho_{\mathrm{pre}}$
may depend on $\calO$. The dealer and the parties make at most $\poly(\secpar)$
queries to $\calO$. The protocol then runs for
a fixed number of rounds, in each of which every party applies a quantum
operation to its registers and sends message registers of fixed dimensions
to the other parties. The communication is the sum of $\log \dim$ over all
message registers, and so does not depend on the inputs; classical protocols
with correlated randomness are the special case of classical registers. When
the parties have access to an oracle, the \emph{cryptographic qubit
operations} of a party are the qubits in the input and output registers of
the oracle calls it makes in the online phase; as the operation of each round
is fixed, this number does not depend on the inputs either. The
protocol has error probability $\epsilon$ if, for every classical input $x$
and every $a \in [n]$, measuring $Y_a$ in the computational basis at the end
of the protocol yields $f_a(x)$ with probability at least $1 - \epsilon$. It
is \emph{without preprocessing} if $\rho_{\mathrm{pre}}$ is a pure product
state.

Quantum preprocessing with classical communication subsumes quantum
communication. Teleporting each online qubit through a preprocessed EPR
pair~\cite{PRL:BBCJPW93} turns a protocol that sends $Q$ qubits and $c$
classical bits into one whose input-independent preprocessing holds $Q$
additional EPR pairs and whose online phase sends $2Q + c$ classical bits.

\heading{Purified protocols.} Every quantum operation is an isometry followed
by discarding a register (Appendix~\ref{app:prelim}). The \emph{purified}
protocol $\bar\Pi$ replaces every operation of every party by this isometry,
and the party keeps the register it would otherwise discard. No party
measures or discards anything, so if the inputs and $\rho_{\mathrm{pre}}$ are
pure then the joint state of all parties stays pure. Discarding the kept
registers recovers $\Pi$, so $\bar\Pi$ has the same message registers, the
same communication and the same error probability as $\Pi$.

\subsection{Classical Secure MPC and PIR}
\label{sec:prelim_classical}

\heading{Semi-honest MPC.} A semi-honest adversary corrupts up to $t$
parties, which keep following the protocol, and learns their views. The
\emph{view} of a party consists of its input, its correlated randomness,
its random tape, the messages it sends and receives and the answers to its
oracle queries. An \emph{adaptive} adversary chooses the parties it
corrupts as the execution proceeds, and a \emph{non-adaptive} (or static)
adversary chooses them before the execution starts. For $S \subseteq [n]$
we write $\View_S(x)$ for the views of the parties in $S$ in an execution
on input $x$, and $x_S$ and $f_S(x)$ for their inputs and outputs. In the
random oracle model, the simulator answers the oracle queries of the
adversary, that is, it may program $\calO$.

\begin{definition}[Privacy in the random oracle model]
\label{def:rom_private}
\label{def:adaptive_private}
A protocol $\Pi$ for $f$ is \emph{$(Q, \delta)$-private against $t$
adaptive corruptions} if there is a simulator $\Sim$ such that, for every
input $x$ and every computationally unbounded adversary $\calA$ that makes
at most $Q$ oracle queries and corrupts at most $t$ parties, $\calA$
outputs $1$ in the following two executions with probabilities, over
$\calO$ and all random tapes, that differ by at most $\delta$. In the
\emph{real execution}, $\Pi$ runs on $x$ with $\calO$, and $\calA$ may at
any point query $\calO$ or corrupt a party $p$, upon which it receives the
view of $p$ produced so far and, as the execution continues, the rest of
it. In the \emph{ideal execution}, $\Sim$ answers the queries of $\calA$ in
place of $\calO$, receives $(x_p, f_p(x))$ when $\calA$ corrupts $p$, and
provides what $\calA$ receives in place of $\Pi$. The protocol is
\emph{$(Q, \delta)$-private against $t$ non-adaptive corruptions} if this holds
for every $\calA$ that corrupts all its parties before the execution
starts.
\end{definition}

Privacy against adaptive corruptions implies privacy against non-adaptive
corruptions. A protocol is \emph{$\delta$-private} against $t$ corruptions
if it is $(Q, \delta)$-private for every $Q = \poly(\secpar)$, and
\emph{secure} if it is $\negl(\secpar)$-private. That $\Sim$ receives
$f_p(x)$ when $p$ is corrupted is the usual convention for semi-honest
adversaries~\cite{JC:Canetti00}. For protocols without oracle that are
perfectly private, that is, $(Q, 0)$-private for every $Q$, the two notions
coincide (see the proof of Lemma~\ref{lem:composition}), and we omit the
qualifier. Our lower bounds hold against non-adaptive corruptions, and our
protocols are private against adaptive corruptions.

\heading{Programming the oracle.} Letting the simulator answer the oracle
queries is the weaker requirement, as a simulator with oracle access to
$\calO$ can be run in its place on an oracle it samples itself, so lower
bounds against it are the stronger statement. Our lower bounds use privacy
only in the following form. For a fixed set $S$ and inputs $x, x'$ with
$x_S = x'_S$ and $f_S(x) = f_S(x')$, the ideal executions on $x$ and on
$x'$ are identical, so a distinguisher making $Q$ queries to $\calO$ tells
$\View_S(x)$ and $\View_S(x')$ apart with advantage at most $2\delta$.

\heading{Hybrid model.} Let $\calF_1, \ldots, \calF_k$ be deterministic
$n$-party functions. A protocol in the \emph{$(\calF_1, \ldots,
\calF_k)$-hybrid model} may, in addition to sending messages, \emph{call}
any $\calF_i$: every party hands $\calF_i$ an input and receives its output,
and the call is not counted as communication. The view of a party
includes its inputs and outputs in every call. For protocols
$\rho_1, \ldots, \rho_k$ for $\calF_1, \ldots, \calF_k$, let
$\Pi^{\rho_1, \ldots, \rho_k}$ be the protocol in which every call to
$\calF_i$ is replaced by an execution of $\rho_i$ on the same inputs, after
which each party continues with its output of $\rho_i$. The following is
the semi-honest composition theorem~\cite{JC:Canetti00} for our notion of
privacy.

\begin{lemma}[Composition]
\label{lem:composition}
\label{lem:adaptive_composition}
Let $\Pi$ be a protocol for $f$ in the $(\calF_1, \ldots, \calF_k)$-hybrid
model that is perfectly correct and $(Q, \delta)$-private against $t$
adaptive corruptions, and let every $\rho_i$ be a perfectly correct
protocol for $\calF_i$ without oracle that is perfectly private against $t$
corruptions. Then $\Pi^{\rho_1, \ldots, \rho_k}$ is a perfectly correct
protocol for $f$ that is $(Q, \delta)$-private against $t$ adaptive
corruptions. Its communication is that of $\Pi$ plus that of $\rho_i$ for
every call to $\calF_i$. The same holds with non-adaptive corruptions in place
of adaptive ones.
\end{lemma}
\begin{proof}
As every $\rho_i$ is perfectly correct, every party receives the same
outputs from the calls in both protocols, which gives correctness and the
communication.

For privacy, note first that every $\rho_i$ has a perfect adaptive
simulator that makes no queries, as observed in~\cite{EC:CDDIM01}. As
$\rho_i$ does not query $\calO$, everything $\calA$ receives is a function
of the views of the parties it has corrupted so far. The simulator produces
each piece by sampling it from its conditional distribution in the real
execution given everything produced so far, and by perfect privacy for the
set $S$ corrupted so far, this distribution depends on $x$ only through
$(x_S, f_S(x))$.

Let $T$ be the interactive map that passes the queries and corruptions of
$\calA$ through to $\Pi$ and, for every call to some $\calF_i$, runs this
simulator of $\rho_i$ on the inputs and outputs of the corrupted parties in
the call, giving $\calA$ its output in place of the call and the output of
the call only when the simulated execution of $\rho_i$ ends. Let $\calA'$
be the adversary against $\Pi$ that runs $\calA$ and $T$; it makes at most
$Q$ queries and $t$ corruptions.

In $\Pi^{\rho_1, \ldots, \rho_k}$ every execution of $\rho_i$ uses fresh
random tapes and no oracle, and the parties continue only with its outputs,
so what $\calA$ receives from it, whenever it corrupts, is distributed as
the output of this simulator. Hence $\calA$ against $\Pi^{\rho_1, \ldots,
\rho_k}$ is distributed as $\calA'$ against $\Pi$, and the simulator for
$\Pi^{\rho_1, \ldots, \rho_k}$ runs $T$ on top of the simulator of $\Pi$.
\end{proof}

\heading{Perfectly private protocols.} The calls will be instantiated with
the protocol of Ben-Or, Goldwasser and Wigderson~\cite{STOC:BenGolWig88},
run over an extension field of $\FF$ with more than $n$ elements,
whose elements have $O(\log q + \log n)$ bits. In the semi-honest analysis
of Asharov and Lindell~\cite{JC:AshLin17}, sharing an input or
reconstructing an output costs $n$ field elements, an addition is free, and
a multiplication costs $n^2$ field elements.

\begin{theorem}[\cite{STOC:BenGolWig88,JC:AshLin17}]
\label{thm:bgw}
Let $t < n/2$, and let $g$ be a deterministic $n$-party function computed
by an arithmetic circuit over $\FF$ with $I$ inputs, $G$
multiplication gates and $U$ outputs. There is a perfectly correct protocol
for $g$ without oracle that is perfectly private against $t$ corruptions
and communicates $O((nI + n^2 G + nU)(\log q + \log n))$ bits.
\end{theorem}

\heading{PIR with correlated randomness.} We focus on single-server PIR with
correlated randomness between two parties denoted the client and the server.
The server receives the input database $D \in \bits^N$, and the client
receives the input index $i \in [N]$ and should output $D[i]$. In terms of
privacy, the adversarial server observes the server's input database, the
server's portion of correlated randomness and the communication transcript
during the PIR query execution.

\begin{definition}[PIR with Correlated Randomness]
\label{def:pir}
A single-server private information retrieval (PIR) with correlated randomness protocol is a tuple of efficient algorithms with access to a random oracle $\calO$ denoted by $\Pi^{\calO} = (\init^\calO, \query^\calO)$ parameterized by the database size $N = \poly(\secpar)$ as follows:
\begin{itemize}
    \item $\init^{\calO}(1^\secpar) \rightarrow (r_c; r_s \mid \bot)$: The initialization algorithm receives the security parameter and outputs the correlated randomness to each party.
    The client receives randomness $r_c$ and the server receives randomness $r_s$.
    The adversary observes no information denoted by $\bot$.
    \item $\query^{\calO}(r_c, i; r_s, D) \rightarrow (\ans; \bot \mid \tx)$: The query algorithm is a (potentially interactive) two-party protocol between the client and the server. The client receives its correlated randomness $r_c$ and query index $i \in [N]$. The server receives its correlated randomness $r_s$ and the input database $D \in \bits^N$. At the end of the protocol, the client receives $\ans \in \bits$ and the server receives no output. The adversary observes the communication transcript $\tx \in \bits^*$ between the client and the server during the protocol's execution; its length may depend on the random tapes, and the scheme has \emph{expected online communication} $c$ if $\Ex[|\btx|] \le c$ for every $i$ and $D$.
\end{itemize}
We say that $\Pi^{\calO}$ has error probability at most $\epsilon$ if for all $\secpar, i \in [N]$ and $D \in \bits^N$, the following holds:
$$
\Pr\left[\bans \ne D[i] \middle\vert
\begin{array}{l}
(\brc; \brs \mid \bot) \getsr \init^{\calO}(1^\secpar)\\
(\bans; \bot \mid \btx) \getsr \query^{\calO}(\brc, i; \brs, D)\\
\end{array}
\right] \le \epsilon.
$$
We say that $\Pi^{\calO}$ is $(Q, \delta)$-private if for all $\secpar$, databases $D \in \bits^N$, pairs of queries $i_0, i_1 \in [N]$ and computationally unbounded adversaries $\calA$ issuing at most $Q$ queries to $\calO$, the following holds:
$$
\Pr\left[\calA^{\calO}(1^\secpar, \brs, D, \btx) = \mathbf{b} \middle\vert
\begin{array}{l}
(\brc; \brs \mid \bot) \getsr \init^{\calO}(1^\secpar)\\
\mathbf{b} \getsr \bits\\
(\bans; \bot \mid \btx) \getsr \query^{\calO}(\brc, i_{\mathbf{b}}; \brs, D)\\
\end{array}
\right] \le \frac{1}{2} + \delta.
$$
It is \emph{$\delta$-private} if it is $(Q, \delta)$-private for every $Q = \poly(\secpar)$.
\end{definition}

As the correlated randomness is sampled before either party receives its
input, it is independent of both parties' inputs. In particular, the
client's randomness is independent of the database.

\subsection{Quantum Secure MPC and PIR}
\label{sec:prelim_quantum}

\heading{QCCC security.} A QCCC PIR with correlated randomness is as in
Definition~\ref{def:pir}, except that $\init^\calO$ and $\query^\calO$ are
quantum algorithms. The error probability is over $\calO$, the
coins of $\init^\calO$ and $\query^\calO$, and the parties' measurements, and
the scheme is \emph{perfectly correct} if it is $0$ for all $\secpar$, $i$ and
$D$. Privacy is against quantum adversaries that receive the classical
$(r_s, D, \tx)$, as before. We do not give the adversary the server's residual
quantum registers, which only strengthens the lower bound. QCCC MPC with
preprocessing is defined analogously, with classical correlated randomness and
messages, error probability over $\calO$, the coins and the parties'
measurements as above, and semi-honest security against quantum adversaries,
where the view of $S$ consists of the inputs, correlated randomness and the
classical messages sent and received by the parties in $S$. The preprocessing
may be run by a trusted party, but its $\poly(\secpar)$ oracle queries count.

\heading{Quantum MPC.}
Defining semi-honest security for quantum protocols requires some care. Since
quantum states on their own can't be cloned in general, an honest-but-curious
server by default wouldn't be able to even read or remember a quantum protocol
in general. So, the semi-honest analog in quantum computing is known as
\emph{specious} security~\cite{C:DupNieSal10}, where
corrupted parties follow the protocol and their view must be simulatable from
their inputs, preprocessing and outputs (see~\cite{EC:ABCGLS19} and the
references therein for further discussion). A specious
adversary follows the
protocol, but may defer every measurement and keep every register it is told
to discard. Its view is therefore the state of its registers in the purified
protocol. Inputs may be entangled with an arbitrary reference
system $R$, which plays the role of the environment. For $S \subseteq [n]$
write $X_S = \bigotimes_{a \in S} X_a$, and likewise $W_S$ and $Y_S$.

\begin{definition}[Statistical specious security]
\label{def:specious}
Let $\Pi$ be an $n$-party protocol with quantum preprocessing for $f$, let $S
\subseteq [n]$ be the set of corrupted parties, and let $\rho_{\mathrm{in}}$
be an input state on $X_{[n]} \otimes R$.
\begin{itemize}[nosep]
    \item The \emph{view} of $S$ after round $j$, written
      $\View^j_S(\rho_{\mathrm{in}})$, is the state of all registers of the
      parties in $S$, together with $R$, after round $j$ of the purified
      protocol run on $\rho_{\mathrm{in}} \otimes \rho_{\mathrm{pre}}$. We
      write $\View_S(\rho_{\mathrm{in}})$ for the view after the last round.
    \item The \emph{ideal view} of $S$, written
      $\Ideal_S(\rho_{\mathrm{in}})$, is the state of the inputs $X_S$, the
      preprocessing $W_S$ and the outputs $Y_S$ of the parties in $S$,
      together with $R$, where the outputs are computed from the inputs by
      the unitary that maps $\ket{x}_{X_{[n]}} \ket{0}_{Y_S}$ to $\ket{x}
      \bigotimes_{a \in S} \ket{f_a(x)}_{Y_a}$.
\end{itemize}
The protocol $\Pi$ is \emph{$\delta$-private for $S$} if for every round $j$
there is a simulator $\Sim^j_S$, a quantum operation that acts on $X_S W_S
Y_S$ and not on $R$, such that for every $\rho_{\mathrm{in}}$,
$$
\Delta\big(\View^j_S(\rho_{\mathrm{in}}),\
\Sim^j_S(\Ideal_S(\rho_{\mathrm{in}})) \big) \le \delta,
$$
where $\Delta$ is the trace distance. It is \emph{$\delta$-private against
$t$ corruptions} if it is $\delta$-private for every $S \subseteq [n]$ with
$|S| \le t$, and \emph{statistically secure} if $\delta = \negl(\secpar)$.
We write $\Sim_S$ for the simulator of the last round.
\end{definition}

Simulating the view after every round, and not only after the last one, is
what makes the definition meaningful. A party that returns every register it
received holds nothing at the end, so the protocol in which one party sends
its input register and the other applies $f$ coherently and returns every
register would otherwise be private~\cite{C:DupNieSal10}. Our lower bounds
use only the view after the last round, which is the weaker requirement, so
they hold for every notion that implies it. In our lower bounds the corrupted
parties have no output. The ideal view is then just their part of the input
and of the preprocessing, and privacy takes the following form, which is the
one our proofs use.

\begin{lemma}[Privacy without outputs]
\label{lem:no_output}
Let $\Pi$ be $\delta$-private for $S$, where the parties in $S$ have no
output. Then there is a quantum operation $\Sim'_S$ acting on $X_S$ such
that, for every input state $\rho_{\mathrm{in}}$ with reduced state
$\rho_{X_S R}$ on $X_S R$, the view $\View_S(\rho_{\mathrm{in}})$ is
$\delta$-close to $\Sim'_S(\rho_{X_S R})$, that is, within trace distance
$\delta$ of it. In particular:
\begin{enumerate}[nosep]
    \item \label{item:no_out_same} If two input states have the same reduced
      state on $X_S R$, then the views of $S$ on them are $2\delta$-close.
    \item \label{item:no_out_prod} If $\rho_{X_S R} = \rho_{X_S} \otimes
      \rho_R$, then $\View_S(\rho_{\mathrm{in}})$ is $\delta$-close to
      $\sigma \otimes \rho_R$ for the state $\sigma = \Sim'_S(\rho_{X_S})$.
\end{enumerate}
\end{lemma}
\begin{proof}
Without outputs, $\Ideal_S(\rho_{\mathrm{in}})$ is $\rho_{X_S R}$ together
with the reduced state of $\rho_{\mathrm{pre}}$ on $W_S$, so we let
$\Sim'_S$ append that state and apply $\Sim_S$. The first item is the
triangle inequality, and the second holds as $\Sim'_S$ does not act on $R$.
\end{proof}

\heading{Privacy relative to an oracle.} In Sections~\ref{sec:qmpc}
and~\ref{sec:qcqc_lb} the parties have quantum access to $\calO$, and the
views depend on $\calO$. There, $\Pi$ is \emph{$(Q, \delta)$-private for
$S$ relative to $\calO$} if for every round $j$ there is a simulator
$\Sim^j_S$ making at most $\poly(\secpar)$ queries to $\calO$ such that, for every $\rho_{\mathrm{in}}$
and every computationally unbounded distinguisher $\calA$ making at most $Q$
queries to $\calO$,
$$
\left| \Pr\left[\calA^\calO\big(\View^j_S(\rho_{\mathrm{in}})\big) = 1\right] -
\Pr\left[\calA^\calO\big(\Sim^j_S(\Ideal_S(\rho_{\mathrm{in}}))\big) = 1\right]
\right| \le \delta,
$$
where the probabilities are over $\calO$ and the protocol, and it is
\emph{secure} if it is $(Q, \negl(\secpar))$-private for every $Q =
\poly(\secpar)$, as for classical protocols. For $Q = 0$ this
is Definition~\ref{def:specious} for the states averaged over $\calO$, and
$(Q, \delta)$-privacy implies $(0, \delta)$-privacy.

\heading{Quantum PIR.} A single-server QPIR with entangled preprocessing for
databases in $\bits^N$ is the case $n = 2$, $t = 1$ and $f(D, i) = (\bot,
D[i])$ with the server $P_1$ corrupted.

\subsection{Blackbox Constructions}

Throughout our work, we will consider blackbox constructions using the definitions directly taken from~\cite{TCC:ReiTreVad04}.

\begin{definition}[Blackbox Constructions]
Let $\mathcal{P}_1$ and $\mathcal{P}_2$ be two primitives. A blackbox construction of $\mathcal{P}_2$
from $\mathcal{P}_1$ consists of the tuple of PPT algorithms $(P_2,S)$ satisfying the following requirements:
\begin{itemize}[noitemsep]
\item For any oracle $P_1$ implementing $\mathcal{P}_1$, the algorithm $P_2^{P_1}$ implements $\mathcal{P}_2$.
\item For any oracle $P_1$ implementing $\mathcal{P}_1$ and any computationally unbounded oracle
adversary $A$ that breaks the security of $P_2^{P_1}$, then the oracle adversary $S^{P_1,A}$ breaks the security of $P_1$.
\end{itemize}
\end{definition}

We define the existence of a primitive relative to an oracle as follows.

\begin{definition}
A primitive $\mathcal{P}$ exists relative to an idealized oracle $\mathcal{O}$ if there exists
an implementation $P$ for $\mathcal{P}$ computable by a PPT algorithm
with access to $\mathcal{O}$. Furthermore, there exists no (computationally unbounded) adversary that makes at most polynomial number of queries to idealized oracle $\mathcal{O}$
that breaks the security of $P$.
\end{definition}

Lastly, we define relativization as follows.

\begin{theorem}[Relativization]
If there is a blackbox construction of primitive $\mathcal{Q}$ from primitive $\mathcal{P}$, then
\begin{itemize}[noitemsep]
\item If $\mathcal{P}$ exists relative to idealized oracle $\mathcal{O}$, then $\mathcal{Q}$ also exists relative to $\mathcal{O}$.
\item If there exists a blackbox construction of $\mathcal{P}$ from idealized oracle $\mathcal{O}$,
then there also exists a blackbox construction of $\mathcal{Q}$ from idealized oracle $\mathcal{O}$.
\end{itemize}
\end{theorem}

The above is useful for our lower bounds proven in the random oracle and crypto oracle models. In particular, if some primitive $\mathcal{P}$ cannot exist relative to any oracle, then this immediately implies that there does not exist a blackbox construction for $\mathcal{P}$ built from any primitive that can be built from the oracle. For the random oracle model, this means our lower bounds imply that no blackbox constructions exist from any symmetric-key cryptography that may be built from a random oracle such as one-way functions, pseudorandom functions, symmetric-key encryption, etc. Similarly, any lower bounds in the crypto oracle model rule out blackbox constructions from a wide-range of assumptions including generic groups models, virtual blackbox obfuscation and fully homomorphic encryption (see~\cite{EC:LinMooWic25,FOCS:HooPerYeo26}).

\subsection{Useful Prior Results}
\label{sec:prelim_prior}

\heading{Key agreement.} A core technical lemma that we utilize is the
result of Barak and Mahmoody~\cite{C:BarMah09}, which presents an explicit
attack for eavesdropping on any two-party protocol in the random oracle
model that enables both parties to agree upon the same output (and, thus, a
potential key exchange protocol). The first blackbox separation between key
exchange and random oracles was presented by Impagliazzo and
Rudich~\cite{STOC:ImpRud89}, but their explicit attack was less efficient
(requiring more random oracle queries).

\begin{theorem}[Due to \cite{C:BarMah09}]
\label{thm:ke_attack}
Let $\Pi^\calO$ be a two-party protocol in the random oracle model
such that both parties make at most $Q$ queries to the random oracle
and both party outputs are identical with probability $p$.
Then, for every $0<\delta<p$, there is an explicit algorithm
$\calA_\delta^\calO$ that, on input the communication transcript between the two parties, makes at most $O(Q^2/\delta^2)$ queries
to the random oracle and produces an output that agrees
with one party's output with probability at least $p-\delta$.
\end{theorem}
Note that the above theorem makes no claim regarding the
running time of $\calA$ but only that it makes a number
of calls that is quadratic in $Q$ and $1/\delta$.

In the QCCC model, we use the following key-agreement attack of Li, Li, Li
and Liu~\cite{ARXIV:LLLL26}.

\begin{theorem}[\cite{ARXIV:LLLL26}]
\label{thm:qccc_ke_attack}
Let $\Pi$ be a two-party QCCC protocol without inputs in the QROM, in which the
parties start in a product state, make at most $d$ superposition queries in
total, and output the same key with probability $1$. There is an explicit
algorithm that, given a transcript of $\Pi$, makes at most $O((dm)^5)$ classical
queries to $\calO$ and outputs the agreed key with probability $1$.
\end{theorem}

This is Theorem~3.1 of~\cite{ARXIV:LLLL26}, stated there for a Boolean oracle
and with a query bound independent of the oracle's domain. Our oracle
$\calO_m: \bits^{\le p} \to \bits^m$ is simulated bitwise by the Boolean oracle
$(x, j) \mapsto \calO_m(x)_j$: each superposition query to $\calO$ is replaced
by $m$ sequential queries, one for each output bit $j \in [m]$, which
multiplies $d$ by $m$, and each classical query of the attacker is answered
by one classical query to $\calO$. It
supersedes the attacks of~\cite{C:ACCFLM22}, which need a party that queries
$\calO$ only classically or the Polynomial Compatibility Conjecture, and the
two-message attack of~\cite{EPRINT:LLLL25}.

\heading{Quantum PIR and oblivious transfer.} Prior work~\cite{EC:ABCGLS19} proved a linear communication lower bound for
information-theoretic QPIR against specious servers with prior entanglement,
which extends the bounds without prior entanglement
of~\cite{FOCS:Nayak99,FOCS:JaiRadSen02,JC:BauBro15}.

\begin{theorem}[{\cite[Corollary~4.6]{EC:ABCGLS19}}]
\label{thm:abcgls}
Let $\Pi$ be a single-server QPIR with entangled preprocessing for databases
in $\bits^N$ with error probability $\epsilon \le N^{-4}/100$ that is
$\delta$-private for $\delta \le N^{-8}/100$. Then the server sends at least
$(N-1)/2$ qubits.
\end{theorem}

The proof in~\cite{EC:ABCGLS19} runs the purified honest server, which is
$0$-specious in their terminology, and uses privacy only for its final state,
so it applies to Definition~\ref{def:specious}. As written it samples a
uniformly random classical database, and the decoding unitary it obtains from
Lemma~\ref{lem:uhlmann} then depends on the database. With the database
register instead maximally entangled with $R$, which
Definition~\ref{def:specious} allows and which we do in the proof of
Lemma~\ref{lem:three_party}, the unitary is independent of the database and
the proof is otherwise unchanged. The bound is on the server's
communication alone, which is why prior entanglement does not help.

On the positive side, we use single-server PIR with quantum communication and
quantum oblivious transfer. We state both as implied from post-quantum one-way functions. Both use the
one-way function only through a pseudorandom generator, the first via a
distributed point function and the second via Naor's
commitment~\cite{JC:Naor91}, and every computational step of their proofs
is a blackbox reduction to its security. Hence, taking $\calO$ as the
pseudorandom generator, both are secure relative to $\calO$ against
computationally unbounded adversaries making $\poly(\secpar)$ queries, as
observed for the latter in~\cite[Section~1.1]{C:BCKM21b}. We do not use the
concurrent construction of~\cite{EC:GLSV21}, which makes non-blackbox use
of the one-way function through zero-knowledge proofs and does not
relativize.

\begin{theorem}[Due to~\cite{EPRINT:AnaBhaGul26}]
\label{thm:qmpc_cpir}
Assuming post-quantum one-way functions, there is a perfectly correct
single-server PIR for databases $D \in \bits^N$ with $O(\secpar \log N)$ qubits
of communication that is private against malicious servers.
\end{theorem}

The protocol is one round and evaluates the two keys of a distributed point
function~\cite{EC:GilIsh14,EC:BoyGilIsh15} in superposition. We also use that
the honest client's final state is a function of its index $i$, its coins and
$D[i]$ alone.

\begin{theorem}[Due to~\cite{C:BCKM21b}]
\label{thm:qmpc_ot}
Assuming post-quantum one-way functions, there is maliciously simulation-secure
oblivious transfer with $\poly(\secpar)$ qubits of communication. Hence, by
the protocols of~\cite{STOC:GolMicWig87,STOC:Kilian88}, which are perfectly
private given ideal oblivious transfer, every classical circuit $g$ has a
semi-honest secure two-party protocol with $\poly(\secpar, |g|)$ qubits of
communication.
\end{theorem}

\heading{Simulating random oracles.} The compressed oracle technique of
Zhandry~\cite{C:Zhandry19} allows efficient simulation of a random oracle
even after a polynomial number of superposition queries.

\begin{theorem}[\cite{C:Zhandry19}]
\label{thm:ro_sim}
There exists an efficient quantum algorithm that perfectly simulates a quantum-accessible random oracle $\calO: \bits^a \rightarrow\bits^b$ after $Q$ queries using $O(Q(a+b))$ qubits of storage.
\end{theorem}

\section{MPC with Classical Communication Lower Bounds}
In this section, we prove lower bounds for secure multi-party computation (MPC) in the preprocessing model of Section~\ref{sec:prelim_models}, with security as in Section~\ref{sec:prelim_classical}.
Recall that the adversary observes nothing of the preprocessing step, and
our result will lower bound only
the online communication.
Therefore, our lower bound holds even if one assumes
stronger computational assumptions (such as public-key cryptography),  
inefficient communication or the existence of a trusted third party during the preprocessing step, as long as the preprocessing makes $\poly(\secpar)$ queries to the random oracle.
In the online phase, however, we assume
that the MPC protocol operates in the random oracle model (or uses a one-way function in a blackbox manner). To start, we will prove lower bounds for single-server PIR with correlated randomness that we will later utilize to prove our MPC with preprocessing lower bounds.

\subsection{Single-Server PIR with Correlated Randomness}
\label{sec:pir}

In this section, we study single-server PIR with correlated randomness
(Definition~\ref{def:pir}) in the random oracle model.
As a note, this notion (and stronger variants with symmetric privacy known as symmetric PIR or SPIR) have been used to build multi-party computation in the past (see~\cite{EC:IshKus04,TCC:BoyCouMey22,EC:BoyCouMey23,C:CouKum24} and references therein as some examples).
We will prove a linear $\Omega(N)$ lower bound
on the online communication between the client and the server
during the execution of the PIR query algorithm.
Later, we will build on top of this result to prove a communication lower bound for MPC with preprocessing.

\begin{theorem}
\label{thm:main_pir}
Relative to a random oracle $\calO$, there does not exist a single-server PIR for an $N$-bit database with correlated randomness with error probability at most $\epsilon$ that is $\delta$-private, where $2\epsilon + 4\delta \le 2/5$, with $o(N)$ expected online communication for a single query.
\end{theorem}

In other words, we unconditionally prove that there does not exist any single-server PIR with correlated randomness making blackbox use of one-way functions that obtains sublinear communication.
As a comparison, this lower bound is much stronger than prior works that study PIR with database-dependent preprocessing where the best results are
sublinear amortized communication lower bounds
for a sufficiently large number of queries in restricted models~\cite{EC:CorKog20,SODA:PerYeo22,EC:CorHenKog22,EC:Yeo23,C:IshShiWic24,FOCS:HooPerYeo26}.
The independence of the correlated randomness from the database is critical, as there exist PIR with database-dependent client preprocessing that obtain very small communication for a single query using one-way functions~\cite{SP:ZPZS24,CCS:RenMugSun24,EC:HPPY25,EC:WanRen25}.
For MPC, it is critical we prove lower bounds for a single query as otherwise the circuit grows too large with respect to our communication lower bound.
Finally, if one wishes for more powerful cryptographic primitives beyond random oracles, there are many known constructions of PIR (without even needing correlated randomness) that obtain sublinear communication from cryptographic assumptions that imply public-key cryptography (see~\cite{FOCS:KusOst97,EC:Paillier99,PKC:DamJur01,PKC:OstSke07} and references therein as examples).

\heading{Proof Overview.} Our proof of the communication lower bound for PIR with correlated randomness using a random oracle proceeds in the following three steps:

\begin{enumerate}
    \item As the first step, we show that an eavesdropper of the PIR query protocol with server's correlated randomness $r_s$ and the client's query index $i \in [N]$ can retrieve database entry $D[i]$ using the communication transcript $\tx$ of the PIR query protocol when the client queries for index $i$. In particular, the eavesdropper successfully computes $D[i]$ without receiving either the client's correlated randomness $r_c$ or the server's input database $D$.
    \item Secondly, we show that the eavesdropper algorithm may be enhanced such that the eavesdropper may pick any query index $i' \in [N]$ to enable retrieving database entry $D[i']$ using only the server's correlated randomness $r_s$ and a single communication transcript $\tx$.
    Critically, we show that the eavesdropper may still successfully decode $D[i']$ even if the communication transcript $\tx$ was the output
    of the PIR query protocol when the client queried for a different index $i \ne i'$.
    \item Finally, we use the above eavesdropper algorithm to show that the server's correlated randomness and a single query communication transcript may act as a good encoding of a random database obtaining our linear communication lower bound.
\end{enumerate}

\heading{Eavesdropper Retrieval for Queried Index.}
For the first step of our proof, we will introduce
the notion of {\em eavesdropper retrieval} in the context of PIR with correlated randomness. 

An eavesdropper $\calE^\calO$ for a PIR with correlated
randomness $\Pi^\calO$ is a third party that has access to the 
random oracle $\calO$ and receives in input
the transcript $T$ of the query algorithm of $\Pi^\calO$.
In addition, the eavesdropper receives 
the server's correlated randomness $r_s$ 
and the client index input $i$.
The task of the eavesdropper is to guess $D[i]$.
We stress that the eavesdropper does not receive
the client's correlated randomness $r_c$,
nor the server's input database $D$.
Clearly, if we let the eavesdropper learn $D$, then 
it will easily guess $D[i]$. 
We also note that knowledge of  
the client's correlated randomness $r_c$ would 
enable the eavesdropper to compute $D[i]$
by acting as the client in transcript $T$.
We next show that,
for every PIR $\Pi^\calO$,
there exists an eavesdropper $\calE^\calO$ that,
even without access to $r_c$ or to $D$,
computes $D[i]$ by making $\poly(\secpar)$ calls to $\calO$.

We formalize the eavesdropper retrieval in this setting using the 
following game $\G_1$ for a PIR with correlated randomness scheme $\Pi^{\calO} = (\init^{\calO}, \query^{\calO})$ and 
an eavesdropper $\calE^\calO$. 
The game takes as input database $D \in \bits^N$ and index $i \in [N]$. 
The eavesdropper $\calE$ 
receives as input the server's correlated randomness $r_s$, 
a communication transcript $\tx$, and a query index $i\in[N]$  
and  has access to the random oracle $\calO$ used to 
compute $r_s$ and $\tx$.
The game is defined as follows:

\medskip\noindent{$\G_1^{\Pi,\calE,\calO}( D, i):$}
\begin{enumerate}
    \item Generate correlated randomness $(\brc, \brs \mid \bot) \leftarrow \init^{\calO}(1^\secpar)$.
    \item Execute PIR query protocol to obtain $(\bans, \bot \mid \btx) \leftarrow\query^{\calO}(\brc, i; \brs, D)$.
    \item Compute eavesdropper answer as 
    $\bans' \leftarrow \calE^\calO(\brs,\btx,i)$.
    \item Output 1 if and only if $\bans' = D[i]$.
\end{enumerate}
We say that the eavesdropper $\calE^{\calO}$ wins the game $\G_1$
if $\G_1$ outputs 1  
(that is, if $\calE$ correctly outputs $D[i]$).

The next lemma says that, for every PIR with correlated randomness,
there exists an eavesdropper that makes $\poly(\secpar)$ 
queries to the random oracle and wins $\G_1$
with a probability that can be made arbitrarily close to the 
probability that the PIR gives the correct answer.
If this were not the case, 
the PIR could be used to construct a
key agreement protocol in the random oracle model that is 
secure against an adversary that makes polynomial number of 
calls to the random oracle, thus contradicting the classic result
of Impagliazzo and Rudich~\cite{STOC:ImpRud89}
(see also~\cite{C:BarMah09}).

More specifically, for a fixed $i \in [N]$,
we consider the following two-party protocol $\Pi_i$ without inputs, 
in which $i$ is hardwired into both parties. 
The client executes $\init^{\calO}$ locally, keeps $r_c$ private,
and sends $r_s$ to the 
server as its first message, which is permitted as the server has no privacy guarantees.
The server samples a uniformly random database $D \getsr \bits^N$.
The two parties then execute the PIR query algorithm $\query^\calO$ 
on client inputs $(r_c, i)$ and server inputs $(r_s, D)$. 
Finally, the client outputs $\ans$ and the server outputs $D[i]$. 
The key observation is that the two parties output the same value
$D[i]$ with probability $1-\epsilon$ (where $\epsilon$ is the
error probability of the PIR). 
Assuming by contradiction that no eavesdropper that receives
$i,T,$ and $r_s$ can guess $D[i]$, we have constructed
a key agreement protocol in the random oracle model that is secure
against adversaries that make only $\poly(\secpar)$ queries
to $\calO$.

\begin{lemma}
\label{lem:query_index}
Let  $\Pi^{\calO}$  be a PIR with correlated randomness 
with error probability $\leq\epsilon$ in which
the parties make $\poly(\secpar)$ queries to $\calO$.
For any constant $0<\gamma<1$,
there exists an eavesdropper algorithm 
$\calE_\gamma$ 
making at most $\poly(\secpar,1/\gamma)$ queries to the random oracle $\calO$ 
such that for all indices $i \in [N]$, the following holds:
$$
\Pr_{\bD \getsr \bits^N}
\left[
\G_1^{\Pi,\calE_\gamma,\calO}(\bD, i) = 1
\right] 
\ge 1 - \epsilon - \gamma.
$$
\end{lemma}
\begin{proof}
For every $\gamma>0$,
we will construct an explicit eavesdropper $\calE_\gamma$
that wins the game $\G_1$ with probability at 
least $1-\epsilon-\gamma$. 
To do this, we fix $i\in [N]$ and consider the two-party protocol $\Pi_i$ without inputs described above.
As we observed above, by the correctness of $\Pi^\calO$,
the two parties of $\Pi_i$ give $D[i]$ in output with probability 
at least $1-\epsilon$. 
Therefore by Theorem~\ref{thm:ke_attack}, for every $0<\gamma<1$,
there exists an adversary $\calA_\gamma$ that makes 
$\poly(\secpar,1/\gamma)$ queries to the random oracle and outputs
$D[i]$ with probability at least $1-\epsilon-\gamma$.
It is straightforward to see that $\calE_\gamma^\calO(r_s, \tx, i) := \calA_\gamma^\calO(r_s, \tx)$,
where $\calA_\gamma$ is the adversary for $\Pi_i$, is an eavesdropper that wins $\G_1$
with probability at least $1-\epsilon-\gamma$, 
where the probability is taken over the choices
of $D$ and over the randomness of $\calO$.
\end{proof}

\begin{remark}[The database as randomness]
\label{rem:uniform_db}
The eavesdropper of Theorem~\ref{thm:ke_attack} is built from the description of a protocol without inputs. It samples fake views of one party conditioned on the transcript, so it depends on the code of both parties. Hardwiring $D$ into the server would therefore give an eavesdropper depending on $D$. Such an eavesdropper is useless below, as Lemma~\ref{lem:fano} uses $\calE^\calO$ to decode a random database. Sampling $D$ on the server's random tape instead yields a single $\calE^\calO$ that succeeds on average over a uniform database, which is all that Lemmas~\ref{lem:any_index} and~\ref{lem:fano} require. The index $i$ is hardwired, so the guarantee holds for each $i$ separately.
\end{remark}

\heading{Eavesdropper Retrieval for Any Index.}
In this section, we expand the retrieval capabilities of the eavesdropper 
by showing that the eavesdropper may actually choose the index $i \in [N]$ of the entry that it wishes to decode. In particular, we show that the eavesdropper is successful,
that is, guesses $D[i]$, even if the observed communication 
transcript $\tx$ was generated by a client querying for a 
different index $i' \ne i$.
To formalize this, we use the following game $\G_2$ 
that is similar to the prior game $\G_1$
with the main difference being that the communication transcript is now always generated by the client querying a fixed index,
say $1$. 
Yet, for the eavesdropper to win the game, it still must decode the database entry $D[i]$ for the index $i$ received as 
input to the game. 
We formally present the game below and highlight the only difference in {\color{red} red}.

\medskip\noindent{$\G_2^{\Pi,\calE,\calO}(D,i):$}
\begin{enumerate}
    \item Generate correlated randomness $(\brc, \brs \mid \bot) \leftarrow \init^{\calO}(1^\secpar)$.
    \item Execute PIR query protocol to obtain $(\bans, \bot \mid \btx) \leftarrow \query^{\calO}(\brc, {\color{red} 1}; \brs, D)$.
    \item Compute eavesdropper answer as $\bans' \leftarrow \calE^\calO(\brs, \btx, i)$.
    \item Output 1 if and only if $\bans' = D[i]$.
\end{enumerate}

To show that an eavesdropper may retrieve any entry of the database
from the communication transcript 
(and not only for the client-queried index), 
we will leverage the privacy guarantees of the underlying PIR with correlated randomness scheme $\Pi^{\calO}$. 
While the eavesdropper does not obtain the database as input, 
we note that
an adversary that tries to break the privacy of $\Pi$ has
access to the database (as per Def.~\ref{def:pir}).
As a result, an adversary may execute the eavesdropper algorithm using different indices and
check the correctness of the eavesdropper algorithm using its database. If the correctness of the eavesdropper algorithm differs significantly for different indices, then we will show that the adversary can compromise the privacy
of the underlying PIR scheme $\Pi^{\calO}$.

\begin{lemma}
\label{lem:any_index}
Suppose the PIR with correlated randomness scheme $\Pi^{\calO}$ has error probability $\epsilon$ and is $\delta$-private.
For any constant $0 < \gamma < 1$,
there exists an eavesdropper algorithm $\calE_\gamma$ making at 
most $\poly(\secpar,1/\gamma)$ 
queries to the random oracle $\calO$ such that for all indices $i \in [N]$, the following holds:
$$
\Pr_{\bD \getsr \bits^N}
\left[
\G_2^{\Pi,\calE_\gamma,\calO}(\bD, i)
= 1
\right] \ge 1 - 2\epsilon - 2\gamma - 4\delta.
$$
\end{lemma}
\begin{proof}
Fix $0<\gamma<1$ and let $\calE_\gamma$ be the 
eavesdropper algorithm described in Lemma~\ref{lem:query_index}.
Towards a contradiction, suppose that there exists some 
$i^\star\in[N]$ such that 
$$
\Pr_{\bD \getsr \bits^N}
\left[
\G_2^{\Pi,\calE_\gamma,\calO}(\bD,i^\star) = 1
\right]
<1-2\epsilon- 2\gamma-4\delta.$$
First, we note that $i^\star\ne 1$, since,
as it is not hard to see,
$$
\Pr_{\bD \getsr \bits^N}
\left[
\G_2^{\Pi,\calE_\gamma,\calO}(\bD,1) = 1
\right]=
\Pr_{\bD \getsr \bits^N}
\left[
\G_1^{\Pi,\calE_\gamma,\calO}(\bD,1) = 1
\right]\ge
1 - \epsilon - \gamma>1-2\epsilon-2\gamma-4\delta,$$
where the first inequality follows from Lemma~\ref{lem:query_index}.
Next, we construct an adversary $\calA$
for the privacy game of the PIR with correlated randomness scheme $\Pi$.
We consider the privacy game for the first query index 
set to $i_0 = 1$ and
the second query index set to $i_1=i^\star\ne 1$.
We also consider 
a uniformly random database $\bD \getsr \bits^N$.
We note that since $\delta$-privacy holds for every database, 
$\calA$'s success probability averaged over 
$\bD$ is also at most $1/2 + \delta$.

Adversary $\calA$, as per Def.~\ref{def:pir},
receives the server's randomness $r_s$, the database $D$ and 
a communication transcript $\tx$ resulting from the 
execution of the PIR query algorithm when the client received 
one chosen at random of either $i_0 = 1$ or $i_1=i^\star$.
Moreover, $\calA$ is given access to the random oracle $\calO$
used to compute $r_s$ and $\tx$, 
and executes the following steps.

\medskip\noindent
{$\calA^{\calO}(1^\secpar, \rs, D, \tx):$}
\begin{enumerate}
\item Compute $\ans_0\leftarrow\calE_\gamma^{\calO}(\rs,\tx,1)$.
\item Compute $\ans_1\leftarrow\calE_\gamma^{\calO}(\rs,\tx,i^\star)$.
\item 
Set $\mathtt{Cond}=(\ans_0 = D[1]) \land (\ans_1 \ne D[i^\star])$.
\item\label{step:if} 
If $\mathtt{Cond}=\mathsf{True}$, return $b' = 0$.
\item Else, return a random bit $b'\getsr \bits$.
\end{enumerate}
Next, we calculate the probability that the adversary outputs the right 
bit $b'$ to predict whether the transcript was produced by executing 
$i_0$ or $i_1$. 

We start with the case when $b=0$, 
and thus the communication transcript received in input 
by $\calA$  is for query $i_0=1$.
In this case, we note that $\Pr[\bans_0 \ne D[1]] \le \epsilon+\gamma$ by Lemma~\ref{lem:query_index}. Furthermore, by our assumption toward contradiction, we know that
$\Pr[\bans_1 = D[i^\star]]<1-2\epsilon-2\gamma-4\delta$. 
Therefore, the probability that $\mathtt{Cond}$ is 
$\mathsf{True}$ at Step~\ref{step:if} is
at least
$$\Pr[\mathtt{Cond}|b=0]\geq\epsilon+\gamma+4\delta.$$
Altogether, the probability that the adversary outputs the correct bit $b'=0$ when the communication transcript is generated using client input $i_0 = 1$ is at least:
\begin{align*}
\Pr&\left[\calA^{\calO}(1^\secpar, \brs, D, \btx) = 0\  \middle\vert
\begin{array}{l}
(\brc; \brs \mid \bot) \getsr \init^{\calO}(1^\secpar)\\
(\bans; \bot \mid \btx) \getsr \query^{\calO}(\brc, i_0 = 1; \brs, D)\\
\end{array}
\right]\\ 
&=\Pr[\mathtt{Cond}|b=0] + \frac{1}{2}\cdot\Pr[\neg\mathtt{Cond}|b=0]\\
&=\frac{1}{2}+\frac{1}{2}\Pr[\mathtt{Cond}|b=0]\\
&> 1/2 + \epsilon/2 + \gamma/2+ 2\delta .
\end{align*}
Next, we calculate the probability that $\calA$ outputs $1$ 
when $b=1$ and thus
the transcript is generated using client input $i_1=i^\star$.
\begin{align*}
\Pr&\left[\calA^{\calO}(1^\secpar, \brs, D, \btx) = 1\ \middle\vert
\begin{array}{l}
(\brc; \brs \mid \bot) \getsr \init^{\calO}(1^\secpar)\\
(\bans; \bot \mid \btx) \getsr
    \query^{\calO}(\brc, i_1 = i^\star; \brs, D)\\
\end{array}
\right]\\ 
&=\frac{1}{2}\Pr[\neg{\mathtt{Cond}} \mid b=1]\\
&\ge\frac{1}{2}\Pr[\bans_1 = D[i^\star] \mid b = 1]\\
&= \frac{1}{2}\Pr[\G_1^{\Pi,\calE_\gamma,\calO}(D,i^\star) = 1]\\
&\ge 1/2 - \epsilon/2 - \gamma/2
\end{align*}
where we use Lemma~\ref{lem:query_index} to bound the probability that 
$\bans_1=D[i^\star]$ as the experiment is exactly the game 
$\G_1^{\Pi,\calE_\gamma,\calO}(\bD,i^\star)$.
Finally, we put the probabilities together to obtain
\begin{align*}
\Pr&\left[\calA^{\calO}(1^\secpar, \brs, D, \btx) = \mathbf{b} \middle\vert
\begin{array}{l}
(\brc; \brs \mid \bot) \getsr \init^{\calO}(1^\secpar)\\
\mathbf{b} \getsr \bits\\
(\bans; \bot \mid \btx) \getsr \query^{\calO}(\brc, i_{\mathbf{b}}; \brs, D)\\
\end{array}
\right]\\
&> \frac{1}{2}\left(\frac{1}{2}+\epsilon/2 + \gamma/2+ 2\delta\right)+
    \frac{1}{2}\left(\frac{1}{2}-\epsilon/2-\gamma/2\right)\\
&= \frac{1}{2}+\delta
\end{align*}
that contradicts the $\delta$-privacy of the underlying PIR scheme $\Pi^{\calO}$ to complete the proof.
\end{proof}

In other words, we have proven that the eavesdropper algorithm $\calE^{\calO}$ is able
to retrieve any database entry $D[i]$ for any $i \in [N]$ using a communication transcript
that is generated from a PIR query protocol execution where the client queries for index $1$.

\heading{Communication Lower Bound.}
As the last step, we show that the
fact that the eavesdropper may retrieve any database
entry $D[i]$ with $i \in [N]$ implies that
the communication transcript must be linear in the database. First, we show that the communication transcript must have large mutual information
with each individual database entry.

\begin{lemma}
\label{lem:fano}
Let $\bD$ be a uniformly random database drawn from $\bits^N$, let $(\brc; \brs \mid \bot) \getsr \init^\calO(1^\secpar)$, and let $\btx$ be the communication transcript of $\query^\calO(\brc, 1; \brs, \bD)$, that is, with client query index $1$.
If $\Pi^{\calO}$ is a PIR with correlated randomness scheme with error probability at most $\epsilon$ and $\delta$-private with $2\epsilon + 4\delta \le 2/5$,
then for all $i \in [N]$,
$$
H(\bD[i] \mid \btx, \brs, \calO) \le 1 - c_0
$$
for the constant $c_0 = 0.022$, where $\calO$ is restricted to the polynomially bounded inputs and outputs used by the algorithms. That is, the entropy of $\bD[i]$ is reduced by a constant amount when conditioned on the transcript, the server's randomness and the random oracle.
\end{lemma}
\begin{proof}
Fix $i \in [N]$ and let $\calE^\calO$ be the eavesdropper of Lemma~\ref{lem:any_index} with $\gamma = 1/200$. Its output $\hat{\bD}_i \leftarrow \calE^{\calO}(\brs, \btx, i)$ is computed from $(\btx, \brs, \calO)$ and independent coins, so $\bD[i] \to (\btx, \brs, \calO) \to \hat{\bD}_i$ is a Markov chain, and by Lemma~\ref{lem:any_index}
$$
p_e = \Pr[\hat{\bD}_i \ne \bD[i]] \le 2\epsilon + 2\gamma + 4\delta \le 41/100.
$$
By Fano's inequality (Lemma~\ref{lem:entropy}(\ref{item:ent_dpi}) and~(\ref{item:ent_fano})), we obtain that
$$
H(\bD[i] \mid \btx, \brs, \calO) \le H_b(p_e) \le H_b(41/100) \le 0.978
$$
where $H_b$ is the binary entropy function, which is increasing on $[0, 1/2]$,
meaning we can pick our constant as $c_0 = 0.022$.
\end{proof}

Finally, we prove our communication lower bound for PIR with correlated randomness in Theorem~\ref{thm:main_pir} by combining 
Lemma~\ref{lem:fano} across all $N$ database entries.

\begin{proof}[Proof of~\thmref{thm:main_pir}]
We will prove that the mutual information between
a uniformly random database $\bD \in \bits^N$ and
a communication transcript $\btx$, given the server's randomness $\brs$ and the oracle $\calO$, is linear in the
database size. Note, we see that
$$
I(\bD; \btx \mid \brs, \calO) = H(\bD \mid \brs, \calO) - H(\bD \mid \btx, \brs, \calO) \ge N - \sum\limits_{i = 1}^N H(\bD[i] \mid \btx, \brs, \calO)
\ge N - N(1-c_0) = \Omega(N)$$
where we use that $\bD$ is independent of $(\brs, \calO)$ and the sub-additivity of entropy (Lemma~\ref{lem:entropy}(\ref{item:ent_ssa})) for
the first inequality and Lemma~\ref{lem:fano} for the second inequality with the constant $c_0 = 0.022 > 0$. Finally, we note that
$I(\bD; \btx \mid \brs, \calO) \le H(\btx \mid \brs, \calO) \le H(\btx) \le |\btx|$ meaning that $|\btx| = {\Omega(N)}$ bits completing the linear communication lower bound proof. Suppose finally that the transcript length is not fixed but has expectation $c$. The same argument applies with $H(\btx) \le H(|\btx|) + H(\btx \mid |\btx|) \le \log(e(c+1)) + c$, so the theorem also holds for expected online communication $c = o(N)$. This is the form used in Section~\ref{sec:mpc}.
\end{proof}

Next, we show that this extends to databases with $N$ entries larger than a single bit. If we consider databases of $N$ entries each with $b$ bits, we show that single-server PIR with correlated randomness still requires $\Omega(N \cdot b)$ communication.

\begin{theorem}
\label{thm:main_pir_large_entries}
    Relative to a random oracle $\calO$, there does not exist a single-server PIR for an $N$ $b$-bit entry database with correlated randomness with error probability at most $\epsilon$ that is $\delta$-private, where $2\epsilon + 4\delta \le 2/5$, with $o(N \cdot b)$ expected online communication for a single query.
\end{theorem}
\begin{proof}
    Note, we can build a single-server PIR $\Pi'$ for a database with $N \cdot b$ $1$-bit entries using a single-server PIR $\Pi$ for $N$ $b$-bit entries. To run a query for $\Pi'$ for index $i \in [N \cdot b]$, we simply execute $\Pi$ with index $\ceil{i / b}$. By Theorem~\ref{thm:main_pir}, there cannot exist $\Pi'$ with $o(N \cdot b)$ communication. Since $\Pi'$ uses the same communication as $\Pi$, we immediately obtain that $\Pi$ also must use $\Omega(N \cdot b)$ communication.
\end{proof}

\subsection{Lower Bound for Classical MPC with Preprocessing}
\label{sec:mpc}

To obtain our proof, we will use the framework introduced by Damg{\aa}rd, Larsen
and Nielsen~\cite{C:DamLarNie19} that was used to prove communication lower
bounds for statistically-secure MPC with preprocessing. We note that the proof
remains nearly identical with the exception that we extend the result to the
random oracle model that requires reducing to our communication lower bounds for
PIR with correlated randomness in the random oracle model that we proved in
Section~\ref{sec:pir}.

\begin{theorem}
\label{thm:main_mpc}
Let $n \ge 2$, $1 \le t \le n-1$, $\ell \ge \log n$ and $2\epsilon + 6\delta \le
2/5$.
For any field $\mathbb{F}_q$ of size $q$, there is an $n$-party function $f_\ell$ with a circuit $C$ over $\mathbb{F}_q$ of size
$|C| = O(n \cdot \ell)$ such that, relative to a random oracle $\calO$, no
$n$-party protocol for $f_\ell$ in the preprocessing model with error
probability at most $\epsilon$ that is $\delta$-private against $t$ non-adaptive corruptions
has expected online communication
$$o\left(|C| \cdot \log q \cdot\frac{t}{n-t}\right).$$ 
\end{theorem}
\begin{proof}
We prove our lower bound for the $n$-party function $f_\ell$, defined as
follows, for integer $\ell \ge 1$. Let $m = \lceil \log (n \ell) \rceil$. Party
$P_i$, for $i \in [n]$, will receive three inputs: a tuple of $\ell$ field elements $d_i \in
\mathbb{F}_q^\ell$, an index $z_i \in \bits^m$ and a bit $b_i \in \bits$. To define the
output of $f_\ell$, we consider the virtual database of $2^m$ entries of field elements $D = d_1 \|
d_2 \| \ldots \| d_n \| 0^{2^m - n\ell}$ obtained by concatenating the $n$
$\ell$-tuple inputs $d_1,\ldots,d_n$ and padding with zeros, and we index
its entries by $\bits^m$. The output for party $P_i$ is defined as $b_i \cdot
D[Z]$, where $Z = z_1 \oplus z_2 \oplus \ldots \oplus z_n$.

We note that the function $f_\ell$ can be computed by a circuit $C$ of
size $|C| = O(n\ell)$. To see this, we note that there exists an $O(nm) = O(n
\log (n\ell))$ circuit for computing $Z$. Next, the circuit for the storage
access function that, on input database $D$, returns $D[Z]$ has size $O(|D|) =
O(n\ell)$. Finally, computing $b_i \cdot D[Z]$ for all parties $i$ given $D[Z]$
requires an $O(n)$ circuit. Thus, $f_\ell$ has a circuit of size $O(n\ell + n
\log(n\ell)) = O(n\ell)$, as $\ell \ge \log n$.

Next we consider $k = \floor{n/(n-t)}$ disjoint groups of players, each of size
exactly $n-t$. Note, there might exist some additional parties not in any of the
$k$ groups. Fix any one of these groups of $n-t$ parties. For convenience,
suppose that this group consists of parties $P_{t+1},\ldots,P_{n}$; for any
other group, the inputs of the $t$ parties outside the group occupy $t\ell$
known positions of $D$, one party of the group carries the index in place of
$P_n$, and the argument below is unchanged.
We consider the following inputs for the $n$ parties that will simulate a PIR
with parties $P_{1},\ldots,P_{t}$ playing the role of the PIR server with
database $D'=d_1 \| \cdots \| d_t$ and the parties, $P_{t+1},\ldots,P_{n}$, of
the group playing the role of the PIR client. Specifically, we set
$b_1=b_2=\ldots=b_t=0$ to ensure that the server parties receive $0$ as output;
and we set $b_{t+1} = \ldots = b_n = 1$ so that the client parties receive
$D[Z]$. The client parties use $d_a = 0^\ell$ for $a \in \{t+1, \ldots, n\}$. Moreover, by setting $z_1 = z_2 = \ldots = z_{n-1} = 0^m$ we have the
effect of setting $Z=z_n$; that is, $z_n$ is the index of the database entry
that is received by the client. The correlated randomness of the PIR is that of
the preprocessing, messages inside the server or inside the client are internal
computation, and the transcript $\tx$ consists of the messages between $P_1,
\ldots, P_t$ and $P_{t+1}, \ldots, P_n$.

The resulting single-server PIR with correlated randomness has error probability
at most $\epsilon$. It is also $\delta$-private. Indeed, for any two indices
$i_0, i_1$ and any fixed $D'$, the inputs and outputs of the server parties $S =
\{P_1, \ldots, P_t\}$ are the same, so the simulator $\Sim_S$ outputs the same
distribution for both. An adversary making $Q$ queries that sees $(r_s, D',
\tx)$, which is part of $\View_S$, therefore distinguishes the two views with
advantage at most $2\delta$, that is, it guesses the index with probability at
most $1/2 + \delta$. Hence, by Lemma~\ref{lem:any_index} with $\gamma = 1/200$,
there is an eavesdropper $\calE$ making $\poly(\secpar)$ queries to $\calO$ such
that, for the transcript $\btx$ of index $1$, the correlated randomness $\brs$
of the server parties and a uniformly random $\bD'$,
\begin{equation}
\label{eq:mpc_eaves}
\Pr\left[\calE^\calO(\brs, \btx, i) = \bD'[i]\right] \ge 1 - 2\epsilon - 2\gamma
- 4\delta \qquad \text{for every } i \in [t\ell].
\end{equation}

The above bound is obtained in an input configuration that depends on the group.
To add the bounds over all groups, we fix a single input $x^*$ in which $b_a =
0$ and $z_a = 0^m$ for every party $a$ and every $d_a$ is uniformly random. For
every fixed $D'$, the inputs and outputs of the server parties $S$ are the same
under $x^*$ as in the configuration above with index $1$, so their simulated
views coincide. As the event in~\eqref{eq:mpc_eaves} is decided by a
$\poly(\secpar)$-query algorithm run on $\View_S$, its probability under $x^*$
is at least
$$
1 - 2\epsilon - 2\gamma - 6\delta \ge \frac{59}{100}.
$$
By Fano's inequality, as in Lemma~\ref{lem:fano}, under $x^*$ we have
$$
H(\bD'[i] \mid \btx, \brs, \calO) \le H_b(41/100) \le 0.978 \qquad
\text{for every } i \in [t\ell],
$$
for the case of $q = 2$ as done in Theorem~\ref{thm:main_pir}. Then, following the proof of Theorem~\ref{thm:main_pir_large_entries} for arbitrary $q \ge 2$ then shows that the expected
communication between $P_1, \ldots, P_t$ and $P_{t+1}, \ldots, P_n$ under $x^*$
is $\Omega(t \ell \log q)$.

Finally, the input $x^*$ does not depend on the group, so the above applies to
each of the $k$ groups and its complement. Each message has one sender and one
recipient, each of which lies in at most one group, so a message is counted for
at most two groups. Summing over the $k \ge n/(2(n-t))$ groups, the expected
communication of the protocol under $x^*$ is at least
$$
\frac{1}{2} \cdot k \cdot \Omega(t\ell\log q) = \Omega\left(n \cdot \ell \cdot \log q \cdot
\frac{t}{n-t}\right),
$$
and so this holds for some fixed input.
\end{proof}

\subsection{Lower Bound for Classical MPC with an Honest Majority}

We also prove a lower bound for MPC protocols in the honest majority setting. However, we are only able to prove a lower bound of $\Omega(|C|)$ that is smaller than our above $\Omega(|C| \cdot t /(n-t))$ lower bound for MPC with preprocessing along with the prior ones proved in~\cite{C:DamLarNie19}. This turns out to be inherent as we will later present a construction for the inner product functionality with $O(|C|)$ communication matching our lower bound. The construction extends to every circuit of multiplicative depth $1$ whose multiplication gates read the inputs of $O(1)$ parties, for $n = O(\log \secpar)$ (see Section~\ref{sec:counter_quad}).

\begin{theorem}
\label{thm:honest_majority_mpc_lb}
Let $n \ge 2$, $1 \le t \le n/2 - 1$, $\ell \ge \log n$ and $2\epsilon + 6\delta \le 2/5$. For any field $\mathbb{F}_q$ of size $q$, there is an $n$-party function $f_\ell$ with a circuit $C$ over $\mathbb{F}_q$ of size $|C| = O(t \cdot \ell)$ such that, relative to a random oracle $\calO$, no $n$-party protocol for $f_\ell$ in the preprocessing model with error probability at most $\epsilon$ that is $\delta$-private against $t$ non-adaptive corruptions has expected online communication $o(|C| \cdot \log q)$.
\end{theorem}
\begin{proof}
    We define function $f_\ell$ such that only $t+1$ parties obtain inputs and only one party obtains the output. The $t$ parties $P_1, \ldots, P_t$ each obtain an $\ell$-tuple of field elements $x_i \in \mathbb{F}_q^\ell$. Party $P_n$ obtains an index $i \in [t \cdot \ell]$. Then, party $P_n$ should receive the output $(x_1 \ldots x_t) \cdot \mathbb{1}_i = (x_1 \ldots x_t)[i]$ that is the $i$-th entry in the tuple of $t \cdot \ell$ field elements obtained by concatenating the $t$ input tuples of $\ell$ field elements of the parties $P_1,\ldots,P_t$.

    We can immediately obtain an instance of single-server PIR with correlated randomness by having the adversary non-adaptively corrupt the $t$ parties $P_1, \ldots, P_t$. These $t$ parties act as the server with the database of $t \cdot \ell$ field elements. Party $P_n$ acts as the client with input $i \in [t \cdot \ell]$. Note, the resulting protocol has the error probability $\epsilon$ and $\delta$-privacy as the MPC protocol. Therefore, we can apply Theorem~\ref{thm:main_pir_large_entries} to obtain there must be $\Omega(t \cdot \ell \cdot \log q)$ communication leaving the set of parties acting as the server $P_1,\ldots,P_t$. The argument follows nearly identically to the proof in Theorem~\ref{thm:main_mpc}.
\end{proof}

We note this lower bound is stronger than the one that we could immediately obtain from plugging in honest majority values of $t$ and $n$ into Theorem~\ref{thm:main_mpc}. As an example, if we consider $t = 1$, Theorem~\ref{thm:main_mpc} obtains a lower bound of $\Omega(\ell \log q)$ where $|C| = O(n \cdot \ell)$, which does not obtain an $\Omega(|C| \log q)$ lower bound if $n \gg t$. In contrast, our lower bound would prove an $\Omega(t \cdot \ell \cdot \log q) = \Omega(|C|\log q)$ lower bound as we consider smaller circuits of $|C| = O(t \cdot \ell)$ whose size depends on $t$.

\subsection{Lower Bound for MPC with Quantum Computation}
\label{sec:qccc_lb}

Theorems~\ref{thm:main_pir} and~\ref{thm:main_mpc} model the parties as
classical machines with classical oracle access. However, almost every step of
the proof is statistical. This means that we can extend our results to
protocols (and adversaries) which support \emph{quantum computation}. We note,
however, that we cannot obtain quantum preprocessing nor quantum communication,
as we show a counterexample to such bounds in Section~\ref{sec:qmpc}. This
section is instead dedicated to extending our bounds to the quantum-computation
classical-communication (QCCC) model of~\cite{C:ACCFLM22} (with classical
preprocessing) in the quantum random oracle model (QROM)~\cite{AC:BDFLSZ11}, as
defined in Sections~\ref{sec:prelim_models} and~\ref{sec:prelim_quantum}.

The proofs of Section~\ref{sec:pir} use the classical queries of the parties in
one place only: the key-agreement attack of Theorem~\ref{thm:ke_attack}, invoked
in Lemma~\ref{lem:query_index}. Every other step runs the eavesdropper on the
classical strings $(r_s, \tx, i)$ or reasons about them
information-theoretically, without rewinding.

Unfortunately, the impossibility of key agreement in the QCCC model is not fully
resolved, as it is famously in the classical setting, since the classical
attacks reconstruct the parties' oracle queries from the transcript, which is
not defined for parties that query in superposition. However, there has been some notable progress in showing the
QCCC barrier. In particular, we use the recent QCCC key-agreement attack of
Li, Li, Li and Liu~\cite{ARXIV:LLLL26} (Theorem~\ref{thm:qccc_ke_attack}). We
may also import any other barriers, such as~\cite{ARXIV:KNVY26,C:LLLL24} (see
Remark~\ref{rem:qccc_imperfect}), and any future lower bounds for QCCC
key-agreement shown in the future will immediately strengthen the stated
corollary here.

\begin{corollary}
\label{cor:qccc}
Relative to a quantum-accessible random oracle $\calO$,
Theorem~\ref{thm:main_pir} for $\epsilon = 0$ (perfectly correct schemes with
$\delta \le 1/10$) and Theorem~\ref{thm:main_mpc} for perfectly correct
protocols (with $\delta \le 1/15$) hold for QCCC protocols with classical preprocessing.
\end{corollary}
\begin{proof}
We follow the proof of Theorem~\ref{thm:main_pir}, replacing
Theorem~\ref{thm:ke_attack} by Theorem~\ref{thm:qccc_ke_attack} in
Lemma~\ref{lem:query_index}. Recall the input-free protocol $\Pi_i$ used there.
Alice runs $\init^\calO$, sends $r_s$, and runs the client on $(r_c, i)$. Bob
samples $D \getsr (\bits^b)^N$ as part of his random tape, runs the server on
$(r_s, D)$ and outputs $D[i]$, while Alice outputs $\ans$. The parties start in
a product state, every message is classical, and they make $d = \poly(\secpar)$
queries in total. Perfect correctness holds for every database, so the two
outputs agree with probability $1$. Let $\calE^\calO(r_s, \tx, i)$ run the
attacker of Theorem~\ref{thm:qccc_ke_attack} on $(r_s, \tx)$. Then, for every
$i \in [N]$,
$$
\Pr\left[\calE^\calO(\brs, \btx, i) = \bD[i]\right] = 1
$$
over $\calO$, the coins and measurements of $\init^\calO$ and $\query^\calO$,
and $\bD \getsr (\bits^b)^N$ with $b = O(\log|\mathbb{F}_q|) = O(\log q)$. This is Lemma~\ref{lem:query_index} with
$\epsilon = 0$ and any $\gamma$ (see Remark~\ref{rem:uniform_db}).

The remaining steps are unchanged. The adversary of Lemma~\ref{lem:any_index}
runs $\calE$ twice on the classical input $(r_s, \tx)$, so no rewinding is
involved, and it makes $\poly(\secpar)$ classical queries. Lemma~\ref{lem:fano}
uses only that the distribution of $\calE$'s output depends on nothing beyond
$\calO$ and $(r_s, \tx, i)$, and that $\bD$ is independent of $(\brs, \calO)$;
with $\epsilon = 0$ it gives $p_e \le 4\delta \le 2/5$. For MPC, the $n-t$
parties of a group form a single QCCC client and the remaining $t$ parties a
single QCCC server, with $\init$ the MPC preprocessing. Messages inside a group
are classical and become internal computation, and the query budgets add. As
$(r_s, D, \tx)$ is part of the corrupted parties' classical view, semi-honest
security gives $\delta$-privacy for $\delta = \negl(\secpar)$. The rest of the
proof of Theorem~\ref{thm:main_mpc} is unchanged.
\end{proof}

\begin{remark}[Imperfect correctness]
\label{rem:qccc_imperfect}
Perfect correctness in the above corollary is inherited directly
from~\cite{ARXIV:LLLL26}. Lemma~\ref{lem:query_index} only needs an eavesdropper
that recovers Bob's output with probability $1 - \epsilon - \gamma$ for every
constant $\gamma > 0$. Such attacks are known for key agreements of specific
forms, so Theorem~\ref{thm:main_pir} additionally is known to hold with
$\epsilon > 0$ for PIR schemes whose protocol $\Pi_i$ has one of the following:
\begin{itemize}[nosep]
    \item every oracle query made before Bob computes his last message is
      classical, while Bob's last move and Alice's final computation may query
      in superposition~\cite[Theorem~5.14]{ARXIV:KNVY26}; this includes the case
      where both parties are classical-query~\cite{C:BarMah09};
    \item the protocol has two messages and Alice's final computation makes no
      oracle queries~\cite[Theorem~5.4]{C:LLLL24}.
\end{itemize}
\end{remark}

\section{Honest Majority MPC for Degree-2 Polynomials}
\label{sec:counter_example}
\label{sec:counter_quad}

Theorem~\ref{thm:honest_majority_mpc_lb} shows that, relative to a random
oracle, honest majority MPC for some circuits $C$ over a field $\FF$
communicates $\Omega(|C|)$ elements of $\FF$, even with preprocessing. In
this section, we show that this bound is tight for degree-2 polynomials
when $n = O(\log \secpar)$. For every polynomial $g$ of degree at most
2 over a finite field $\FF$ and every $n = 2t + s$ with $t, s \ge 1$, we
give a protocol $\Prep$ in the random oracle model without preprocessing
that is private against $t$ adaptive corruptions
(Theorem~\ref{thm:rom_quad}). For $n = O(\log \secpar)$, it communicates
$O(|C|)$ elements of $\FF$ plus $\poly(\secpar)$ bits for every circuit
$C$ for $g$ in which each multiplication gate reads the inputs of $O(1)$
parties (Corollary~\ref{cor:rom_quad_circuit}). The protocol uses the oracle only as
a pseudorandom generator.

Notably, the function $f$, the sum of $n/2$ inner products, is degree-2
and was used in the proof of the statistical
$\Omega(|C| \cdot t/s)$ bound~\cite[Theorem~1]{C:DamLarNie19}. For
$n = O(\log \secpar)$, $\Prep$ computes it with $n\ell$
elements of $\FF$ plus $\poly(\secpar)$ bits (Section~\ref{sec:counter_dln}).
The random oracle thus saves a factor of $t/s$ over every statistically
secure protocol, even one with quantum communication
(Theorem~\ref{thm:statistical_nopre}). For
$n = 3$, the protocol shows that the three-party lemma behind that bound
fails relative to a random oracle (Section~\ref{sec:counter_three}).

\heading{Structure of the protocol.} The protocol $\Prep$ evaluates every
bilinear part of $g$ between a pair of parties by the circuit randomization
of Beaver~\cite{C:Beaver91b}, with the correlated randomness dealt by the
other $n - 2$ parties through the pseudorandom secret sharing of Cramer,
Damg{\aa}rd and Ishai~\cite{TCC:CraDamIsh05}. For a pair holding $y, z \in
\FF^\ell$, it has four phases, which the figure marks. In the
\emph{Deal} phase the two parties obtain pads $r, r' \in
\FF^\ell$ along with shares $C + C' = \langle
r, r' \rangle$ of the inner product of the pads; nothing in this
phase depends on the inputs. In the \emph{Mask} phase the two parties
exchange the masked inputs $u = y - r$ and $v =
z - r'$, the only messages whose length is that of the inputs. In the
\emph{Share} phase the two parties compute shares $s_A,
s_B \in \FF$ such that
\begin{equation}
\label{eq:ce_shares}
s_A + s_B = \big( \langle r, v \rangle +
C \big) + \big( \langle u, z \rangle + C' \big) =
\langle y, z \rangle.
\end{equation}
In the \emph{Add} phase the shares of all pairs are added by a cheap
honest majority protocol. At its core, $\Prep$ beats the bound
of~\cite{C:DamLarNie19} because the dealers never send the pads. They send
a $\secpar$-bit seed per pad, which is not possible statistically. Cramer et
al.\ also evaluate degree-2 polynomials in this
way~\cite[Section~4.1]{TCC:CraDamIsh05}, from keys distributed in advance
and with every masked input sent to every party, which on $f$ costs
$\Theta(n^2 \ell)$ elements and does not beat the bound
of~\cite{C:DamLarNie19}. In $\Prep$, each masked input is sent to one party,
and the parties deal the keys themselves at a cost independent of $\ell$.

\heading{Conventions.} Throughout this section, we consider classical
$n$-party protocols without preprocessing in the random oracle model
(Section~\ref{sec:prelim_models}), with privacy against adaptive corruptions
as in Definition~\ref{def:rom_private}. For $m \ge 1$ we write
$\calO^{\FF}_m(z) \in \FF^m$ for $m$ elements of $\FF$ read from $\calO_{mk}(z)$: the
$j$-th is the integer with the bits $(j-1)k + 1, \ldots, jk$ of $\calO_{mk}(z)$
reduced modulo $q$, for $k = \lceil \log q \rceil + \secpar$, where
we identify $\FF$ with $\{0, \ldots, q - 1\}$ as a set. Each element is
within $2^{-\secpar}$ of uniform in statistical distance, and exactly uniform
when $q$ is a power of two and $k = \log q$. The protocol calls the
ideal functionality $\Fsum$ of Figure~\ref{fig:ce_functions}, which
Lemma~\ref{lem:composition} replaces by a perfectly private protocol. In the
privacy proof, the simulator runs the corrupted parties itself, so we
describe their views by the values they receive from honest parties.

\subsection{The Protocol $\Prep$}
\label{sec:counter_n}
\label{sec:prep}

\heading{The function.} Let $g \colon (\FF^\ell)^n \to \FF$ be a polynomial
of degree at most 2 in the entries of $x_1, \ldots, x_n \in \FF^\ell$.
Grouping its monomials by the parties they involve,
$$
g(x_1, \ldots, x_n) = \sum_{p \in [n]} g_p(x_p) +
\sum_{\{p, p'\} \in \calP} g_{p,p'}(x_p, x_{p'}),
$$
where $g_p$ has degree at most 2, $g_{p,p'}$ is bilinear and $\calP$ is
the set of pairs $\{p, p'\}$ with a monomial $x_p[a] \cdot x_{p'}[b]$ in
$g$. For $e = \{p, p'\} \in \calP$ with $p < p'$ we write $g_e(a, b) =
a^\top M_e \, b$ for a matrix $M_e \in \FF^{\ell \times \ell}$ of rank
$\ell_e \le \ell$, and factor $M_e = L_e^\top R_e$ with $L_e, R_e \in
\FF^{\ell_e \times \ell}$, so that $g_e(a, b) = \langle L_e \, a, R_e \, b
\rangle$. We let $f_g$ be the $n$-party function in which party $P_p$ holds
$(x_p, b_p) \in \FF^\ell \times \bits$ and receives $b_p \cdot g(x_1,
\ldots, x_n)$ (Figure~\ref{fig:ce_functions}); as in
Theorem~\ref{thm:main_mpc}, the bit $b_p$ selects whether $P_p$ receives
the output. A degree-2 polynomial is what an arithmetic circuit of
multiplicative depth 1 computes, and we say degree 2 throughout.

\begin{figure}[t]
\centering
\input{figures/functionalities}
\caption{The function $f_g$ and the functionality $\Fsum$ called by $\Prep$
(Section~\ref{sec:counter_n}).}
\label{fig:ce_functions}
\end{figure}

\heading{Pairs, helpers and seed sets.} For every pair $e = \{p, p'\} \in
\calP$ with $p < p'$, we call $A_e := P_p$ and $B_e := P_{p'}$ the parties
of the pair, holding $y_e := L_e \, x_p$ and $z_e := R_e \, x_{p'}$ in
$\FF^{\ell_e}$, and the \emph{helpers} $H_e := [n] \setminus e$ of the pair
are the other $n - 2$ parties. The \emph{seed sets} $\calJ_e$ of the pair
are all subsets of $H_e$ of size $n - t - 1$. We fix an order on the parties
and let $H(J)$ be the first party of $J \in \calJ_e$ and $H(J, J')$ the
first party of $J \cap J'$ for $J, J' \in \calJ_e$. We use two facts about
seed sets.
\begin{enumerate}
    \item \label{item:ce_meet} Any $J, J' \in \calJ_e$ intersect, as
      $2(n - t - 1) - (n - 2) = s > 0$. So $H(J, J')$ is well defined.
    \item \label{item:ce_miss} Every set of at most $t - 1$ helpers is
      disjoint from some $J \in \calJ_e$, as $(n - 2) - (t - 1) = n - t -
      1$. So if one party of the pair and $t - 1$ helpers are corrupted,
      there is a pad of the other party that they do not know.
\end{enumerate}

\heading{The protocol.} The protocol $\Prep$ is given in
Figure~\ref{fig:pin}. For every pair $e \in \calP$, the parties run the
Deal, Mask and Share phases with $A_e$ holding $y_e$ and $B_e$ holding
$z_e$ as the two parties above, and with the helpers $H_e$ jointly as the
dealer.
In the Deal phase, for every seed set $J$, the party $H(J)$ selects the
seeds $k_J$ and $k'_J$ of one pad of $A_e$ and one pad of
$B_e$ and sends them to all parties in $J$, the first also to $A_e$ and the
second also to $B_e$ (line~\ref{line:pin_seeds}). The pads are the outputs
of $\calO$ on the seeds, labeled by the pair and the seed set, and the pad
of each party is the sum of its $|\calJ_e|$ pads
(line~\ref{line:pin_pads}). Every cross term $\langle r_J,
r'_{J'} \rangle$ of the two pads is computed by the helper $H(J, J')$,
which lies in both $J$ and $J'$ and so holds both seeds. Every helper
splits the sum $c_h$ of its cross terms into a uniform share for
$A_e$ and the complementary share for $B_e$ (line~\ref{line:pin_cross}).
The seeds, pads and shares of lines~\ref{line:pin_seeds}
to~\ref{line:pin_cross} are separate for each pair, and we omit the index
$e$ on them. In the Add phase, every party adds its local terms $g_p(x_p)$
to its shares of all its pairs and calls $\Fsum$ with its output bit
(line~\ref{line:pin_out}).

\begin{figure}[t]
\centering
\input{figures/pi_rep}
\caption{The $n$-party protocol $\Prep$ for the function $f_g$ of
Figure~\ref{fig:ce_functions}. The strings $e$ and $J$ in the inputs of
$\calO$ are fixed encodings of the pair and of the seed set.}
\label{fig:pin}
\label{fig:Prep}
\end{figure}

\begin{lemma}
\label{lem:fsum}
For $t < n/2$, there is a perfectly correct protocol for $\Fsum$ without
oracle that is perfectly private against $t$ corruptions and communicates
$O(n^3 (\log q + \log n))$ bits.
\end{lemma}
\begin{proof}
$\Fsum$ is computed by an arithmetic circuit over $\FF$ with $2n$ inputs,
$n$ multiplications and $n$ outputs, so
Theorem~\ref{thm:bgw}~\cite{STOC:BenGolWig88,JC:AshLin17} applies.
\end{proof}

\subsection{Analysis}
\label{app:adaptive}

\begin{theorem}
\label{thm:rom_quad}
Let $n = 2t + s$ with $t, s \ge 1$, let $\secpar \ge 1$, let $\FF$ be a
finite field and let $g$, $\calP$ and $\ell_e$ be as above, with $L :=
\sum_{e \in \calP} \ell_e$. The protocol $\Prep$ is a protocol for $f_g$ in
the $\Fsum$-hybrid model that is perfectly correct and, for every $Q$, is
$(Q, (Q + L) \cdot 2^{-\secpar})$-private against $t$ adaptive corruptions,
where the term $L \cdot 2^{-\secpar}$ vanishes when $q$ is a power of
two. It communicates
$$
\sum_{e \in \calP} \big( 2\ell_e + 2(n - 2) \big) \ \text{ elements of~~}
\FF \quad \text{and} \quad |\calP| \cdot 2 \secpar (n - t - 1)
\binom{n-2}{t-1} \ \text{ bits.}
$$
\end{theorem}

\begin{proof}
We establish correctness, privacy, and the communication in turn.

\heading{Correctness.} Fix a pair $e \in \calP$. By
fact~\ref{item:ce_meet}, every $(J, J') \in \calJ_e^2$ is counted by
exactly one helper in line~\ref{line:pin_cross}, so
\begin{equation}
\label{eq:ce_cross}
C_e + C'_e = \sum_{h \in H_e} \big( c_{1,h} + c_{2,h} \big) = \sum_{J, J'}
\langle r_J, r'_{J'} \rangle = \langle r_e, r'_e \rangle.
\end{equation}
Hence
$$
s_{A_e} + s_{B_e} = \langle r_e, z_e - r'_e \rangle + \langle y_e - r_e, z_e
\rangle + \langle r_e, r'_e \rangle = \langle y_e, z_e \rangle = g_e(x_p,
x_{p'}),
$$
which is~\eqref{eq:ce_shares}. Summing over $\calP$ and adding the local
terms $g_p(x_p)$, the inputs to $\Fsum$ sum to $g(x)$, and party $p$
receives $b_p \cdot g(x)$.

\heading{Privacy.} Fix an adversary $\calA$ that makes at most $Q$ queries
to $\calO$ and corrupts at most $t$ parties. As the simulator runs the
corrupted parties itself, the view of a corrupted party is determined by
its input, the values it receives from honest parties and its output of
$\Fsum$. For a pair $e \in \calP$, the values that $A_e$ receives from
honest parties are the seeds $k_J$ sent by honest parties $H(J)$ in
line~\ref{line:pin_seeds}, the elements $c_{1,h}$ sent by honest helpers
$h$ in line~\ref{line:pin_cross} and, while $B_e$ is honest, the masked
input $v_e$ of line~\ref{line:pin_mask}. For $B_e$ they are the seeds
$k'_J$, the elements $c_{2,h}$ and the masked input $u_e$, and a helper
receives only the seeds $k_J, k'_J$ of the seed sets $J$ that contain it.
In what follows we consider $A_e$; the case of $B_e$ is argued in the same
way with the roles of $k_J, r_J, c_{1,h}, u_e, y_e$ and $k'_J, r'_J,
c_{2,h}, v_e, z_e$ exchanged.

The seeds and the elements $c_{1,h}$ are uniform and independent of the
inputs, so the only value received by a corrupted party that depends on
the input of an honest party is the masked input
$$
v_e = z_e - r'_e = z_e - \sum_{J \in \calJ_e} \calO^{\FF}_{\ell_e}(k'_J \| 2 \| e \| J)
$$
that a corrupted $A_e$ receives from an honest $B_e$. We say that a seed
is \emph{hidden} while no corrupted party holds it. While $B_e$ is honest,
at most $t - 1$ helpers of the pair are corrupted, and once $B_e$ is
corrupted at most $t - 2$ are, so by fact~\ref{item:ce_miss} there is a
seed set $J^\star \in \calJ_e$ none of whose helpers is corrupted before
$B_e$ is, or at all if $B_e$ stays honest. We call $J^\star$ the
\emph{special subset} of the pair, $k'_{J^\star}$ its \emph{special seed}
and $r'_{J^\star}$ its \emph{special pad} (if there is more than one such
seed set, we pick one arbitrarily). The special seed is hidden until $B_e$
is corrupted, and the special pad masks $v_e$. The simulator, which does
not know $z_e$, will therefore send a uniform $v_e$; if $B_e$ is corrupted
later, so that $\calA$ learns $z_e$ and the special seed, it will program
$\calO$ at the point of the special seed so that $r'_e = z_e - v_e$.

The simulator $\Sim$ answers every query of $\calA$ with an independent
uniform string, keeping a table so that a repeated query receives the same
answer, and it evaluates $\calO$ through the same table whenever it runs a
party. At the start, it samples all seeds $k_J, k'_J$ of all pairs
uniformly, as the parties $H(J)$ do in line~\ref{line:pin_seeds}. When
$\calA$ corrupts a party $p$, the simulator receives $x_p$ and $f_p(x)$ and
runs $p$ as specified by $\Prep$ on $x_p$ and these seeds, computing the
pads through its table. This gives the view of $p$ up to the values that
$p$ receives from honest parties, which for a pair $e$ with $p \in e$ the
simulator produces as follows.
\begin{enumerate}
\item \label{item:sim_mask} The masked input that the first corrupted
  party of the pair receives from the other party, while that party is
  honest, is uniform in $\FF^{\ell_e}$.
\item \label{item:sim_prog} When the second party of the pair is
  corrupted after the first has received this masked input, say $B_e$
  after $A_e$ has received $v_e$, the simulator chooses the special
  subset $J^\star$ and enters into its table, at the point $k'_{J^\star}
  \| 2 \| e \| J^\star$, a uniformly random string among those that
  $\calO^{\FF}_{\ell_e}$ maps to
  $$
  z_e - v_e - \sum_{J \ne J^\star} r'_J,
  $$
  with the pads $r'_J$ evaluated through the table, so that $r'_e = z_e -
  v_e$. If the table already has an entry at this point, the simulator
  aborts.
\item \label{item:sim_cross} Of the elements $c_{1,h}$ and $c_{2,h}$ of
  an honest helper $h$, the first that $\calA$ receives is uniform in
  $\FF$, and the second, if $\calA$ receives it, is such that the two sum
  to $c_h$, computed as in line~\ref{line:pin_cross} through the table,
  after step~\ref{item:sim_prog}, from the seeds of the pair, all of
  which are by then held by corrupted parties.
\end{enumerate}
Finally, $p$ calls $\Fsum$ with the $(s_p, b_p)$ computed from its view
and receives $f_p(x) = b_p \cdot g(x)$, and a party corrupted after the
execution shows this call in its view.

We compare the ideal execution with the real one, in which we also sample
$\calO$ lazily, so that a point receives a uniform value the first time it
is evaluated by a party or queried by $\calA$. In both executions, consider
the event
\begin{center}
$\mathsf{Bad}$: $\calA$ queries $\calO$ at the point of a seed that is
hidden at the time of the query.
\end{center}
We first show that, unless $\mathsf{Bad}$ occurs, the two executions are
within $L \cdot 2^{-\secpar}$ of each other. The seeds are uniform in both.
In the real execution, the special pad enters what $\calA$ receives only
through $v_e$. The cross terms containing it are computed by the helpers
$H(J', J^\star)$ for $J' \in \calJ_e$, which lie in $J^\star$ and are
honest, and each such $c_h$ is split into $c_h = c_{1,h} + c_{2,h}$ with
$c_{1,h}$ uniform. Of these, $A_e$ receives $c_{1,h}$ and $B_e$ receives
$c_{2,h}$. Unless $\mathsf{Bad}$ occurs, the point of the special seed is
not queried, so the special pad is a fresh output of $\calO^{\FF}_{\ell_e}$,
which is within $\ell_e \cdot 2^{-\secpar}$ of uniform by the convention
of this section. Hence $v_e$ is independent of the rest of the view, as in
step~\ref{item:sim_mask}. The
elements $c_{1,h}$ of honest helpers are uniform while $B_e$ is honest and
sum with $c_{2,h}$ to $c_h$ once both parties of the pair are corrupted,
as in step~\ref{item:sim_cross}. When $B_e$ is corrupted after $A_e$ has received $v_e$, the string at the
point of the special seed is, in both executions, a uniform preimage under
$\calO^{\FF}_{\ell_e}$ of the special pad. The pad is exactly uniform in the ideal
execution and within $\ell_e \cdot 2^{-\secpar}$ of uniform in the real
one, so the two executions are within $\ell_e \cdot 2^{-\secpar}$ of each
other here as well. The simulator
aborts in step~\ref{item:sim_prog} only if $\mathsf{Bad}$ has occurred.
The output of $\Fsum$ is $f_p(x)$ in both executions, and every other
value in the views is computed by the corrupted parties from these. In
each pair at most one masked input is received by a corrupted party from
an honest one, and it is masked by the special pad, so the distances add
up to at most $\sum_{e \in \calP} \ell_e \cdot 2^{-\secpar} = L \cdot
2^{-\secpar}$.

It remains to bound the probability of $\mathsf{Bad}$ in the ideal
execution. There, the point of a hidden seed is never evaluated. The
simulator runs only corrupted parties, which hold their seeds, and
computes the cross terms of an honest helper in step~\ref{item:sim_cross}
only when both parties of the pair are corrupted and no seed of the pair
is hidden. It programs the point of a special seed in
step~\ref{item:sim_prog} at the moment this seed ceases to be hidden. So
the hidden seeds are uniform and independent of everything that $\calA$
has received. We observe that a query specifies the pair, the seed set and
one of the labels $1$ and $2$, so each point queried by $\calA$ is the
point of at most one seed. This implies that every time $\calA$ makes a
query, the query has probability at most $2^{-\secpar}$ of being the point
of a hidden seed, despite the fact that there could be as many as $2
|\calP| \binom{n-2}{t-1}$ hidden seeds. Therefore
$$
\Pr[\mathsf{Bad}] \le Q \cdot 2^{-\secpar},
$$
and the outputs of $\calA$ in the two executions differ by at most $Q
\cdot 2^{-\secpar} + L \cdot 2^{-\secpar}$. We can conclude that $\Prep$ is
$(Q, (Q + L) \cdot 2^{-\secpar})$-private against $t$ adaptive
corruptions.

\heading{Communication.} We count by phase for a pair $e$. In the Deal
phase, each of the $\binom{n-2}{t-1}$ seed sets costs $2\secpar(n - t - 2)$
bits to its other parties and $2\secpar$ bits to the pair
(line~\ref{line:pin_seeds}), and each of the $n - 2$ helpers sends $2$
elements (line~\ref{line:pin_cross}). In the Mask phase the pair exchanges
$2 \ell_e$ elements (line~\ref{line:pin_mask}).
\end{proof}

\begin{corollary}
\label{cor:rom_quad}
Let $n$, $t$, $s$, $\secpar$, $\FF$, $g$ and $L$ be as in
Theorem~\ref{thm:rom_quad}. For every $Q$, the function $f_g$ has a protocol
in the random oracle model without preprocessing that is $(Q, (Q + L)
\cdot 2^{-\secpar})$-private against $t$ adaptive corruptions and
communicates $2L + O(n^2 |\calP|)$ elements of $\FF$ and $|\calP| \cdot 2
\secpar (n - t - 1) \binom{n-2}{t-1} + O(n^3 \log n)$ bits, which is $2L$
elements plus $\poly(\secpar)$ bits for $n = O(\log \secpar)$.
\end{corollary}
\begin{proof}
Replace the call to $\Fsum$ by the protocol of Lemma~\ref{lem:fsum} and
apply Lemma~\ref{lem:composition} to Theorem~\ref{thm:rom_quad}. For $n =
O(\log \secpar)$, we have $|\calP| \le n^2$ and $\binom{n-2}{t-1} < 2^n =
\poly(\secpar)$.
\end{proof}

\begin{corollary}
\label{cor:rom_quad_circuit}
Let $C$ be an arithmetic circuit over $\FF$ of multiplicative depth $1$
computing $g$, that is, every multiplication gate of $C$ multiplies two
affine functions of the inputs, and suppose that every multiplication gate
reads the inputs of at most $k$ parties. Then the protocol of
Corollary~\ref{cor:rom_quad} communicates at most $2k(k-1) |C|$ elements of
$\FF$ plus the terms of Corollary~\ref{cor:rom_quad} that do not depend on
$\ell$, which is $O(|C|)$ elements plus $\poly(\secpar)$ bits for $k = O(1)$
and $n = O(\log \secpar)$.
\end{corollary}
\begin{proof}
A multiplication gate $(\sum_p a_p(x_p) + \alpha) \cdot (\sum_p b_p(x_p) +
\beta)$ with $a_p, b_p$ linear adds $a_p b_{p'} + a_{p'} b_p$, a bilinear
form of rank at most $2$, to $g_{p,p'}$ for each of its $\binom{k}{2}$
pairs, so $L \le k(k-1) \cdot |C|$.
\end{proof}

For gates that read the inputs of $\Theta(n)$ parties the count is
$\Theta(n^2)$ elements per gate, no better than the information-theoretic
protocols, and for degree-3 polynomials we have no protocol below their
cost. For a polynomial map $g = (g^1, \ldots, g^m)$ with $m$ outputs, the
protocol is run once per output and the shares are added by one call to
$\Fsum$ with $m$-element inputs, which multiplies the communication by $m$;
this $\Fsum$ costs $O(n^3 m (\log q + \log n))$ bits by
Theorem~\ref{thm:bgw}.

\heading{Non-adaptive corruptions and pseudorandom generators.} Against
non-adaptive corruptions the simulator of Theorem~\ref{thm:rom_quad} never programs the
oracle: its third exception arises only when a party is corrupted after
sending its masked input. It then answers the queries of $\calA$ by lazy
sampling, that is, it does not program $\calO$, so $\Prep$ is also
private against $t$ non-adaptive corruptions when the simulator may only query
$\calO$. This simulator uses the oracle only to expand each $\secpar$-bit
seed into one pad, so any pseudorandom generator can replace it, giving a protocol in the standard model that is private against
polynomial-time adversaries and $t$ non-adaptive corruptions.

\heading{Quantum adversaries.} The proof of Theorem~\ref{thm:rom_quad} uses
the oracle in two ways. The pad of a hidden seed is uniform unless its
point is queried, and the special pad of a pair is programmed once. Both
steps carry over to adversaries that query $\calO$ in superposition, as in
the QCCC model of Section~\ref{sec:prelim_models}, at the cost of a square
root in the bound. The simulator answers superposition queries by
simulating a random function (Theorem~\ref{thm:ro_sim}). For the hidden
seeds, the one-way to hiding lemma of~\cite{C:AmbHamUnr19} bounds the
advantage of an adversary making $Q$ superposition queries by $O(Q \sqrt{m}
\cdot 2^{-\secpar/2})$, where $m$ is the number of seeds, in place of $Q
\cdot 2^{-\secpar}$. For the programmed point, whose seed is uniform and
unknown to the adversary at that time, the adaptive reprogramming lemma
of~\cite{AC:GHHM21} bounds the advantage by $O(\sqrt{Q} \cdot
2^{-\secpar/2})$ per pair. So $\Prep$ is private against such adversaries
with $\delta = \poly(Q, \secpar) \cdot 2^{-\secpar/2}$ for $n = O(\log
\secpar)$, which is still negligible. We only sketch the argument here. In
the standard model, the protocol above is private against quantum
polynomial-time adversaries and $t$ non-adaptive corruptions whenever the
pseudorandom generator is post-quantum secure.

\heading{Rings.} The Deal, Mask and Share phases use only additions and
multiplications, so $\Prep$ runs over any finite commutative ring $R$ whose
elements can be sampled from random bits. The bilinear parts $g_e$ are
then split into their monomials, or gate by gate as in
Corollary~\ref{cor:rom_quad_circuit}, in place of the rank factorization,
which need not exist over a ring. The functionality $\Fsum$ has a perfectly
private honest majority protocol over any finite ring whose cost does not
depend on $\ell$~\cite{EC:CFIK03}. We present the protocol over fields
because our lower bounds are for fields, where they are strongest.

\subsection{Application to Inner Products and a Proof Barrier}
\label{sec:counter_dln}

Let $n$ be even and consider the $n$-party function $f$
of~\cite[Section~3.6]{C:DamLarNie19}, also used in
Theorem~\ref{thm:statistical_nopre}: the parties are $P_{j,i}$ for $j \in
\{1, 2\}$ and $i \in [n/2]$, party $P_{j,i}$ holds $x_{j,i} \in \FF^\ell$
and $b_{j,i} \in \bits$ and receives $b_{j,i} \cdot \sum_{i \in [n/2]}
\langle x_{1,i}, x_{2,i} \rangle$. This is $f_g$ for $g(x) = \sum_i
\langle x_{1,i}, x_{2,i} \rangle$. The pairs in $\calP$ are $\{P_{1,i},
P_{2,i}\}$, every $g_p$ is zero, and $L_e$ and $R_e$ are the identity, so
$\ell_e = \ell$ and $L = n\ell/2$. The function $f$ has an arithmetic
circuit of size $|C| = O(n\ell)$ with $n\ell/2$ multiplication gates.

\begin{corollary}
\label{thm:rom_nopre}
Let $n = 2t + s$ with $s \ge 2$ even, let $\ell, \secpar \ge 1$ and let
$f$ be as above. For every $Q$, the function $f$ has a protocol in the
random oracle model without preprocessing that is $(Q, (Q + n\ell/2) \cdot
2^{-\secpar})$-private against $t$ adaptive corruptions and communicates
$n\ell + n(n-2)$ elements of $\FF$ and $\secpar \cdot n (n - t - 1)
\binom{n-2}{t-1} + O(n^3 \log n)$ bits, that is, $2$ elements per
multiplication gate and $n\ell$ elements plus $\poly(\secpar)$ bits for $n
= O(\log \secpar)$.
\end{corollary}
\begin{proof}
Corollary~\ref{cor:rom_quad} with $|\calP| = n/2$ and $\ell_e = \ell$.
\end{proof}

Every statistically secure protocol for $f$ without preprocessing
communicates $\Omega(n \ell \cdot t/s)$ field elements
(Theorem~\ref{thm:statistical_nopre}, which extends~\cite[Theorem~1]{C:DamLarNie19}
to quantum communication), so
for $n = O(\log \secpar)$ and $\ell = \poly(\secpar)$ large enough the
random oracle saves a factor of $t/s$, which is $\Theta(n)$ for $s = O(1)$.

\heading{Three parties.}
\label{sec:counter_three}
\label{lem:three_party_rom}
The bound of~\cite{C:DamLarNie19} is proved through a three-party lemma: in
every statistically secure protocol in which $P_1$ and $P_2$ hold $x_1, x_2
\in \FF^\ell$ and $P_3$ receives $\langle x_1, x_2 \rangle$, the party
$P_3$ communicates $\Omega(\ell \log q)$
bits~\cite[Theorem~3]{C:DamLarNie19} (Lemma~\ref{lem:three_party}). This
function is $f_g$ for $n = 3$, $g = \langle x_1, x_2 \rangle$, $b_1 = b_2 =
0$ and $b_3 = 1$, and for $t = 1$ the protocol $\Prep$ has a single seed set, $J =
\{P_3\}$. The party $P_3$ deals the two seeds and the two shares of
$\langle r, r' \rangle$ and takes part in $\Fsum$, so it communicates
$2\secpar + O(\log q)$ bits, while $P_1$ and $P_2$ exchange $2\ell$
elements. So the three-party lemma fails relative to a random oracle, and
with it the route to a bound growing with $t/s$. This is why
Theorem~\ref{thm:honest_majority_mpc_lb} is proved from PIR, and
Corollary~\ref{thm:rom_nopre} shows that for $f$ and $n = O(\log \secpar)$
no bound growing with $t/s$ holds.

\heading{Larger $t/s$.} The communication of $\Prep$ that does not depend
on $\ell$ is exponential in $n$, as in the pseudorandom secret sharing
of~\cite{TCC:CraDamIsh05}. Over $\mathbb{F}_2$ and for $t/s = \omega(\secpar)$, the two parties of a
pair can instead pick their pads themselves and compute the shares of
$\langle r, r' \rangle$ by OT extension~\cite{C:IKNP03} in the random oracle
model, at $O(\secpar)$ bits per pad bit. This is the mechanism by which the
protocols of~\cite{C:GLOS25,EC:SonYe25} evaluate any circuit at a cost per
gate independent of $n$, and on $f$ it beats the bound
of~\cite{C:DamLarNie19} whenever $t/s = \omega(\secpar)$. For $n =
\omega(\log \secpar)$ with $t/s = O(\secpar)$, the best protocols for $f$
communicate $O(|C| \cdot \min(t/s, \secpar))$ bits, and whether $O(|C|)$
bits suffice there remains open.

\section{MPC with Quantum Communication}
\label{sec:quantum}

In this section, we present our results for MPC with quantum communication. In particular, we present a simple two-party protocol from post-quantum one-way functions whose communication may be sublinear in the circuit size and is analogous to the trivial FHE solution. Afterwards, we show that (quantum-accessible) random oracles may reduce communication, but the computational costs of blackbox protocols must remain high. Finally, we show that
quantum communication alone (without a random oracle) is not sufficient to reduce communication by presenting lower bounds
for information-theoretic protocols.

    \subsection{Two-Party Computation with Quantum Preprocessing}
\label{sec:qmpc}

Notice that in Corollary~\ref{cor:qccc}, our result allows quantum
\emph{computation} but does not allow for quantum communication or
preprocessing, both of which are kept classical. Here we show this
restriction is, in fact, essential to obtain a communication lower bound
using our techniques.
In particular, there is, in the QROM,
a PIR protocol with $O(\secpar \log N)$ communication~\cite{EPRINT:AnaBhaGul26}, and we build
a symmetric PIR protocol~\cite{STOC:GIKM98}
with $\poly(\secpar, \log N)$ communication,
using either quantum communication or quantum preprocessing.
This directly implies a secure two-party computation protocol in the QROM
for any circuit $C$ where one input is of length $O(\log \secpar)$, which
shows that Theorem~\ref{thm:main_pir} does not extend to quantum
communication or quantum preprocessing (Corollary~\ref{cor:qmpc_pir}). A
protocol tailored to the hard instance of Theorem~\ref{thm:main_mpc}, in
which both inputs are long, shows the same for that theorem
(Corollary~\ref{cor:qmpc_mpc}).

\heading{Model.}
We use the model of Section~\ref{sec:prelim_models} with $(Q, \delta)$-privacy
relative to $\calO$ (Section~\ref{sec:prelim_quantum}), so that our protocols
are secure against the same unbounded, query-bounded adversaries as the lower
bounds they contradict. We establish privacy for classical inputs and for
honest parties whose coins are classical, which is the setting of the lower
bounds these protocols contradict (Section~\ref{sec:prelim_classical}). We
construct protocols with
quantum communication and convert them to classical communication with quantum
preprocessing by teleportation. A lower bound against semi-honest protocols is
the stronger statement, so semi-honest upper bounds suffice to rule one out.
Our ingredients are single-server PIR with quantum communication
(Theorem~\ref{thm:qmpc_cpir}) and quantum oblivious transfer
(Theorem~\ref{thm:qmpc_ot}).

\begin{corollary}[Symmetric PIR]
\label{cor:qmpc_spir}
Relative to a quantum-accessible random oracle, there is a single-server
symmetric PIR for databases in $(\bits^m)^N$ with $O(m
\secpar \log N) + \poly(\secpar, \log N)$ qubits of communication that is
private against malicious clients and malicious servers.
\end{corollary}
\begin{proof}
The server samples $\kappa_{j,c} \getsr \bits^\secpar$ for $j \in [\log N]$, $c
\in \bits$ and masks each row as $\tilde{D}[w] := D[w] \oplus \bigoplus_{j}
\calO_m(\kappa_{j, w_j} \| j \| w)$ (Section~\ref{sec:prelim_models}); it sends $\kappa_{j, i_j}$ for each $j$ by $\log N$
sequential oblivious transfers (Theorem~\ref{thm:qmpc_ot}), and the client,
which can now compute the mask of row $i$, retrieves the $m$ bits of
$\tilde{D}[i]$ by $m$ executions of
Theorem~\ref{thm:qmpc_cpir}~\cite{STOC:NaoPin99,TCC:FIPR05}. The server's view
is the $\log N$ transfer sender views and the $m$ query states, hence
computationally independent of $i$ by receiver privacy and
Theorem~\ref{thm:qmpc_cpir}, over a hybrid across the $\log N + m$ instances.
For the client, the simulator extracts classical choice bits $i_1, \ldots,
i_{\log N}$ from the oblivious-transfer simulator, queries the ideal
functionality on the index $i$ they define, and runs the honest server of
Theorem~\ref{thm:qmpc_cpir} on the table whose row $i$ is $D[i] \oplus
\bigoplus_j \calO(\kappa_{j,i_j}\|j\|i)$ and whose other rows are uniform. Every
row $w \ne i$ differs from $i$ in some bit $j$ and so carries the term
$\calO(\kappa_{j, 1-i_j} \| j \| w)$, whose key is a uniform $\secpar$-bit
string hidden by the oblivious transfer, and no two rows share such a term; by
one-way to hiding~\cite{C:AmbHamUnr19} a client making $Q$ oracle queries
distinguishes the two tables with advantage $O(Q \sqrt{\log N} \cdot
2^{-\secpar/2})$, whatever states it submits to the $m$ executions. Masking
whole rows at once binds those executions to a single row.
\end{proof}

\begin{theorem}
\label{thm:qmpc_2pc}
Let $C$ be a circuit computing $f : \bits^a \times \bits^b \to \bits^m$ with $a
= O(\log \secpar)$ and $|C| = \poly(\secpar)$. Relative to a quantum-accessible
random oracle, there is a perfectly correct, semi-honest secure two-party
protocol for $f$ with $O(m \secpar a) + \poly(\secpar)$ qubits of communication,
independent of $|C|$ and $b$.
\end{theorem}
\begin{proof}
Let Alice hold $x \in \bits^a$ and Bob hold $y \in \bits^b$. Bob samples $r
\getsr \bits^m$ and forms the truth table $T[w] := f(w, y) \oplus r$ for $w \in
\bits^a$, at cost $2^a$ evaluations of $C$. Alice retrieves $v := T[x] = f(x,y)
\oplus r$ as the client of Corollary~\ref{cor:qmpc_spir} on the $2^a \times m$
table $T$. To give the output to Alice, Bob sends $r$ and Alice outputs $v
\oplus r$; to give it to Bob, Alice sends $v$ and Bob outputs $v \oplus r$; for
both, do both. Correctness is perfect. For a corrupted Bob, the client's
messages are computationally independent of $x$, and $v$ is a function of his
own input, randomness and output. For a corrupted Alice, her view is a function
of her input, her coins and $v$, which is uniform when she does not receive $r$
and is $f(x,y) \oplus r$ for a uniform $r$ when she does; the simulator samples
the coins itself. Both views are therefore simulatable from the party's input
and output.
\end{proof}

Against a semi-honest Alice the mask $r$ already does the work of the server
privacy, so Corollary~\ref{cor:qmpc_spir} may be replaced by $m$ executions of
Theorem~\ref{thm:qmpc_cpir}, with no oblivious transfers; the symmetric PIR is
what buys security against a \emph{malicious} client in the variant where only
Alice receives output.

\begin{corollary}
\label{cor:qmpc_pir}
Relative to a quantum-accessible random oracle, there is a perfectly correct
single-server PIR with entangled preprocessing and $O(\secpar \log N)$ bits of
classical online communication that is $\negl(\secpar)$-private against servers
making $\poly(\secpar)$ queries. In particular, Theorem~\ref{thm:main_pir} fails
for entangled preprocessing.
\end{corollary}
\begin{proof}
Apply Theorem~\ref{thm:qmpc_2pc} to $f(i, D) = D[i]$ with $a = \log N$ and $m =
1$, where the truth table is the database itself and the protocol is a single
execution of Theorem~\ref{thm:qmpc_cpir}, and teleport its $O(\secpar \log N)$
qubits. For $N = \omega(\secpar \log \secpar)$, say $N = \secpar^2$, this is
$o(N)$.
\end{proof}

The same is true of the hard instance behind Theorem~\ref{thm:main_mpc}, even
though there both parties hold $\ell$-bit inputs.

\begin{corollary}
\label{cor:qmpc_mpc}
Relative to a quantum-accessible random oracle, the two-party instance of the
functionality in the proof of Theorem~\ref{thm:main_mpc} has a semi-honest
secure protocol with entangled preprocessing and $\poly(\secpar, \log \ell)$
bits of classical online communication. In particular,
Theorem~\ref{thm:main_mpc} fails for entangled preprocessing.
\end{corollary}
\begin{proof}
Recall that $P_i$ holds $(d_i, z_i, b_i) \in \bits^\ell \times \bits^{k+1}
\times \bits$ for $\ell = 2^k$ and outputs $b_i \cdot D[z_1 \oplus z_2]$ for
$D = d_1 \| d_2$, with $|C| = O(\ell)$. Write $z_i = (h_i, w_i) \in \bits
\times \bits^k$ for the bit string $h_i \| w_i$ and identify $\bits^k$ with
$[\ell] = \{0, \ldots, \ell-1\}$; then the queried position is the offset
$w_1 \oplus w_2$ of the chunk selected by $h_1 \oplus h_2$. Each $P_i$ samples
$s_i \getsr \bits$ and publishes nothing, but prepares the shifted, masked chunk
$c_i[u] := d_i[u \oplus w_i] \oplus s_i$ for $u \in [\ell]$. Party $P_1$
retrieves $\alpha := c_2[w_1] = d_2[w_1 \oplus w_2] \oplus s_2$ and $P_2$
retrieves $\beta := c_1[w_2] = d_1[w_1 \oplus w_2] \oplus s_1$, each as the
client of Theorem~\ref{thm:qmpc_cpir} on an index it holds, using $O(\secpar
\log \ell)$ qubits. They then run Theorem~\ref{thm:qmpc_ot} on the constant-size
circuit $g$ that takes $(\alpha, s_1, h_1, b_1)$ and $(\beta, s_2, h_2, b_2)$,
sets $D[z_1 \oplus z_2] := \beta \oplus s_1$ if $h_1 \oplus h_2 = 0$ and $\alpha
\oplus s_2$ otherwise, and outputs $b_i \cdot D[z_1 \oplus z_2]$ to $P_i$, at
cost $\poly(\secpar)$. Correctness is immediate. Each party's retrieved bit is
uniform by the other's mask, each query hides its index by
Theorem~\ref{thm:qmpc_cpir}, and the view of $g$ is simulatable from inputs and
outputs, so each party's view is simulatable. Teleporting all qubits gives
$\poly(\secpar, \log \ell)$ classical online communication, which is $o(|C|)$
for $\ell = \secpar^{c}$ with $c$ exceeding the degree of the $\poly(\secpar)$
in Theorem~\ref{thm:qmpc_ot}.
\end{proof}

    \subsection{Computation Lower Bound for MPC with Quantum Preprocessing}\label{sec:qcqc_lb}

In this subsection, we give a new lower bound for MPC with quantum preprocessing. In particular, we first show that any computationally-secure QPIR constructed relative to a quantum-accessible random oracle with sublinear $o(N)$ communication requires the server to perform $\Omega(N)$ cryptographic qubit operations, showing that privacy can only be achieved through heavy use of the post-quantum one-way function.
In fact, we show that our result holds even when given quantum access to the
stronger crypto oracle (Definition~\ref{def:crypto_oracle}). Our proof uses
techniques from~\cite{EC:DujHaj24} along with prior linear communication lower
bounds for information-theoretic QPIR~\cite{JC:BauBro15,EC:ABCGLS19}. Moreover,
the result is tight up to $\poly(\secpar)$ factors as it is matched by recent
work~\cite{EPRINT:AnaBhaGul26} that constructs a QPIR making $O(N)$ evaluations
of a post-quantum one-way function. From this lower bound, we will also derive a
computation lower bound for MPC with quantum communication.

\subsubsection{Lower Bounds for Single-Server PIR with Quantum Communication}

To obtain our lower bound, we will rely upon the following two results from prior works (Section~\ref{sec:prelim_prior}). The first is a linear communication lower bound for information-theoretic QPIR even with quantum preprocessing as presented in~\cite{EC:ABCGLS19} (see Theorem~\ref{thm:abcgls}). The second is the compressed oracle technique of Zhandry~\cite{C:Zhandry19} (see Theorem~\ref{thm:ro_sim}). To start, we show that the lower bound in~\cite{EC:ABCGLS19} extends to QPIR for databases with more than $1$-bit entries.

\begin{theorem}
\label{thm:itqpir_large_entries}
Let $\Pi$ be a single-server QPIR with entangled preprocessing for databases
in $(\bits^b)^N$ with error probability $\epsilon \le (Nb)^{-4}/100$ that is
$\delta$-private for $\delta \le (Nb)^{-8}/100$. Then the server sends at least
$(Nb-1)/2$ qubits.
\end{theorem}
\begin{proof}
    The proof follows identically to the proof in Theorem~\ref{thm:main_pir_large_entries}. Note, one can build a single-server PIR for a database with $N \cdot b$ $1$-bit entries from $\Pi$ with the same communication cost as $\Pi$. By Theorem~\ref{thm:abcgls}, $\Pi$ must have the server send at least $(Nb - 1)/2$ qubits.
\end{proof}

At a high level, our proof compiles any
computationally-secure QPIR with a quantum-accessible random oracle into an information-theoretic QPIR.
To do this, we follow the blueprint of~\cite{EC:DujHaj24} where each server's
(superposition) query to the random oracle
is replaced by an oracle simulated by the client.
The resulting communication cost of the server becomes its original
communication cost plus the size of its random oracle queries.
Afterwards, we apply the prior linear communication lower bounds for information-theoretic QPIR~\cite{EC:ABCGLS19} to prove
that either the server's original communication must be linear or the size of
the server's random oracle queries must be linear in the
database size.

\begin{theorem}
\label{thm:ro_comp_lb}
    Relative to a quantum-accessible random oracle, there does not exist a
    single-server PIR for a database with $N$ $b$-bit entries with quantum preprocessing and quantum communication with
    at most $\epsilon$ error probability that is $(0, \delta)$-private relative
    to $\calO$ (Section~\ref{sec:prelim_quantum}) such that $\epsilon \le
    (Nb)^{-4}/100$ and $\delta \le (Nb)^{-8}/100$ in which the server sends $o(N \cdot b)$
    qubits online and performs $o(N \cdot b)$ cryptographic qubit operations. In fact,
    the number of qubits sent by the server plus its cryptographic qubit
    operations is at least $(Nb-1)/2$.
\end{theorem}
\begin{proof}
As the first step, we compile any single-server QPIR with a quantum-accessible
random oracle into an information-theoretic QPIR (without any random oracle). To
do this, we will replace any queries to the random oracle to one that is
simulated by the client using Theorem~\ref{thm:ro_sim}. Note, this trivially can
replace any client queries to the random oracle, which are made to the same
simulator. For server queries to the random oracle, the server will send the
query to the client instead.
The client will simulate the random oracle query locally and send back to the server both the query and the oracle output.
If the dealer queries $\calO$ when preparing $\rho_{\mathrm{pre}}$, it does so through the same simulator and hands the simulator's register to the client together with the client's preprocessing register, so that the client continues the same simulation in the online phase.
As the random oracle simulation is perfect,
we note the correctness of the QPIR protocol is unaffected. For privacy, the simulation is perfect, so the joint state of the server's registers and $R$ at the end of the compiled protocol is the state of the original protocol averaged over $\calO$, and a distinguisher of the compiled protocol makes no oracle queries. Hence $(0, \delta)$-privacy relative to $\calO$ gives $\delta$-privacy in the sense of Definition~\ref{def:specious}, as Theorem~\ref{thm:abcgls} requires.
Therefore, if the original QPIR has $\epsilon$ error probability and is $(0, \delta)$-private relative to $\calO$, then our information-theoretic QPIR also has $\epsilon$ error probability and is $\delta$-private.
The qubits sent by the server become the qubits sent by the server in the
computationally-secure QPIR protocol along with the inputs and outputs of all
server random oracle queries (that is, the server's cryptographic qubit
operations). Note, each query is also sent back to the server, but this
communication is sent by the client and does not count towards the server's
communication. Finally, we apply the lower bound of $(Nb-1)/2$ qubits sent by the
server for information-theoretic QPIR from Theorem~\ref{thm:itqpir_large_entries}. This
immediately implies that the number of qubits sent by the server in the original
protocol plus its cryptographic qubit operations (inputs and outputs to the
random oracle) is at least $(Nb-1)/2$, so one of the two must be $\Omega(Nb)$.
\end{proof}

Next, we show that the above lower bound on the server cryptographic operations holds in a stronger model of quantum-accessible crypto oracles (Definition~\ref{def:crypto_oracle}).
First, we show that it is possible to perfectly simulate a quantum-accessible crypto oracle efficiently by adapting the random oracle simulation of Zhandry~\cite{C:Zhandry19}. To do this,
we adapt the algorithm $\mathcal{B}$ to be
a quantum circuit $\mathcal{B}^R$ with access to the private quantum-accessible
random function $R$.

\begin{theorem}
\label{thm:co_sim}
There exists an efficient quantum algorithm that perfectly simulates a quantum-accessible crypto oracle $\mathcal{B}^R$ after $Q$ queries using polynomial storage. 
\end{theorem}
\begin{proof}
    For any query $\ket{x}\ket{y}$ to $\mathcal{B}^R$, we execute $\mathcal{B}^R$ on
    $\ket{x}$, XOR the output into the register $\ket{y}$ and then execute it in
    reverse to erase intermediate values. Note, if $\mathcal{B}^R$ makes at most
    polynomial queries to $R$ classically, then the above quantum circuit
    execution also makes polynomially many superposition queries to $R$ (note, twice
    as many since we execute it in reverse as well).
    For every query to random function $R$, we use Theorem~\ref{thm:ro_sim} to
    efficiently and perfectly simulate the quantum-accessible private random function $R$. The compressed oracle is stateful, but it is equivalent, up to an isometry on its database register, to the oracle $\sum_R \ket{R}\bra{R} \otimes O_R$ that applies the query unitary $O_R$ of the function $R$ held in superposition in the database register. As $\mathcal{B}^R$ is deterministic for every fixed $R$, running $\mathcal{B}$ forward and backward with $O_R$ returns its work registers to $\ket{0}$ for every $R$ in the superposition at once, so they are not left entangled with the database register. As the total number of queries to $R$ is polynomial, the simulation algorithm only needs polynomial qubits of storage.
\end{proof}

\begin{theorem}
\label{thm:qpir_crypto_oracle}
    Relative to a quantum-accessible crypto oracle, there does not exist a
    single-server PIR for a database with $N$ $b$-bit entries with quantum preprocessing and quantum communication with
    at most $\epsilon \le (Nb)^{-4}/100$ error probability that is $(0,
    \delta)$-private relative to $\mathcal{B}^R$ (Section~\ref{sec:prelim_quantum}) for
    $\delta \le (Nb)^{-8}/100$ in which the server sends $o(N \cdot b)$ qubits online and
    performs $o(N \cdot b)$ cryptographic qubit operations. In fact, the number of
    qubits sent by the server plus its cryptographic qubit operations is at
    least $(Nb-1)/2$.
\end{theorem}
\begin{proof}
    The proof follows identically to Theorem~\ref{thm:ro_comp_lb} with the exception that we replace the client simulation of the random oracle with the efficient and perfect client simulation of the crypto oracle using Theorem~\ref{thm:co_sim}.
\end{proof}

\subsubsection{Lower Bounds for MPC with Quantum Communication}

Next, we prove a lower bound on the number of cryptographic operations that are required for MPC with small communication with quantum access to a crypto oracle. In particular, we show that if any MPC protocol uses $o(|C|\log q)$ communication even against an adversary with a single corruption, then the total cryptographic operations performed by all parties must be at least $\Omega(|C| \log q)$.
To prove this, we will use the same framework as~\cite{C:DamLarNie19} that was also used in Theorem~\ref{thm:main_mpc}. The main change is that we replace the communication lower bounds of PIR with the cryptographic operation lower bounds proved in Theorem~\ref{thm:qpir_crypto_oracle}.

\begin{theorem}
\label{thm:qmpc_comp_lb}
    Relative to a quantum-accessible crypto oracle, for every field $\mathbb{F}_q$, there is a circuit
    $C$ over $\mathbb{F}_q$ of size $|C| = O(n \cdot \ell)$ such that there does not exist an
    $n$-party secure multi-party computation protocol for $C$ using quantum
    communication and preprocessing with at most $\epsilon$ error probability
    that is $(0, \delta)$-private relative to the crypto oracle against $t = 1$
    non-adaptively corrupted party satisfying both the following efficiency measures:
    \begin{itemize}
        \item Total online communication of $o(|C| \log q)$.
        \item Total of $o(|C| \log q)$ cryptographic qubit operations across
          all $n$ parties during the online phase.
    \end{itemize}
The above holds for any choice of $n \ge 2$, 
$\ell \ge \log n$, $\epsilon \le (\ell\log q)^{-4}/100$ and $\delta \le (\ell\log q)^{-8}/100$.
\end{theorem}
\begin{proof}
We prove our lower bound for the same $n$-party function $f_\ell$ as described
in Theorem~\ref{thm:main_mpc}. To obtain our lower bound, we will prove that
each of the $n$ parties must send $\Omega(\ell\log q)$ qubits online or perform
$\Omega(\ell\log q)$ cryptographic qubit operations. More precisely, we show that for
every party $P_i$, the number of qubits sent by $P_i$ plus its cryptographic
qubit operations is at least $(\ell\log q - 1)/2$. We note this immediately implies
that the total online communication must be $\Omega(n \cdot \ell \cdot \log q)$ or the total
cryptographic qubit operations across all parties must be $\Omega(n \cdot \ell \cdot \log q)$
by the following argument. Recall that the message registers and the oracle
calls of each party are fixed by the protocol (Section~\ref{sec:prelim_models}),
so neither quantity depends on the inputs. Summing over all $n$ parties counts
each message exactly once, by its sender, so the total online communication plus
the total cryptographic qubit operations is at least $n(\ell\log q-1)/2$, and one of
the two must be $\Omega(n \cdot \ell \cdot \log q)$.

Consider any party $P_i$ with $i \in [n]$ that has input of $\ell$-tuple of elements
$d_i \in \mathbb{F}_q^\ell$. We will consider $P_i$ as the server for PIR with an
$\ell$-entry database where each entry has $O(\log q)$ bits and the remaining $n - 1$ parties as the PIR client. To do
this, we set the input of party $P_i$ to be $b_i = 0$ and $z_i = 0^m$, where $m
= \lceil \log (n \ell) \rceil$ is the length of the indices. Pick any $j \ne i$
and set the input of party $P_j$ to be $b_j = 1$ and $z_j$ to the position in $D
= d_1 \| \cdots \| d_n$ of the entry of $d_i$ that the client wishes to
retrieve. Finally, for all $k \ne j$ and $k \ne i$, we set $b_k = 1$ and $z_k =
0^m$, and every party $P_k$ with $k \ne i$ uses $d_k = 0^\ell$. Note, the output of party $P_j$ will be exactly $b_j \cdot D[z_1 \oplus
\ldots \oplus z_n] = D[z_j]$ since $b_j = 1$, which is the chosen entry of
$d_i$. In other words, the client of the remaining $n - 1$ parties obtains any
entry of its choice of the $\ell$-entry database $d_i$ held by $P_i$ acting as the
PIR server. As $P_i$ has output $0$ on every input, the $(0, \delta)$-privacy of
the protocol for $S = \{P_i\}$ gives $(0, \delta)$-privacy of this PIR against
the server. By Theorem~\ref{thm:qpir_crypto_oracle}, we immediately prove that
the number of qubits sent by party $P_i$ plus its cryptographic qubit operations
is at least $(\ell\log q - 1)/2$.
\end{proof}

    \subsection{Statistical Communication Bounds for Quantum MPC}
\label{sec:statistical_lb}

Damg{\aa}rd, Larsen and Nielsen~\cite{C:DamLarNie19} prove
Theorem~\ref{thm:main_mpc} for statistically secure protocols with classical
correlated randomness. There, the PIR step is the fact that the communication
of a statistically secure PIR with correlated randomness has entropy at least
that of the database~\cite[Lemma~5]{C:DamLarNie19}. Here we extend their
result and show that it also applies in the quantum setting. 

First, Theorem~\ref{thm:statistical_mpc} shows that there is an $n$-party
circuit $C$, such that any quantum protocol with statistical
specious
security (Definition~\ref{def:specious}) requires $\Omega(|C| \log q \cdot
t / (n-t))$ qubits of communication, even with arbitrary quantum
preprocessing.
The result follows the same template as our main result without
a random oracle. However, the PIR
impossibility used at the end of the proof appeals directly to the result from~\cite{EC:ABCGLS19} for specious servers with prior
entanglement (Theorem~\ref{thm:abcgls}).

Theorem~\ref{thm:statistical_nopre}
further extends the no preprocessing lower bound of 
Damg{\aa}rd, Larsen and Nielsen~\cite{C:DamLarNie19} in the honest majority
setting. We obtain the full $\Omega(|C| \log q \cdot t / (n-2t))$ bound for
MPC with statistical security and quantum communication.

\begin{theorem}
\label{thm:statistical_mpc}
Let $n \ge 2$, $1 \le t \le n-1$ and $\ell \ge \log n$, and let $f$ be the
$n$-party function in the proof of Theorem~\ref{thm:main_mpc}, which has a
circuit over $\FF$ of size $|C| = O(n \ell)$. Every $n$-party protocol with
quantum preprocessing and quantum communication for $f$ with error
probability $\epsilon \le (t\ell \floor{\log q})^{-4}/100$ that is
$\delta$-private against $t$ non-adaptive corruptions for $\delta \le (t\ell
\floor{\log q})^{-8}/100$ communicates at least
$$
\floor*{\frac{n}{n-t}} \cdot \frac{t\ell \floor{\log q} - 1}{2}
$$
qubits, which is $\Omega(|C| \log q \cdot t / (n-t))$ for $t\ell
\floor{\log q} \ge 2$. In particular, this holds for statistically secure
protocols, where $\epsilon, \delta = \negl(\secpar)$ and $t\ell \log q =
\poly(\secpar)$.
\end{theorem}
\begin{proof}
Let $k = \floor{n/(n-t)}$ and fix disjoint groups $G_1, \ldots, G_k \subseteq
[n]$ of size $n-t$. Fix a group $G$ with complement $S$, so $|S| = t$, and a party $c \in G$,
and let $\Pi_G$ be the following two-party protocol. The server holds $D
\in \bits^{t\ell \floor{\log q}}$ and reads it in blocks of $\floor{\log
q}$ bits, each an integer below $2^{\floor{\log q}} \le q$ and thus an
element of $\FF$, identified with $\{0, \ldots, q-1\}$ as a set; it runs
the parties in $S$ on the $d_a \in \FF^\ell$ whose concatenation is this
sequence of elements, with $z_a = 0$ and $b_a = 0$. The client holds $i \in
[t\ell \floor{\log q}]$ and runs the parties in $G$ on $d_a = 0^\ell$, $b_a
= 1$, $z_a = 0$ for $a \ne c$ and $z_c$ the position in $d_1 \| \cdots \|
d_n$ of the element containing $D[i]$, and outputs the bit of $Y_c$ at the
position of $D[i]$ in that element. Messages inside $S$ or $G$ are local operations,
$\rho_{\mathrm{pre}}$ is a state on $W_S \otimes W_G$, and the messages of
each round from $S$ to $G$ and from $G$ to $S$ form the server's and the
client's message registers, so $\Pi_G$ has the form
of~\cite[Definition~2.1]{EC:ABCGLS19} and the server sends exactly the qubits
that $\Pi$ sends from $S$ to $G$.

As $f_c(x)$ is the element containing $D[i]$, $\Pi_G$ has error probability
$\epsilon$. As $b_a = 0$ for
$a \in S$, the outputs of the parties in $S$ are $0$ on every input. So the
simulator $\Sim_S$ of $\Pi$, applied after embedding $D$ into $X_S$ and
setting $Y_S$ to $0$, simulates the view of the server for every input state
of $\Pi_G$, including those entangled with $R$. Hence
$\Pi_G$ is a $\delta$-private QPIR with entangled preprocessing for $N =
t\ell \floor{\log q}$, and by Theorem~\ref{thm:abcgls} the parties in $S$
send at least $(t\ell \floor{\log q} - 1)/2$ qubits to the parties in $G$,
for every input of $\Pi$ since
message registers have fixed dimensions. Each message register has one
recipient, which lies in at most one group, so summing over the $k$ disjoint
groups gives the bound.
\end{proof}

\subsubsection{Honest Majority without Preprocessing}
\label{sec:statistical_nopre}

Damg{\aa}rd, Larsen and Nielsen~\cite{C:DamLarNie19} also prove an $\Omega(n \cdot t \cdot \ell / s)$ bound for
$n = 2t + s$ parties without preprocessing~\cite[Theorem~1]{C:DamLarNie19}.
At the proof's core is a three-party inner product protocol in which the
output party
must communicate a lot. Here, we show that the argument carries over
to quantum information theory.
We use the entropy notation and facts of Appendix~\ref{app:prelim}, and let
$\kappa(\epsilon, \delta) = 1 - 2\delta - (H_b(\epsilon + 2\sqrt{\delta}) + (\epsilon + 2\sqrt{\delta}) \log (q -
1))/\log q$.

\begin{lemma}
\label{lem:three_party}
Let $\Pi$ be a three-party protocol without preprocessing in which $P_1$ and
$P_2$ hold $x_1, x_2 \in \FF^m$ and have no output, and $P_3$ has no input
and outputs the inner product $\langle x_1, x_2 \rangle$ over $\FF$.
Suppose $\Pi$ has error probability $\epsilon$ and is $\delta$-private for $S
= \{1\}$ and for $S = \{2\}$, where $\epsilon + 2\sqrt{\delta} \le 1/2$. Then
the message registers sent or received by $P_3$ comprise at least $(m \log q \cdot
\kappa(\epsilon, \delta) - 2)/2$ qubits.
\end{lemma}
\begin{proof}
We analyze the purified protocol, which has the same message registers and
error probability as $\Pi$ and is the protocol that
Definition~\ref{def:specious} refers to. Let $c_3$ be the number of qubits
sent or received by $P_3$.

Run the purified protocol with a classical $x_2 \in \FF^m$ and with $X_1$
maximally entangled with a reference $R$, that is, in the state
$$
q^{-m/2} \sum_{x \in \FF^m} \ket{x}_{X_1} \ket{x}_R.
$$
Let $\ket{\psi^{x_2}}$ be
the final (pure) state on $R A_1 A_2 A_3$, where $A_a$ holds all registers
of $P_a$. Let $X$ be the outcome
of measuring $R$ in the computational basis. The protocol does not act on
$R$, so $X$ is uniform on $\FF^m$ and, conditioned on $X = x$, the state of
$A_1 A_2 A_3$ is the final state of the protocol on the classical input $(x,
x_2)$. All entropies below are of the state of $X A_1 A_2 A_3$ obtained in
this way from $\psi^{0^m}$. We show that
\begin{align}
H(X \mid A_2 A_3) &\le m \cdot \big( H_b(\epsilon + 2\sqrt{\delta}) +
  (\epsilon + 2\sqrt{\delta}) \log (q - 1) \big),
  \label{eq:tp_decode} \\
H(X \mid A_2) &\ge m\log q - 2\delta m\log q - 2,
  \label{eq:tp_private} \\
H(X \mid A_2) - H(X \mid A_2 A_3) &\le 2 c_3.
  \label{eq:tp_comm}
\end{align}
That is, the input of $P_1$ is determined by the joint view of $P_2$ and
$P_3$, it is hidden from $P_2$ alone, and the difference is paid for by the
communication of $P_3$. Together they give $2 c_3 \ge m\log q \cdot
\kappa(\epsilon, \delta) - 2$.

First, we establish~\eqref{eq:tp_decode} using correctness and privacy for
$S = \{1\}$. Fix $j \in [m]$ and let $e_j$ be the $j$-th unit vector. The
inputs with $x_2 = 0^m$ and with $x_2 = e_j$ have the same reduced state on
$X_1 R$, and $P_1$ has no output. So by
Lemma~\ref{lem:no_output}(\ref{item:no_out_same}) the views
$\psi^{0^m}_{A_1 R}$ and $\psi^{e_j}_{A_1 R}$ of $P_1$ are $2\delta$-close,
and by Lemma~\ref{lem:uhlmann} there is a unitary $U_j$ on $A_2 A_3$ such
that $U_j \ket{\psi^{0^m}}$ and $\ket{\psi^{e_j}}$ are
$2\sqrt{\delta}$-close, as $\sqrt{2\delta(2 - 2\delta)} \le 2\sqrt{\delta}$. On input $(x, e_j)$ the output is $\langle x, e_j
\rangle = x[j]$, so in $\psi^{e_j}$ measuring $Y_3$ yields $X[j]$ with
probability at least $1 - \epsilon$. Hence applying $U_j$ to the registers
$A_2 A_3$ of $\psi^{0^m}$ and measuring $Y_3$ yields an element $G_j$ of
$\FF$ with
$\Pr[G_j \ne X[j]] \le \epsilon + 2\sqrt{\delta}$. By
Lemma~\ref{lem:entropy}(\ref{item:ent_dpi}) and~(\ref{item:ent_fano}),
$$
H(X[j] \mid A_2 A_3) \le H(X[j] \mid G_j) \le H_b(\epsilon + 2\sqrt{\delta}) + (\epsilon + 2\sqrt{\delta})\log(q - 1),
$$
and summing over $j \in [m]$ with
Lemma~\ref{lem:entropy}(\ref{item:ent_ssa}) gives~\eqref{eq:tp_decode}.

Next, we establish~\eqref{eq:tp_private} using privacy for $S = \{2\}$. On
input $x_2 = 0^m$ the register $X_2$ is in a product state with $R$, whose
reduced state is maximally mixed, and $P_2$ has no output. So by
Lemma~\ref{lem:no_output}(\ref{item:no_out_prod}) the view
$\psi^{0^m}_{A_2 R}$ of $P_2$ is $\delta$-close to the product of a fixed
state on $A_2$ and the maximally mixed state on $R$. Measuring $R$ does not
increase the distance, so the state of $X A_2$ is $\delta$-close to one in
which $X$ is uniform and independent of $A_2$, where $H(X \mid A_2) = m\log q$.
Now~\eqref{eq:tp_private} follows from
Lemma~\ref{lem:entropy}(\ref{item:ent_cont}).

Finally, we establish~\eqref{eq:tp_comm}, which does not use security. The
left-hand side is $I(X ; A_3 \mid A_2) = H(A_3 \mid A_2) - H(A_3 \mid A_2
X)$, which is at most $2 H(A_3)$ by
Lemma~\ref{lem:entropy}(\ref{item:ent_sub}). The state of $A_3$ does not
depend on whether $R$ is measured. As $P_3$ has no input and the protocol is
without preprocessing, the registers of $P_3$ start in a pure state, so
$H(A_3) \le c_3$ by Lemma~\ref{lem:comm_entropy}.
\end{proof}

\begin{theorem}
\label{thm:statistical_nopre}
Let $n = 2t + s$ with $s$ even and $s \mid n$, let $\ell \ge 1$, and let $f$
be the $n$-party function of~\cite[Section~3.6]{C:DamLarNie19}: the parties
are $P_{j,i}$ for $j \in \{1, 2\}$ and $i \in [n/2]$, and $P_{j,i}$ holds
$x_{j,i} \in \FF^\ell$ and $b_{j,i} \in \bits$ and outputs $b_{j,i} \cdot
\langle x_1, x_2 \rangle$ for $x_j = x_{j,1} \| \cdots \| x_{j,n/2}$, which
has a circuit over $\FF$ of size $|C| = O(n \ell)$. As in Theorem~\ref{thm:main_mpc},
the bit $b_{j,i}$ selects whether $P_{j,i}$ receives the output. Every
$n$-party protocol without
preprocessing for $f$ with error probability $\epsilon$ that is
$\delta$-private against $t$ non-adaptive corruptions, where $\epsilon + 2\sqrt{\delta}
\le 1/2$, communicates at least
$$
\frac{n}{4s} \cdot \big(t \cdot \ell \cdot \log q \cdot \kappa(\epsilon, \delta) - 2\big)
$$
qubits. In particular, statistically secure protocols communicate $\Omega(|C| \log q
\cdot t / s)$ qubits.
\end{theorem}
\begin{proof}
Let $h = s/2$. For $g \in [n/s]$, let $G_g$ consist of the parties $P_{j,i}$
with $(g-1)h < i \le gh$. Fix $G = G_g$ and a party $c \in G$, and let $S_j$
be the set of parties $P_{j,i}$ outside $G$, so $|S_j| = n/2 - h = t$. Let
$\Pi_G$ be the three-party protocol in which $P_1$ runs $S_1$ on $x_1 \in
\FF^{t\ell}$ split among its parties with all $b_{1,i} = 0$, $P_2$ runs
$S_2$ likewise on $x_2$, and $P_3$ runs $G$ on $x_{j,i} = 0^\ell$ and
$b_{j,i} = 0$ except for $b_c = 1$, and outputs $Y_c$. Then $\Pi_G$ is
without preprocessing and computes the inner product on $m = t\ell$ elements
with error probability $\epsilon$. As $S_1$ and $S_2$ have $t$ parties whose
outputs are $0$ on every input, the simulator $\Sim_{S_j}$ of $\Pi$, applied
after embedding $x_j$ and setting the outputs to $0$, shows that $\Pi_G$ is
$\delta$-private for $S = \{1\}$ and for $S = \{2\}$. By
Lemma~\ref{lem:three_party}, the message registers of $\Pi$ between $G$ and
the other parties comprise at least $(t\ell \log q \cdot \kappa(\epsilon,
\delta) - 2)/2$ qubits. Each message register has one sender and one recipient, so it
is counted for at most two of the $n/s$ groups, which gives the bound.
\end{proof}

\section*{Acknowledgments}\label{sec:acks} 
The work of G. Persiano was partially supported by a Google Sponsored Research Agreement and by Diagonal Argument.

\section*{AI Disclosure}\label{sec:ai}
The authors made use of Claude, by Anthropic, and Gemini, by Google, in the
preparation of the manuscript. These tools were used to assist with writing
and editing, helping to catch errors and refine proofs.

Most of the results themselves were discovered by the authors prior to
AI use. However, the notable exception is the protocol
presented in Section~\ref{sec:counter_example}, which was found with the
assistance of Claude Fable 5.1. The authors worked to understand and clarify the
presentation of this result and take responsibility for their veracity.

{\small
\bibliographystyle{alpha}
\bibliography{references/others,references/abbrev0,references/crypto}

\newcommand{\etalchar}[1]{$^{#1}$}
\begin{thebibliography}{GHHM21}

\bibitem[ABC{\etalchar{+}}19]{EC:ABCGLS19}
Dorit Aharonov, Zvika Brakerski, Kai-Min Chung, Ayal Green, Ching-Yi Lai, and Or~Sattath.
\newblock On quantum advantage in information theoretic single-server {PIR}.
\newblock In Yuval Ishai and Vincent Rijmen, editors, {\em Advances in Cryptology -- {EUROCRYPT}~2019, Part~III}, volume 11478 of {\em Lecture Notes in Computer Science}, pages 219--246, Darmstadt, Germany, May~19--23, 2019. Springer, Cham, Switzerland.

\bibitem[ABG26]{EPRINT:AnaBhaGul26}
Prabhanjan Ananth, Divyanshu Bhardwaj, and Aditya Gulati.
\newblock A note on single-server {QPIR} from one-way functions.
\newblock Cryptology ePrint Archive, Report 2026/1524, 2026.

\bibitem[ABT19]{EC:AppBraTsa19}
Benny Applebaum, Zvika Brakerski, and Rotem Tsabary.
\newblock Degree 2 is complete for the round-complexity of malicious {MPC}.
\newblock In Yuval Ishai and Vincent Rijmen, editors, {\em Advances in Cryptology -- {EUROCRYPT}~2019, Part~II}, volume 11477 of {\em Lecture Notes in Computer Science}, pages 504--531, Darmstadt, Germany, May~19--23, 2019. Springer, Cham, Switzerland.

\bibitem[ACC{\etalchar{+}}22]{C:ACCFLM22}
Per Austrin, Hao Chung, Kai-Min Chung, Shiuan Fu, Yao-Ting Lin, and Mohammad Mahmoody.
\newblock On the impossibility of key agreements from quantum random oracles.
\newblock In Yevgeniy Dodis and Thomas Shrimpton, editors, {\em Advances in Cryptology -- {CRYPTO}~2022, Part~II}, volume 13508 of {\em Lecture Notes in Computer Science}, pages 165--194, Santa Barbara, CA, USA, August~15--18, 2022. Springer, Cham, Switzerland.

\bibitem[AGH10]{C:AguGabHer10}
Carlos {Aguilar-Melchor}, Philippe Gaborit, and Javier Herranz.
\newblock Additively homomorphic encryption with d-operand multiplications.
\newblock In Tal Rabin, editor, {\em Advances in Cryptology -- {CRYPTO}~2010}, volume 6223 of {\em Lecture Notes in Computer Science}, pages 138--154, Santa Barbara, CA, USA, August~15--19, 2010. Springer Berlin Heidelberg, Germany.

\bibitem[AHU19]{C:AmbHamUnr19}
Andris Ambainis, Mike Hamburg, and Dominique Unruh.
\newblock Quantum security proofs using semi-classical oracles.
\newblock In Alexandra Boldyreva and Daniele Micciancio, editors, {\em Advances in Cryptology -- {CRYPTO}~2019, Part~II}, volume 11693 of {\em Lecture Notes in Computer Science}, pages 269--295, Santa Barbara, CA, USA, August~18--22, 2019. Springer, Cham, Switzerland.

\bibitem[AIK04]{FOCS:AppIshKus04}
Benny Applebaum, Yuval Ishai, and Eyal Kushilevitz.
\newblock Cryptography in {NC}{$^0$}.
\newblock In {\em 45th Annual Symposium on Foundations of Computer Science}, pages 166--175, Rome, Italy, October~17--19, 2004. {IEEE} Computer Society Press.

\bibitem[AL17]{JC:AshLin17}
Gilad Asharov and Yehuda Lindell.
\newblock A full proof of the {BGW} protocol for perfectly secure multiparty computation.
\newblock {\em Journal of Cryptology}, 30(1):58--151, January 2017.

\bibitem[ARS24]{EC:AbrRoySch24}
Damiano Abram, Lawrence Roy, and Peter Scholl.
\newblock Succinct homomorphic secret sharing.
\newblock In Marc Joye and Gregor Leander, editors, {\em Advances in Cryptology -- {EUROCRYPT}~2024, Part~VI}, volume 14656 of {\em Lecture Notes in Computer Science}, pages 301--330, Zurich, Switzerland, May~26--30, 2024. Springer, Cham, Switzerland.

\bibitem[AV21]{JC:AppVas21}
Benny Applebaum and Prashant~Nalini Vasudevan.
\newblock Placing conditional disclosure of secrets in the communication complexity universe.
\newblock {\em Journal of Cryptology}, 34(2):11, April 2021.

\bibitem[BB84]{IEEE:BenBra84}
Charles~H. Bennett and Gilles Brassard.
\newblock Quantum cryptography: Public key distribution and coin tossing.
\newblock In {\em Proceedings of the {IEEE} International Conference on Computers, Systems and Signal Processing}, pages 175--179, Bangalore, India, 1984.
\newblock Reprinted in Theoretical Computer Science 560 (2014), 7--11.

\bibitem[BB15]{JC:BauBro15}
{\"A}min Baumeler and Anne Broadbent.
\newblock Quantum private information retrieval has linear communication complexity.
\newblock {\em Journal of Cryptology}, 28(1):161--175, January 2015.

\bibitem[BBC{\etalchar{+}}93]{PRL:BBCJPW93}
Charles~H. Bennett, Gilles Brassard, Claude Cr{\'e}peau, Richard Jozsa, Asher Peres, and William~K. Wootters.
\newblock Teleporting an unknown quantum state via dual classical and {Einstein}-{Podolsky}-{Rosen} channels.
\newblock {\em Physical Review Letters}, 70(13):1895--1899, 1993.

\bibitem[BCG{\etalchar{+}}19]{C:BCGIKS19}
Elette Boyle, Geoffroy Couteau, Niv Gilboa, Yuval Ishai, Lisa Kohl, and Peter Scholl.
\newblock Efficient pseudorandom correlation generators: Silent {OT} extension and more.
\newblock In Alexandra Boldyreva and Daniele Micciancio, editors, {\em Advances in Cryptology -- {CRYPTO}~2019, Part~III}, volume 11694 of {\em Lecture Notes in Computer Science}, pages 489--518, Santa Barbara, CA, USA, August~18--22, 2019. Springer, Cham, Switzerland.

\bibitem[BCG{\etalchar{+}}20]{FOCS:BCGIKS20}
Elette Boyle, Geoffroy Couteau, Niv Gilboa, Yuval Ishai, Lisa Kohl, and Peter Scholl.
\newblock Correlated pseudorandom functions from variable-density {LPN}.
\newblock In {\em 61st Annual Symposium on Foundations of Computer Science}, pages 1069--1080, Durham, NC, USA, November~16--19, 2020. {IEEE} Computer Society Press.

\bibitem[BCGI18]{CCS:BCGI18}
Elette Boyle, Geoffroy Couteau, Niv Gilboa, and Yuval Ishai.
\newblock Compressing vector {OLE}.
\newblock In David Lie, Mohammad Mannan, Michael Backes, and XiaoFeng Wang, editors, {\em ACM CCS 2018: 25th Conference on Computer and Communications Security}, pages 896--912, Toronto, ON, Canada, October~15--19, 2018. {ACM} Press.

\bibitem[BCKM21]{C:BCKM21b}
James Bartusek, Andrea Coladangelo, Dakshita Khurana, and Fermi Ma.
\newblock One-way functions imply secure computation in a quantum world.
\newblock In Tal Malkin and Chris Peikert, editors, {\em Advances in Cryptology -- {CRYPTO}~2021, Part~I}, volume 12825 of {\em Lecture Notes in Computer Science}, pages 467--496, Virtual Event, August~16--20, 2021. Springer, Cham, Switzerland.

\bibitem[BCM22]{TCC:BoyCouMey22}
Elette Boyle, Geoffroy Couteau, and Pierre Meyer.
\newblock Sublinear secure computation from new assumptions.
\newblock In Eike Kiltz and Vinod Vaikuntanathan, editors, {\em TCC~2022: 20th Theory of Cryptography Conference, Part~II}, volume 13748 of {\em Lecture Notes in Computer Science}, pages 121--150, Chicago, IL, USA, November~7--10, 2022. Springer, Cham, Switzerland.

\bibitem[BCM23]{EC:BoyCouMey23}
Elette Boyle, Geoffroy Couteau, and Pierre Meyer.
\newblock Sublinear-communication secure multiparty computation does not require {FHE}.
\newblock In Carmit Hazay and Martijn Stam, editors, {\em Advances in Cryptology -- {EUROCRYPT}~2023, Part~II}, volume 14005 of {\em Lecture Notes in Computer Science}, pages 159--189, Lyon, France, April~23--27, 2023. Springer, Cham, Switzerland.

\bibitem[BDF{\etalchar{+}}11]{AC:BDFLSZ11}
Dan Boneh, {\"O}zg{\"u}r Dagdelen, Marc Fischlin, Anja Lehmann, Christian Schaffner, and Mark Zhandry.
\newblock Random oracles in a quantum world.
\newblock In Dong~Hoon Lee and Xiaoyun Wang, editors, {\em Advances in Cryptology -- {ASIACRYPT}~2011}, volume 7073 of {\em Lecture Notes in Computer Science}, pages 41--69, Seoul, South Korea, December~4--8, 2011. Springer Berlin Heidelberg, Germany.

\bibitem[BDOZ11]{EC:BDOZ11}
Rikke Bendlin, Ivan Damg{\aa}rd, Claudio Orlandi, and Sarah Zakarias.
\newblock Semi-homomorphic encryption and multiparty computation.
\newblock In Kenneth~G. Paterson, editor, {\em Advances in Cryptology -- {EUROCRYPT}~2011}, volume 6632 of {\em Lecture Notes in Computer Science}, pages 169--188, Tallinn, Estonia, May~15--19, 2011. Springer Berlin Heidelberg, Germany.

\bibitem[Bea92]{C:Beaver91b}
Donald Beaver.
\newblock Efficient multiparty protocols using circuit randomization.
\newblock In Joan Feigenbaum, editor, {\em Advances in Cryptology -- {CRYPTO}'91}, volume 576 of {\em Lecture Notes in Computer Science}, pages 420--432, Santa Barbara, CA, USA, August~11--15, 1992. Springer Berlin Heidelberg, Germany.

\bibitem[Bea96]{STOC:Beaver96a}
Donald Beaver.
\newblock Correlated pseudorandomness and the complexity of private computations.
\newblock In {\em 28th Annual {ACM} Symposium on Theory of Computing}, pages 479--488, Philadelphia, PA, USA, May~22--24, 1996. {ACM} Press.

\bibitem[BFKR91]{C:BFKR90}
Donald Beaver, Joan Feigenbaum, Joe Kilian, and Phillip Rogaway.
\newblock Security with low communication overhead.
\newblock In Alfred~J. Menezes and Scott~A. Vanstone, editors, {\em Advances in Cryptology -- {CRYPTO}'90}, volume 537 of {\em Lecture Notes in Computer Science}, pages 62--76, Santa Barbara, CA, USA, August~11--15, 1991. Springer Berlin Heidelberg, Germany.

\bibitem[BGI15]{EC:BoyGilIsh15}
Elette Boyle, Niv Gilboa, and Yuval Ishai.
\newblock Function secret sharing.
\newblock In Elisabeth Oswald and Marc Fischlin, editors, {\em Advances in Cryptology -- {EUROCRYPT}~2015, Part~II}, volume 9057 of {\em Lecture Notes in Computer Science}, pages 337--367, Sofia, Bulgaria, April~26--30, 2015. Springer Berlin Heidelberg, Germany.

\bibitem[BGI16]{C:BoyGilIsh16}
Elette Boyle, Niv Gilboa, and Yuval Ishai.
\newblock Breaking the circuit size barrier for secure computation under {DDH}.
\newblock In Matthew Robshaw and Jonathan Katz, editors, {\em Advances in Cryptology -- {CRYPTO}~2016, Part~I}, volume 9814 of {\em Lecture Notes in Computer Science}, pages 509--539, Santa Barbara, CA, USA, August~14--18, 2016. Springer Berlin Heidelberg, Germany.

\bibitem[BGJK21]{EC:BGJK21}
Gabrielle Beck, Aarushi Goel, Abhishek Jain, and Gabriel Kaptchuk.
\newblock Order-{C} secure multiparty computation for highly repetitive circuits.
\newblock In Anne Canteaut and Fran\c{c}ois-Xavier Standaert, editors, {\em Advances in Cryptology -- {EUROCRYPT}~2021, Part~II}, volume 12697 of {\em Lecture Notes in Computer Science}, pages 663--693, Zagreb, Croatia, October~17--21, 2021. Springer, Cham, Switzerland.

\bibitem[BGN05]{TCC:BonGohNis05}
Dan Boneh, Eu-Jin Goh, and Kobbi Nissim.
\newblock Evaluating 2-{DNF} formulas on ciphertexts.
\newblock In Joe Kilian, editor, {\em TCC~2005: 2nd Theory of Cryptography Conference}, volume 3378 of {\em Lecture Notes in Computer Science}, pages 325--341, Cambridge, MA, USA, February~10--12, 2005. Springer Berlin Heidelberg, Germany.

\bibitem[BGW88]{STOC:BenGolWig88}
Michael {Ben-Or}, Shafi Goldwasser, and Avi Wigderson.
\newblock Completeness theorems for non-cryptographic fault-tolerant distributed computation (extended abstract).
\newblock In {\em 20th Annual {ACM} Symposium on Theory of Computing}, pages 1--10, Chicago, IL, USA, May~2--4, 1988. {ACM} Press.

\bibitem[BI05]{C:BarIsh05}
Omer Barkol and Yuval Ishai.
\newblock Secure computation of constant-depth circuits with applications to database search problems.
\newblock In Victor Shoup, editor, {\em Advances in Cryptology -- {CRYPTO}~2005}, volume 3621 of {\em Lecture Notes in Computer Science}, pages 395--411, Santa Barbara, CA, USA, August~14--18, 2005. Springer Berlin Heidelberg, Germany.

\bibitem[BIK{\etalchar{+}}17]{CCS:BIKMMP17}
Keith Bonawitz, Vladimir Ivanov, Ben Kreuter, Antonio Marcedone, H.~Brendan McMahan, Sarvar Patel, Daniel Ramage, Aaron Segal, and Karn Seth.
\newblock Practical secure aggregation for privacy-preserving machine learning.
\newblock In Bhavani~M. Thuraisingham, David Evans, Tal Malkin, and Dongyan Xu, editors, {\em ACM CCS 2017: 24th Conference on Computer and Communications Security}, pages 1175--1191, Dallas, TX, USA, October~31~--~November~2, 2017. {ACM} Press.

\bibitem[BL18]{EC:BenLin18}
Fabrice Benhamouda and Huijia Lin.
\newblock k-round multiparty computation from k-round oblivious transfer via garbled interactive circuits.
\newblock In Jesper~Buus Nielsen and Vincent Rijmen, editors, {\em Advances in Cryptology -- {EUROCRYPT}~2018, Part~II}, volume 10821 of {\em Lecture Notes in Computer Science}, pages 500--532, Tel Aviv, Israel, April~29~--~May~3, 2018. Springer, Cham, Switzerland.

\bibitem[BM09]{C:BarMah09}
Boaz Barak and Mohammad {Mahmoody-Ghidary}.
\newblock {Merkle} puzzles are optimal - an {$O(n^2)$}-query attack on any key exchange from a random oracle.
\newblock In Shai Halevi, editor, {\em Advances in Cryptology -- {CRYPTO}~2009}, volume 5677 of {\em Lecture Notes in Computer Science}, pages 374--390, Santa Barbara, CA, USA, August~16--20, 2009. Springer Berlin Heidelberg, Germany.

\bibitem[BMR90]{STOC:BeaMicRog90}
Donald Beaver, Silvio Micali, and Phillip Rogaway.
\newblock The round complexity of secure protocols (extended abstract).
\newblock In {\em 22nd Annual {ACM} Symposium on Theory of Computing}, pages 503--513, Baltimore, MD, USA, May~14--16, 1990. {ACM} Press.

\bibitem[BR93]{CCS:BelRog93}
Mihir Bellare and Phillip Rogaway.
\newblock Random oracles are practical: {A} paradigm for designing efficient protocols.
\newblock In Dorothy~E. Denning, Raymond Pyle, Ravi Ganesan, Ravi~S. Sandhu, and Victoria Ashby, editors, {\em ACM CCS 93: 1st Conference on Computer and Communications Security}, pages 62--73, Fairfax, Virginia, USA, November~3--5, 1993. {ACM} Press.

\bibitem[BW92]{PRL:BenWie92}
Charles~H. Bennett and Stephen~J. Wiesner.
\newblock Communication via one- and two-particle operators on {Einstein}-{Podolsky}-{Rosen} states.
\newblock {\em Physical Review Letters}, 69(20):2881--2884, 1992.

\bibitem[BY25]{EC:BieYeo25}
Alexander Bienstock and Kevin Yeo.
\newblock Weakly super-invertible matrices and constant communication dishonest majority {MPC}.
\newblock In Serge Fehr and Pierre-Alain Fouque, editors, {\em Advances in Cryptology -- {EUROCRYPT}~2025, Part~V}, volume 15605 of {\em Lecture Notes in Computer Science}, pages 421--452, Madrid, Spain, May~4--8, 2025. Springer, Cham, Switzerland.

\bibitem[Can00]{JC:Canetti00}
Ran Canetti.
\newblock Security and composition of multiparty cryptographic protocols.
\newblock {\em Journal of Cryptology}, 13(1):143--202, January 2000.

\bibitem[CCD88]{STOC:ChaCreDam88}
David Chaum, Claude Cr{\'e}peau, and Ivan Damg{\r a}rd.
\newblock Multiparty unconditionally secure protocols (extended abstract).
\newblock In {\em 20th Annual {ACM} Symposium on Theory of Computing}, pages 11--19, Chicago, IL, USA, May~2--4, 1988. {ACM} Press.

\bibitem[CCMN24]{TCC:CCMN24}
Pierre Charbit, Geoffroy Couteau, Pierre Meyer, and Reza Naserasr.
\newblock A note on low-communication secure multiparty computation via circuit depth-reduction.
\newblock In Elette Boyle and Mohammad Mahmoody, editors, {\em TCC~2024: 22nd Theory of Cryptography Conference, Part~IV}, volume 15367 of {\em Lecture Notes in Computer Science}, pages 167--199, Milan, Italy, December~2--6, 2024. Springer, Cham, Switzerland.

\bibitem[CCXY18]{C:CCXY18}
Ignacio Cascudo, Ronald Cramer, Chaoping Xing, and Chen Yuan.
\newblock Amortized complexity of information-theoretically secure {MPC} revisited.
\newblock In Hovav Shacham and Alexandra Boldyreva, editors, {\em Advances in Cryptology -- {CRYPTO}~2018, Part~III}, volume 10993 of {\em Lecture Notes in Computer Science}, pages 395--426, Santa Barbara, CA, USA, August~19--23, 2018. Springer, Cham, Switzerland.

\bibitem[CDD{\etalchar{+}}01]{EC:CDDIM01}
Ran Canetti, Ivan Damg{\aa}rd, Stefan Dziembowski, Yuval Ishai, and Tal Malkin.
\newblock On adaptive vs. non-adaptive security of multiparty protocols.
\newblock In Birgit Pfitzmann, editor, {\em Advances in Cryptology -- {EUROCRYPT}~2001}, volume 2045 of {\em Lecture Notes in Computer Science}, pages 262--279, Innsbruck, Austria, May~6--10, 2001. Springer Berlin Heidelberg, Germany.

\bibitem[CDI05]{TCC:CraDamIsh05}
Ronald Cramer, Ivan Damg{\aa}rd, and Yuval Ishai.
\newblock Share conversion, pseudorandom secret-sharing and applications to secure computation.
\newblock In Joe Kilian, editor, {\em TCC~2005: 2nd Theory of Cryptography Conference}, volume 3378 of {\em Lecture Notes in Computer Science}, pages 342--362, Cambridge, MA, USA, February~10--12, 2005. Springer Berlin Heidelberg, Germany.

\bibitem[CFIK03]{EC:CFIK03}
Ronald Cramer, Serge Fehr, Yuval Ishai, and Eyal Kushilevitz.
\newblock Efficient multi-party computation over rings.
\newblock In Eli Biham, editor, {\em Advances in Cryptology -- {EUROCRYPT}~2003}, volume 2656 of {\em Lecture Notes in Computer Science}, pages 596--613, Warsaw, Poland, May~4--8, 2003. Springer Berlin Heidelberg, Germany.

\bibitem[CGB17]{prio}
Henry Corrigan-Gibbs and Dan Boneh.
\newblock Prio: Private, robust, and scalable computation of aggregate statistics.
\newblock In {\em 14th USENIX symposium on networked systems design and implementation (NSDI 17)}, pages 259--282, 2017.

\bibitem[CHK22]{EC:CorHenKog22}
Henry {Corrigan-Gibbs}, Alexandra Henzinger, and Dmitry Kogan.
\newblock Single-server private information retrieval with sublinear amortized time.
\newblock In Orr Dunkelman and Stefan Dziembowski, editors, {\em Advances in Cryptology -- {EUROCRYPT}~2022, Part~II}, volume 13276 of {\em Lecture Notes in Computer Science}, pages 3--33, Trondheim, Norway, May~30~--~June~3, 2022. Springer, Cham, Switzerland.

\bibitem[CHKV25]{EC:CHKV25}
Henry {Corrigan-Gibbs}, Alexandra Henzinger, Yael~Tauman Kalai, and Vinod Vaikuntanathan.
\newblock Somewhat homomorphic encryption from linear homomorphism and sparse {LPN}.
\newblock In Serge Fehr and Pierre-Alain Fouque, editors, {\em Advances in Cryptology -- {EUROCRYPT}~2025, Part~II}, volume 15602 of {\em Lecture Notes in Computer Science}, pages 3--33, Madrid, Spain, May~4--8, 2025. Springer, Cham, Switzerland.

\bibitem[CK89]{STOC:ChoKus89}
Benny Chor and Eyal Kushilevitz.
\newblock A zero-one law for {Boolean} privacy (extended abstract).
\newblock In {\em 21st Annual {ACM} Symposium on Theory of Computing}, pages 62--72, Seattle, WA, USA, May~15--17, 1989. {ACM} Press.

\bibitem[CK93]{ChoKus93}
Benny Chor and Eyal Kushilevitz.
\newblock A communication-privacy tradeoff for modular addition.
\newblock {\em Information Processing Letters}, 45(4):205--210, 1993.

\bibitem[CK20]{EC:CorKog20}
Henry {Corrigan-Gibbs} and Dmitry Kogan.
\newblock Private information retrieval with sublinear online time.
\newblock In Anne Canteaut and Yuval Ishai, editors, {\em Advances in Cryptology -- {EUROCRYPT}~2020, Part~I}, volume 12105 of {\em Lecture Notes in Computer Science}, pages 44--75, Zagreb, Croatia, May~10--14, 2020. Springer, Cham, Switzerland.

\bibitem[CK24]{C:CouKum24}
Geoffroy Couteau and Naman Kumar.
\newblock 10-party sublinear secure computation from standard assumptions.
\newblock In Leonid Reyzin and Douglas Stebila, editors, {\em Advances in Cryptology -- {CRYPTO}~2024, Part~IX}, volume 14928 of {\em Lecture Notes in Computer Science}, pages 39--73, Santa Barbara, CA, USA, August~18--22, 2024. Springer, Cham, Switzerland.

\bibitem[CKY25]{TCC:CouKumYe25}
Geoffroy Couteau, Naman Kumar, and Xiaxi Ye.
\newblock Multiparty homomorphic secret sharing and more from {LPN} and {MQ}.
\newblock In Benny Applebaum and Huijia~(Rachel) Lin, editors, {\em TCC~2025: 23rd Theory of Cryptography Conference, Part~II}, volume 16269 of {\em Lecture Notes in Computer Science}, pages 498--530, Aarhus, Denmark, December~1--5, 2025. Springer, Cham, Switzerland.

\bibitem[CM21]{EC:CouMey21}
Geoffroy Couteau and Pierre Meyer.
\newblock Breaking the circuit size barrier for secure computation under quasi-polynomial {LPN}.
\newblock In Anne Canteaut and Fran\c{c}ois-Xavier Standaert, editors, {\em Advances in Cryptology -- {EUROCRYPT}~2021, Part~II}, volume 12697 of {\em Lecture Notes in Computer Science}, pages 842--870, Zagreb, Croatia, October~17--21, 2021. Springer, Cham, Switzerland.

\bibitem[Cou19]{EC:Couteau19}
Geoffroy Couteau.
\newblock A note on the communication complexity of multiparty computation in the correlated randomness model.
\newblock In Yuval Ishai and Vincent Rijmen, editors, {\em Advances in Cryptology -- {EUROCRYPT}~2019, Part~II}, volume 11477 of {\em Lecture Notes in Computer Science}, pages 473--503, Darmstadt, Germany, May~19--23, 2019. Springer, Cham, Switzerland.

\bibitem[DGJ{\etalchar{+}}20]{EC:DGJMS20}
Yfke Dulek, Alex~B. Grilo, Stacey Jeffery, Christian Majenz, and Christian Schaffner.
\newblock Secure multi-party quantum computation with a dishonest majority.
\newblock In Anne Canteaut and Yuval Ishai, editors, {\em Advances in Cryptology -- {EUROCRYPT}~2020, Part~III}, volume 12107 of {\em Lecture Notes in Computer Science}, pages 729--758, Zagreb, Croatia, May~10--14, 2020. Springer, Cham, Switzerland.

\bibitem[DH24]{EC:DujHaj24}
Jesko Dujmovic and Mohammad Hajiabadi.
\newblock Lower-bounds on public-key operations in {PIR}.
\newblock In Marc Joye and Gregor Leander, editors, {\em Advances in Cryptology -- {EUROCRYPT}~2024, Part~VI}, volume 14656 of {\em Lecture Notes in Computer Science}, pages 65--87, Zurich, Switzerland, May~26--30, 2024. Springer, Cham, Switzerland.

\bibitem[DI05]{C:DamIsh05}
Ivan Damg{\aa}rd and Yuval Ishai.
\newblock Constant-round multiparty computation using a black-box pseudorandom generator.
\newblock In Victor Shoup, editor, {\em Advances in Cryptology -- {CRYPTO}~2005}, volume 3621 of {\em Lecture Notes in Computer Science}, pages 378--394, Santa Barbara, CA, USA, August~14--18, 2005. Springer Berlin Heidelberg, Germany.

\bibitem[DIJL23]{C:DIJL23}
Quang Dao, Yuval Ishai, Aayush Jain, and Huijia Lin.
\newblock Multi-party homomorphic secret sharing and sublinear {MPC} from sparse {LPN}.
\newblock In Helena Handschuh and Anna Lysyanskaya, editors, {\em Advances in Cryptology -- {CRYPTO}~2023, Part~II}, volume 14082 of {\em Lecture Notes in Computer Science}, pages 315--348, Santa Barbara, CA, USA, August~20--24, 2023. Springer, Cham, Switzerland.

\bibitem[DIK10]{EC:DamIshKro10}
Ivan Damg{\aa}rd, Yuval Ishai, and Mikkel Kr{\o}igaard.
\newblock Perfectly secure multiparty computation and the computational overhead of cryptography.
\newblock In Henri Gilbert, editor, {\em Advances in Cryptology -- {EUROCRYPT}~2010}, volume 6110 of {\em Lecture Notes in Computer Science}, pages 445--465, French Riviera, May~30~--~June~3, 2010. Springer Berlin Heidelberg, Germany.

\bibitem[DJ01]{PKC:DamJur01}
Ivan Damg{\aa}rd and Mats Jurik.
\newblock A generalisation, a simplification and some applications of {Paillier}'s probabilistic public-key system.
\newblock In Kwangjo Kim, editor, {\em PKC~2001: 4th International Workshop on Theory and Practice in Public Key Cryptography}, volume 1992 of {\em Lecture Notes in Computer Science}, pages 119--136, Cheju Island, South Korea, February~13--15, 2001. Springer Berlin Heidelberg, Germany.

\bibitem[DLN19]{C:DamLarNie19}
Ivan Damg{\r a}rd, Kasper~Green Larsen, and Jesper~Buus Nielsen.
\newblock Communication lower bounds for statistically secure {MPC}, with or without preprocessing.
\newblock In Alexandra Boldyreva and Daniele Micciancio, editors, {\em Advances in Cryptology -- {CRYPTO}~2019, Part~II}, volume 11693 of {\em Lecture Notes in Computer Science}, pages 61--84, Santa Barbara, CA, USA, August~18--22, 2019. Springer, Cham, Switzerland.

\bibitem[DLS21]{ITC:DamLiSch21}
Ivan~Bjerre Damg{\r a}rd, Boyang Li, and Nikolaj~Ignatieff Schwartzbach.
\newblock More communication lower bounds for information-theoretic {MPC}.
\newblock In Stefano Tessaro, editor, {\em ITC 2021: 2nd Conference on Information-Theoretic Cryptography}, volume 199 of {\em Leibniz International Proceedings in Informatics (LIPIcs)}, pages 2:1--2:18, Seattle, WA, USA, July~23--26, 2021. Schloss Dagstuhl - Leibniz-Zentrum fuer Informatik.

\bibitem[DMO00]{EC:diCMalOst00}
Giovanni {Di Crescenzo}, Tal Malkin, and Rafail Ostrovsky.
\newblock Single database private information retrieval implies oblivious transfer.
\newblock In Bart Preneel, editor, {\em Advances in Cryptology -- {EUROCRYPT}~2000}, volume 1807 of {\em Lecture Notes in Computer Science}, pages 122--138, Bruges, Belgium, May~14--18, 2000. Springer Berlin Heidelberg, Germany.

\bibitem[DN07]{C:DamNie07}
Ivan Damg{\aa}rd and Jesper~Buus Nielsen.
\newblock Scalable and unconditionally secure multiparty computation.
\newblock In Alfred Menezes, editor, {\em Advances in Cryptology -- {CRYPTO}~2007}, volume 4622 of {\em Lecture Notes in Computer Science}, pages 572--590, Santa Barbara, CA, USA, August~19--23, 2007. Springer Berlin Heidelberg, Germany.

\bibitem[DNOR16]{EC:DNOR16}
Ivan Damg{\r a}rd, Jesper~Buus Nielsen, Rafail Ostrovsky, and Adi Ros{\'e}n.
\newblock Unconditionally secure computation with reduced interaction.
\newblock In Marc Fischlin and Jean-S{\'{e}}bastien Coron, editors, {\em Advances in Cryptology -- {EUROCRYPT}~2016, Part~II}, volume 9666 of {\em Lecture Notes in Computer Science}, pages 420--447, Vienna, Austria, May~8--12, 2016. Springer Berlin Heidelberg, Germany.

\bibitem[DNPR16]{C:DNPR16}
Ivan Damg{\r a}rd, Jesper~Buus Nielsen, Antigoni Polychroniadou, and Michael Raskin.
\newblock On the communication required for unconditionally secure multiplication.
\newblock In Matthew Robshaw and Jonathan Katz, editors, {\em Advances in Cryptology -- {CRYPTO}~2016, Part~II}, volume 9815 of {\em Lecture Notes in Computer Science}, pages 459--488, Santa Barbara, CA, USA, August~14--18, 2016. Springer Berlin Heidelberg, Germany.

\bibitem[DNS10]{C:DupNieSal10}
Fr{\'e}d{\'e}ric Dupuis, Jesper~Buus Nielsen, and Louis Salvail.
\newblock Secure two-party quantum evaluation of unitaries against specious adversaries.
\newblock In Tal Rabin, editor, {\em Advances in Cryptology -- {CRYPTO}~2010}, volume 6223 of {\em Lecture Notes in Computer Science}, pages 685--706, Santa Barbara, CA, USA, August~15--19, 2010. Springer Berlin Heidelberg, Germany.

\bibitem[DPP14]{C:DatPraPra14}
Deepesh Data, Manoj Prabhakaran, and Vinod~M. Prabhakaran.
\newblock On the communication complexity of secure computation.
\newblock In Juan~A. Garay and Rosario Gennaro, editors, {\em Advances in Cryptology -- {CRYPTO}~2014, Part~II}, volume 8617 of {\em Lecture Notes in Computer Science}, pages 199--216, Santa Barbara, CA, USA, August~17--21, 2014. Springer Berlin Heidelberg, Germany.

\bibitem[DPSZ12]{C:DPSZ12}
Ivan Damg{\aa}rd, Valerio Pastro, Nigel~P. Smart, and Sarah Zakarias.
\newblock Multiparty computation from somewhat homomorphic encryption.
\newblock In Reihaneh Safavi-Naini and Ran Canetti, editors, {\em Advances in Cryptology -- {CRYPTO}~2012}, volume 7417 of {\em Lecture Notes in Computer Science}, pages 643--662, Santa Barbara, CA, USA, August~19--23, 2012. Springer Berlin Heidelberg, Germany.

\bibitem[EGP{\etalchar{+}}23]{EC:EGPSW23}
Daniel Escudero, Vipul Goyal, Antigoni Polychroniadou, Yifan Song, and Chenkai Weng.
\newblock {SuperPack}: Dishonest majority {MPC} with constant online communication.
\newblock In Carmit Hazay and Martijn Stam, editors, {\em Advances in Cryptology -- {EUROCRYPT}~2023, Part~II}, volume 14005 of {\em Lecture Notes in Computer Science}, pages 220--250, Lyon, France, April~23--27, 2023. Springer, Cham, Switzerland.

\bibitem[EGPS22]{CCS:EGPS22}
Daniel Escudero, Vipul Goyal, Antigoni Polychroniadou, and Yifan Song.
\newblock {TurboPack}: Honest majority {MPC} with constant online communication.
\newblock In Heng Yin, Angelos Stavrou, Cas Cremers, and Elaine Shi, editors, {\em ACM CCS 2022: 29th Conference on Computer and Communications Security}, pages 951--964, Los Angeles, CA, USA, November~7--11, 2022. {ACM} Press.

\bibitem[FIPR05]{TCC:FIPR05}
Michael~J. Freedman, Yuval Ishai, Benny Pinkas, and Omer Reingold.
\newblock Keyword search and oblivious pseudorandom functions.
\newblock In Joe Kilian, editor, {\em TCC~2005: 2nd Theory of Cryptography Conference}, volume 3378 of {\em Lecture Notes in Computer Science}, pages 303--324, Cambridge, MA, USA, February~10--12, 2005. Springer Berlin Heidelberg, Germany.

\bibitem[FY92]{STOC:FraYun92}
Matthew~K. Franklin and Moti Yung.
\newblock Communication complexity of secure computation (extended abstract).
\newblock In {\em 24th Annual {ACM} Symposium on Theory of Computing}, pages 699--710, Victoria, BC, Canada, May~4--6, 1992. {ACM} Press.

\bibitem[Gen09]{STOC:Gentry09}
Craig Gentry.
\newblock Fully homomorphic encryption using ideal lattices.
\newblock In Michael Mitzenmacher, editor, {\em 41st Annual {ACM} Symposium on Theory of Computing}, pages 169--178, Bethesda, MD, USA, May~31~--~June~2, 2009. {ACM} Press.

\bibitem[GGW25]{TCC:GarGunWan25}
Sanjam Garg, Sam Gunn, and Mingyuan Wang.
\newblock Black-box crypto is useless for pseudorandom codes.
\newblock In Benny Applebaum and Huijia~(Rachel) Lin, editors, {\em TCC~2025: 23rd Theory of Cryptography Conference, Part~IV}, volume 16271 of {\em Lecture Notes in Computer Science}, pages 205--224, Aarhus, Denmark, December~1--5, 2025. Springer, Cham, Switzerland.

\bibitem[GHHM21]{AC:GHHM21}
Alex~B. Grilo, Kathrin H{\"o}velmanns, Andreas H{\"u}lsing, and Christian Majenz.
\newblock Tight adaptive reprogramming in the {QROM}.
\newblock In Mehdi Tibouchi and Huaxiong Wang, editors, {\em Advances in Cryptology -- {ASIACRYPT}~2021, Part~I}, volume 13090 of {\em Lecture Notes in Computer Science}, pages 637--667, Singapore, December~6--10, 2021. Springer, Cham, Switzerland.

\bibitem[GI14]{EC:GilIsh14}
Niv Gilboa and Yuval Ishai.
\newblock Distributed point functions and their applications.
\newblock In Phong~Q. Nguyen and Elisabeth Oswald, editors, {\em Advances in Cryptology -- {EUROCRYPT}~2014}, volume 8441 of {\em Lecture Notes in Computer Science}, pages 640--658, Copenhagen, Denmark, May~11--15, 2014. Springer Berlin Heidelberg, Germany.

\bibitem[GIKM98]{STOC:GIKM98}
Yael Gertner, Yuval Ishai, Eyal Kushilevitz, and Tal Malkin.
\newblock Protecting data privacy in private information retrieval schemes.
\newblock In {\em 30th Annual {ACM} Symposium on Theory of Computing}, pages 151--160, Dallas, TX, USA, May~23--26, 1998. {ACM} Press.

\bibitem[GIOZ17]{C:GIOZ17}
Juan~A. Garay, Yuval Ishai, Rafail Ostrovsky, and Vassilis Zikas.
\newblock The price of low communication in secure multi-party computation.
\newblock In Jonathan Katz and Hovav Shacham, editors, {\em Advances in Cryptology -- {CRYPTO}~2017, Part~I}, volume 10401 of {\em Lecture Notes in Computer Science}, pages 420--446, Santa Barbara, CA, USA, August~20--24, 2017. Springer, Cham, Switzerland.

\bibitem[GIP15]{C:GenIshPol15}
Daniel Genkin, Yuval Ishai, and Antigoni Polychroniadou.
\newblock Efficient multi-party computation: From passive to active security via secure {SIMD} circuits.
\newblock In Rosario Gennaro and Matthew J.~B. Robshaw, editors, {\em Advances in Cryptology -- {CRYPTO}~2015, Part~II}, volume 9216 of {\em Lecture Notes in Computer Science}, pages 721--741, Santa Barbara, CA, USA, August~16--20, 2015. Springer Berlin Heidelberg, Germany.

\bibitem[GLOS25]{C:GLOS25}
Vipul Goyal, Junru Li, Rafail Ostrovsky, and Yifan Song.
\newblock Towards building scalable constant-round {MPC} from minimal assumptions via round collapsing.
\newblock In Yael~Tauman Kalai and Seny~F. Kamara, editors, {\em Advances in Cryptology -- {CRYPTO}~2025, Part~IV}, volume 16003 of {\em Lecture Notes in Computer Science}, pages 70--104, Santa Barbara, CA, USA, August~17--21, 2025. Springer, Cham, Switzerland.

\bibitem[GLSV21]{EC:GLSV21}
Alex~B. Grilo, Huijia Lin, Fang Song, and Vinod Vaikuntanathan.
\newblock Oblivious transfer is in {MiniQCrypt}.
\newblock In Anne Canteaut and Fran\c{c}ois-Xavier Standaert, editors, {\em Advances in Cryptology -- {EUROCRYPT}~2021, Part~II}, volume 12697 of {\em Lecture Notes in Computer Science}, pages 531--561, Zagreb, Croatia, October~17--21, 2021. Springer, Cham, Switzerland.

\bibitem[GMPY26]{EPRINT:GMPY26}
Uma Girish, Alex May, Natalie Parham, and Henry Yuen.
\newblock New bounds on private simultaneous quantum message passing.
\newblock Cryptology {ePrint} Archive, Paper 2026/1236, 2026.

\bibitem[GMW87]{STOC:GolMicWig87}
Oded Goldreich, Silvio Micali, and Avi Wigderson.
\newblock How to play any mental game or {A} completeness theorem for protocols with honest majority.
\newblock In Alfred Aho, editor, {\em 19th Annual {ACM} Symposium on Theory of Computing}, pages 218--229, New York City, NY, USA, May~25--27, 1987. {ACM} Press.

\bibitem[GPS21]{C:GoyPolSon21}
Vipul Goyal, Antigoni Polychroniadou, and Yifan Song.
\newblock Unconditional communication-efficient {MPC} via hall's marriage theorem.
\newblock In Tal Malkin and Chris Peikert, editors, {\em Advances in Cryptology -- {CRYPTO}~2021, Part~II}, volume 12826 of {\em Lecture Notes in Computer Science}, pages 275--304, Virtual Event, August~16--20, 2021. Springer, Cham, Switzerland.

\bibitem[GPS22]{C:GoyPolSon22}
Vipul Goyal, Antigoni Polychroniadou, and Yifan Song.
\newblock Sharing transformation and dishonest majority {MPC} with packed secret sharing.
\newblock In Yevgeniy Dodis and Thomas Shrimpton, editors, {\em Advances in Cryptology -- {CRYPTO}~2022, Part~IV}, volume 13510 of {\em Lecture Notes in Computer Science}, pages 3--32, Santa Barbara, CA, USA, August~15--18, 2022. Springer, Cham, Switzerland.

\bibitem[GS22]{EC:GarSri18}
Sanjam Garg and Akshayaram Srinivasan.
\newblock Two-round multiparty secure computation from minimal assumptions.
\newblock {\em Journal of the ACM}, 69(5):1--30, 2022.

\bibitem[HHS08]{TCC:HaiHocSeg08}
Iftach Haitner, Jonathan~J. Hoch, and Gil Segev.
\newblock A linear lower bound on the communication complexity of single-server private information retrieval.
\newblock In Ran Canetti, editor, {\em TCC~2008: 5th Theory of Cryptography Conference}, volume 4948 of {\em Lecture Notes in Computer Science}, pages 445--464, San Francisco, CA, USA, March~19--21, 2008. Springer Berlin Heidelberg, Germany.

\bibitem[HPPY25]{EC:HPPY25}
Alexander Hoover, Sarvar Patel, Giuseppe Persiano, and Kevin Yeo.
\newblock Plinko: Single-server {PIR} with efficient updates via invertible {PRFs}.
\newblock In Serge Fehr and Pierre-Alain Fouque, editors, {\em Advances in Cryptology -- {EUROCRYPT}~2025, Part~VI}, volume 15606 of {\em Lecture Notes in Computer Science}, pages 3--33, Madrid, Spain, May~4--8, 2025. Springer, Cham, Switzerland.

\bibitem[HPY26]{FOCS:HooPerYeo26}
Alexander Hoover, Giuseppe Persiano, and Kevin Yeo.
\newblock Lower bounds for {PIR} with preprocessing from blackbox cryptography.
\newblock In {\em 67th IEEE Symposium on Foundations of Computer Science (to appear)}, 2026.

\bibitem[IK04]{EC:IshKus04}
Yuval Ishai and Eyal Kushilevitz.
\newblock On the hardness of information-theoretic multiparty computation.
\newblock In Christian Cachin and Jan Camenisch, editors, {\em Advances in Cryptology -- {EUROCRYPT}~2004}, volume 3027 of {\em Lecture Notes in Computer Science}, pages 439--455, Interlaken, Switzerland, May~2--6, 2004. Springer Berlin Heidelberg, Germany.

\bibitem[IKM{\etalchar{+}}13]{TCC:IKMOP13}
Yuval Ishai, Eyal Kushilevitz, Sigurd Meldgaard, Claudio Orlandi, and Anat {Paskin-Cherniavsky}.
\newblock On the power of correlated randomness in secure computation.
\newblock In Amit Sahai, editor, {\em TCC~2013: 10th Theory of Cryptography Conference}, volume 7785 of {\em Lecture Notes in Computer Science}, pages 600--620, Tokyo, Japan, March~3--6, 2013. Springer Berlin Heidelberg, Germany.

\bibitem[IKN{\etalchar{+}}20]{ads_conversion}
Mihaela Ion, Ben Kreuter, Ahmet~Erhan Nergiz, Sarvar Patel, Shobhit Saxena, Karn Seth, Mariana Raykova, David Shanahan, and Moti Yung.
\newblock On deploying secure computing: Private intersection-sum-with-cardinality.
\newblock In {\em 2020 IEEE European Symposium on Security and Privacy (EuroS\&P)}, pages 370--389. IEEE, 2020.

\bibitem[IKNP03]{C:IKNP03}
Yuval Ishai, Joe Kilian, Kobbi Nissim, and Erez Petrank.
\newblock Extending oblivious transfers efficiently.
\newblock In Dan Boneh, editor, {\em Advances in Cryptology -- {CRYPTO}~2003}, volume 2729 of {\em Lecture Notes in Computer Science}, pages 145--161, Santa Barbara, CA, USA, August~17--21, 2003. Springer Berlin Heidelberg, Germany.

\bibitem[IKOS07]{STOC:IKOS07}
Yuval Ishai, Eyal Kushilevitz, Rafail Ostrovsky, and Amit Sahai.
\newblock Zero-knowledge from secure multiparty computation.
\newblock In David~S. Johnson and Uriel Feige, editors, {\em 39th Annual {ACM} Symposium on Theory of Computing}, pages 21--30, San Diego, CA, USA, June~11--13, 2007. {ACM} Press.

\bibitem[IPS08]{C:IshPraSah08}
Yuval Ishai, Manoj Prabhakaran, and Amit Sahai.
\newblock Founding cryptography on oblivious transfer - efficiently.
\newblock In David Wagner, editor, {\em Advances in Cryptology -- {CRYPTO}~2008}, volume 5157 of {\em Lecture Notes in Computer Science}, pages 572--591, Santa Barbara, CA, USA, August~17--21, 2008. Springer Berlin Heidelberg, Germany.

\bibitem[IR89]{STOC:ImpRud89}
Russell Impagliazzo and Steven Rudich.
\newblock Limits on the provable consequences of one-way permutations.
\newblock In {\em 21st Annual {ACM} Symposium on Theory of Computing}, pages 44--61, Seattle, WA, USA, May~15--17, 1989. {ACM} Press.

\bibitem[ISW24]{C:IshShiWic24}
Yuval Ishai, Elaine Shi, and Daniel Wichs.
\newblock {PIR} with client-side preprocessing: Information-theoretic constructions and lower bounds.
\newblock In Leonid Reyzin and Douglas Stebila, editors, {\em Advances in Cryptology -- {CRYPTO}~2024, Part~IX}, volume 14928 of {\em Lecture Notes in Computer Science}, pages 148--182, Santa Barbara, CA, USA, August~18--22, 2024. Springer, Cham, Switzerland.

\bibitem[JRS02]{FOCS:JaiRadSen02}
Rahul Jain, Jaikumar Radhakrishnan, and Pranab Sen.
\newblock Privacy and interaction in quantum communication complexity and a theorem about the relative entropy of quantum states.
\newblock In {\em 43rd Annual Symposium on Foundations of Computer Science}, pages 429--438, Vancouver, BC, Canada, November~16--19, 2002. {IEEE} Computer Society Press.

\bibitem[KDW03]{STOC:KerdeWol03}
Iordanis Kerenidis and Ronald De~Wolf.
\newblock Exponential lower bound for 2-query locally decodable codes via a quantum argument.
\newblock In {\em Proceedings of the thirty-fifth annual ACM symposium on Theory of computing}, pages 106--115, 2003.

\bibitem[Kil88]{STOC:Kilian88}
Joe Kilian.
\newblock Founding cryptography on oblivious transfer.
\newblock In {\em 20th Annual {ACM} Symposium on Theory of Computing}, pages 20--31, Chicago, IL, USA, May~2--4, 1988. {ACM} Press.

\bibitem[KLLR16]{QIC:KLLR16}
Iordanis Kerenidis, Mathieu Lauri{\`e}re, Fran{\c c}ois {Le Gall}, and Mathys Rennela.
\newblock Information cost of quantum communication protocols.
\newblock {\em Quantum Information and Computation}, 16(3\&4):181--196, 2016.

\bibitem[KN21]{ITC:KawNis21}
Akinori Kawachi and Harumichi Nishimura.
\newblock Communication complexity of private simultaneous quantum messages protocols.
\newblock In Stefano Tessaro, editor, {\em ITC 2021: 2nd Conference on Information-Theoretic Cryptography}, volume 199 of {\em Leibniz International Proceedings in Informatics (LIPIcs)}, pages 20:1--20:19, Seattle, WA, USA, July~23--26, 2021. Schloss Dagstuhl - Leibniz-Zentrum fuer Informatik.

\bibitem[KNVY26]{ARXIV:KNVY26}
Fuyuki Kitagawa, Ryo Nishimaki, Agi Villanyi, and Takashi Yamakawa.
\newblock Towards the impossibility of imperfectly complete key agreement in the {QROM}.
\newblock arXiv:2608.17610, 2026.

\bibitem[KO97]{FOCS:KusOst97}
Eyal Kushilevitz and Rafail Ostrovsky.
\newblock Replication is {NOT} needed: {SINGLE} database, computationally-private information retrieval.
\newblock In {\em 38th Annual Symposium on Foundations of Computer Science}, pages 364--373, Miami Beach, Florida, October~19--22, 1997. {IEEE} Computer Society Press.

\bibitem[KOS16]{CCS:KelOrsSch16}
Marcel Keller, Emmanuela Orsini, and Peter Scholl.
\newblock {MASCOT}: Faster malicious arithmetic secure computation with oblivious transfer.
\newblock In Edgar~R. Weippl, Stefan Katzenbeisser, Christopher Kruegel, Andrew~C. Myers, and Shai Halevi, editors, {\em ACM CCS 2016: 23rd Conference on Computer and Communications Security}, pages 830--842, Vienna, Austria, October~24--28, 2016. {ACM} Press.

\bibitem[KPR18]{EC:KelPasRot18}
Marcel Keller, Valerio Pastro, and Dragos Rotaru.
\newblock Overdrive: Making {SPDZ} great again.
\newblock In Jesper~Buus Nielsen and Vincent Rijmen, editors, {\em Advances in Cryptology -- {EUROCRYPT}~2018, Part~III}, volume 10822 of {\em Lecture Notes in Computer Science}, pages 158--189, Tel Aviv, Israel, April~29~--~May~3, 2018. Springer, Cham, Switzerland.

\bibitem[Kus92]{Kushelvitz92}
Eyal Kushilevitz.
\newblock Privacy and communication complexity.
\newblock {\em SIAM Journal on Discrete Mathematics}, 5(2):273--284, 1992.

\bibitem[LLLL24]{C:LLLL24}
Longcheng Li, Qian Li, Xingjian Li, and Qipeng Liu.
\newblock How (not) to build quantum {PKE} in {Minicrypt}.
\newblock In Leonid Reyzin and Douglas Stebila, editors, {\em Advances in Cryptology -- {CRYPTO}~2024, Part~VII}, volume 14926 of {\em Lecture Notes in Computer Science}, pages 152--183, Santa Barbara, CA, USA, August~18--22, 2024. Springer, Cham, Switzerland.

\bibitem[LLLL25]{EPRINT:LLLL25}
Longcheng Li, Qian Li, Xingjian Li, and Qipeng Liu.
\newblock Cryptomania v.s. {Minicrypt} in a quantum world.
\newblock Cryptology ePrint Archive, Report 2025/639, 2025.

\bibitem[LLLL26]{ARXIV:LLLL26}
Longcheng Li, Qian Li, Xingjian Li, and Qipeng Liu.
\newblock Impossibility of perfectly complete many-round key agreement in the {QROM}.
\newblock arXiv:2608.03824, 2026.

\bibitem[LMW25]{EC:LinMooWic25}
Wei-Kai Lin, Ethan Mook, and Daniel Wichs.
\newblock Black box crypto is useless for doubly efficient {PIR}.
\newblock In Serge Fehr and Pierre-Alain Fouque, editors, {\em Advances in Cryptology -- {EUROCRYPT}~2025, Part~VI}, volume 15606 of {\em Lecture Notes in Computer Science}, pages 65--93, Madrid, Spain, May~4--8, 2025. Springer, Cham, Switzerland.

\bibitem[Lo97]{Lo97}
Hoi-Kwong Lo.
\newblock Insecurity of quantum secure computations.
\newblock {\em Physical Review A}, 56(2):1154, 1997.

\bibitem[LS25]{TCC:LiSon25}
Junru Li and Yifan Song.
\newblock Honest majority constant-round {MPC} with linear communication from one-way functions.
\newblock In Benny Applebaum and Huijia~(Rachel) Lin, editors, {\em TCC~2025: 23rd Theory of Cryptography Conference, Part~I}, volume 16268 of {\em Lecture Notes in Computer Science}, pages 3--32, Aarhus, Denmark, December~1--5, 2025. Springer, Cham, Switzerland.

\bibitem[LS26]{EPRINT:LiSon26}
Junru Li and Yifan Song.
\newblock Breaking the $\omega(|c|\kappa)$ barrier on garbled circuit size in the random oracle model.
\newblock Cryptology ePrint Archive, Report 2026/1297, 2026.
\newblock {CRYPTO} 2026.

\bibitem[Nao91]{JC:Naor91}
Moni Naor.
\newblock Bit commitment using pseudorandomness.
\newblock {\em Journal of Cryptology}, 4(2):151--158, January 1991.

\bibitem[Nay99]{FOCS:Nayak99}
Ashwin Nayak.
\newblock Optimal lower bounds for quantum automata and random access codes.
\newblock In {\em 40th Annual Symposium on Foundations of Computer Science}, pages 369--377, New York, NY, USA, October~17--19, 1999. {IEEE} Computer Society Press.

\bibitem[NC10]{BOOK:NieChu10}
Michael~A. Nielsen and Isaac~L. Chuang.
\newblock {\em Quantum Computation and Quantum Information: 10th Anniversary Edition}.
\newblock Cambridge University Press, 2010.

\bibitem[NN01]{STOC:NaoNis01}
Moni Naor and Kobbi Nissim.
\newblock Communication preserving protocols for secure function evaluation.
\newblock In {\em 33rd Annual {ACM} Symposium on Theory of Computing}, pages 590--599, Crete, Greece, July~6--8, 2001. {ACM} Press.

\bibitem[NP99]{STOC:NaoPin99}
Moni Naor and Benny Pinkas.
\newblock Oblivious transfer and polynomial evaluation.
\newblock In {\em 31st Annual {ACM} Symposium on Theory of Computing}, pages 245--254, Atlanta, GA, USA, May~1--4, 1999. {ACM} Press.

\bibitem[OS07]{PKC:OstSke07}
Rafail Ostrovsky and William~E. Skeith, III.
\newblock A survey of single-database private information retrieval: Techniques and applications (invited talk).
\newblock In Tatsuaki Okamoto and Xiaoyun Wang, editors, {\em PKC~2007: 10th International Conference on Theory and Practice of Public Key Cryptography}, volume 4450 of {\em Lecture Notes in Computer Science}, pages 393--411, Beijing, China, April~16--20, 2007. Springer Berlin Heidelberg, Germany.

\bibitem[PAD{\etalchar{+}}23]{USENIX:PADBBH23}
Antigoni Polychroniadou, Gilad Asharov, Benjamin~E. Diamond, Tucker Balch, Hans Buehler, Richard Hua, Suwen Gu, Greg Gimler, and Manuela Veloso.
\newblock Prime match: {A} privacy-preserving inventory matching system.
\newblock In Joseph~A. Calandrino and Carmela Troncoso, editors, {\em USENIX Security 2023: 32nd {USENIX} Security Symposium}, pages 6417--6434, Anaheim, CA, USA, August~9--11, 2023. {USENIX} Association.

\bibitem[Pai99]{EC:Paillier99}
Pascal Paillier.
\newblock Public-key cryptosystems based on composite degree residuosity classes.
\newblock In Jacques Stern, editor, {\em Advances in Cryptology -- {EUROCRYPT}'99}, volume 1592 of {\em Lecture Notes in Computer Science}, pages 223--238, Prague, Czech Republic, May~2--6, 1999. Springer Berlin Heidelberg, Germany.

\bibitem[PY22]{SODA:PerYeo22}
Giuseppe Persiano and Kevin Yeo.
\newblock Limits of preprocessing for single-server {PIR}.
\newblock In Joseph~(Seffi) Naor and Niv Buchbinder, editors, {\em 33rd Annual {ACM}-{SIAM} Symposium on Discrete Algorithms}, pages 2522--2548, Virtual Conference / Alexandria, VA, USA, January~9--12, 2022. {ACM-SIAM}.

\bibitem[RMS24]{CCS:RenMugSun24}
Ling Ren, Muhammad~Haris Mughees, and I~Sun.
\newblock Simple and practical amortized sublinear private information retrieval using dummy subsets.
\newblock In Bo~Luo, Xiaojing Liao, Jun Xu, Engin Kirda, and David Lie, editors, {\em ACM CCS 2024: 31st Conference on Computer and Communications Security}, pages 1420--1433, Salt Lake City, UT, USA, October~14--18, 2024. {ACM} Press.

\bibitem[RTV04]{TCC:ReiTreVad04}
Omer Reingold, Luca Trevisan, and Salil~P. Vadhan.
\newblock Notions of reducibility between cryptographic primitives.
\newblock In Moni Naor, editor, {\em TCC~2004: 1st Theory of Cryptography Conference}, volume 2951 of {\em Lecture Notes in Computer Science}, pages 1--20, Cambridge, MA, USA, February~19--21, 2004. Springer Berlin Heidelberg, Germany.

\bibitem[SY25]{EC:SonYe25}
Yifan Song and Xiaxi Ye.
\newblock Honest majority {MPC} with {{\~O}}(|{C}|) communication in minicrypt.
\newblock In Serge Fehr and Pierre-Alain Fouque, editors, {\em Advances in Cryptology -- {EUROCRYPT}~2025, Part~V}, volume 15605 of {\em Lecture Notes in Computer Science}, pages 453--483, Madrid, Spain, May~4--8, 2025. Springer, Cham, Switzerland.

\bibitem[TPY{\etalchar{+}}19]{USENIX:TPYRKI19}
Kurt Thomas, Jennifer Pullman, Kevin Yeo, Ananth Raghunathan, Patrick~Gage Kelley, Luca Invernizzi, Borbala Benko, Tadek Pietraszek, Sarvar Patel, Dan Boneh, and Elie Bursztein.
\newblock Protecting accounts from credential stuffing with password breach alerting.
\newblock In Nadia Heninger and Patrick Traynor, editors, {\em USENIX Security 2019: 28th {USENIX} Security Symposium}, pages 1556--1571, Santa Clara, CA, USA, August~14--16, 2019. {USENIX} Association.

\bibitem[Wil17]{BOOK:Wilde17}
Mark~M. Wilde.
\newblock {\em Quantum Information Theory}.
\newblock Cambridge University Press, 2nd edition, 2017.

\bibitem[Win16]{CMP:Winter16}
Andreas Winter.
\newblock Tight uniform continuity bounds for quantum entropies: Conditional entropy, relative entropy distance and energy constraints.
\newblock {\em Communications in Mathematical Physics}, 347(1):291--313, 2016.

\bibitem[WR25]{EC:WanRen25}
Zhikun Wang and Ling Ren.
\newblock Single-server client preprocessing {PIR} with tight space-time trade-off.
\newblock In Serge Fehr and Pierre-Alain Fouque, editors, {\em Advances in Cryptology -- {EUROCRYPT}~2025, Part~VI}, volume 15606 of {\em Lecture Notes in Computer Science}, pages 94--122, Madrid, Spain, May~4--8, 2025. Springer, Cham, Switzerland.

\bibitem[Yao82]{FOCS:Yao82}
Andrew~C Yao.
\newblock Protocols for secure computations.
\newblock In {\em 23rd annual symposium on foundations of computer science (sfcs 1982)}, pages 160--164. IEEE, 1982.

\bibitem[Yeo23]{EC:Yeo23}
Kevin Yeo.
\newblock Lower bounds for (batch) {PIR} with private preprocessing.
\newblock In Carmit Hazay and Martijn Stam, editors, {\em Advances in Cryptology -- {EUROCRYPT}~2023, Part~I}, volume 14004 of {\em Lecture Notes in Computer Science}, pages 518--550, Lyon, France, April~23--27, 2023. Springer, Cham, Switzerland.

\bibitem[Zha19]{C:Zhandry19}
Mark Zhandry.
\newblock How to record quantum queries, and applications to quantum indifferentiability.
\newblock In Alexandra Boldyreva and Daniele Micciancio, editors, {\em Advances in Cryptology -- {CRYPTO}~2019, Part~II}, volume 11693 of {\em Lecture Notes in Computer Science}, pages 239--268, Santa Barbara, CA, USA, August~18--22, 2019. Springer, Cham, Switzerland.

\bibitem[ZPZS24]{SP:ZPZS24}
Mingxun Zhou, Andrew Park, Wenting Zheng, and Elaine Shi.
\newblock Piano: Extremely simple, single-server {PIR} with sublinear server computation.
\newblock In {\em 2024 {IEEE} Symposium on Security and Privacy}, pages 4296--4314, San Francisco, CA, USA, May~19--23, 2024. {IEEE} Computer Society Press.

\bibitem[ZYFZ26]{EPRINT:ZYFZ26}
Yubo Zeng, Kang Yang, Dengguo Feng, and Min Zhang.
\newblock Unconditionally secure {MPC} for {B}oolean circuits with constant communication.
\newblock Cryptology ePrint Archive, Report 2026/1194, 2026.
\newblock {CRYPTO} 2026.

\end{thebibliography}
}

\appendix
\section{Additional Preliminaries}
\label{app:prelim}

For completeness, we recall the quantum information background and the
entropy facts used in Sections~\ref{sec:pir} and~\ref{sec:statistical_lb}.

\heading{States and operations.} A register $A$ is a finite-dimensional
Hilbert space of dimension $\dim A$, and a qubit is a register of dimension
$2$. A state on $A$ is a density operator $\rho_A$, which is \emph{pure} if
$\rho_A = \ket{\psi}\bra{\psi}$. For a state $\rho_{AB}$ on registers $A B$,
its reduced state on $A$ is $\rho_A = \mathrm{tr}_B(\rho_{AB})$. A register
$X$ is \emph{classical} in $\rho_{XB}$ if $\rho_{XB} = \sum_x p(x)
\ket{x}\bra{x} \otimes \rho_B^x$. The trace distance is $\Delta(\rho, \sigma)
= \frac{1}{2} \| \rho - \sigma \|_1$, and $\rho$ and $\sigma$ are
$\delta$-close if $\Delta(\rho, \sigma) \le \delta$. Trace distance does not
increase under quantum operations, and the probabilities that $\rho$ and
$\sigma$ assign to any measurement outcome differ by at most $\Delta(\rho,
\sigma)$.

A quantum operation from $A$ to $A'$ is a completely positive
trace-preserving map, and a measurement is a quantum operation whose output
is classical. An isometry $V$ from $A$ to $A'$ maps $\rho_A$ to $V \rho_A
V^\dagger$. By Stinespring's dilation theorem, every quantum operation from
$A$ to $A'$ is an isometry from $A$ to $A' E$ followed by discarding $E$.

\begin{lemma}[Uhlmann's theorem~{\cite[Lemma~4.4]{EC:ABCGLS19}}]
\label{lem:uhlmann}
Let $\ket{\phi}$ and $\ket{\psi}$ be pure states on $A B$ whose reduced
states on $A$ are $\delta$-close. Then there is a unitary $U$ on $B$ such
that $(\mathrm{id}_A \otimes U)\ket{\psi}$ and $\ket{\phi}$ are
$\sqrt{\delta(2-\delta)}$-close.
\end{lemma}

\heading{Entropy.} The von Neumann entropy of $A$ in a state $\rho$ is
$H(A)_\rho = -\mathrm{tr}(\rho_A \log \rho_A)$. The conditional entropy is
$H(A \mid B) = H(AB) - H(B)$ and the conditional mutual information is
$$
I(A ; B \mid C) = H(A \mid C) - H(A \mid BC) = H(B \mid C) - H(B \mid AC).
$$
We drop the subscript $\rho$ when the state is clear. On classical registers
these are equivalent to Shannon's entropy notions, and $H_b$ denotes the
binary entropy function. We use the following standard
facts~\cite{BOOK:NieChu10,BOOK:Wilde17}.

\begin{lemma}[Entropy facts]
\label{lem:entropy}
For every state on registers $A B C$:
\begin{enumerate}[nosep]
    \item \label{item:ent_dim} $0 \le H(A) \le \log \dim A$, and $H(A) = 0$
      if and only if the state of $A$ is pure.
    \item \label{item:ent_sub} $|H(A) - H(B)| \le H(AB) \le H(A) + H(B)$.
      Equivalently, $-H(A) \le H(A \mid B) \le H(A)$.
    \item \label{item:ent_ssa} $H(A \mid BC) \le H(A \mid B)$, and hence
      $H(AB \mid C) \le H(A \mid C) + H(B \mid C)$.
    \item \label{item:ent_dpi} If $B'$ is obtained from $B$ by a quantum
      operation, then $H(A \mid B) \le H(A \mid B')$.
    \item \label{item:ent_fano} If $X$ is a classical variable and $G$ is a
      classical variable with $\Pr[G \ne X] \le p \le 1/2$, then $H(X \mid G) \le
      H_b(p) + p \log(|X| - 1)$.
    \item \label{item:ent_cont} If $\rho_{AB}$ and $\sigma_{AB}$ are
      $\delta$-close, then $|H(A \mid B)_\rho - H(A \mid B)_\sigma| \le
      2\delta \log \dim A + 2$~\cite{CMP:Winter16}.
\end{enumerate}
\end{lemma}

\begin{lemma}[Local operations]
\label{lem:local_ops}
Let $\rho$ be a state on registers $A B$.
\begin{enumerate}[nosep]
    \item \label{item:loc_iso} Applying an isometry to $A$, or any quantum
      operation to $B$, does not change $H(A)$.
    \item \label{item:loc_move} If $A = A' Q$, then $|H(A') - H(A)| \le
      \log \dim Q$.
\end{enumerate}
\end{lemma}
\begin{proof}
An isometry preserves the non-zero eigenvalues of $\rho_A$, and an operation
on $B$ does not change $\rho_A$. The second item is
Lemma~\ref{lem:entropy}(\ref{item:ent_sub}) applied to $A' Q$, with $H(Q)
\le \log \dim Q$.
\end{proof}

\begin{lemma}[Communication bounds entropy]
\label{lem:comm_entropy}
Let $G \subseteq [n]$ and let $A_G$ be the registers held by the parties in
$G$ during a run of a purified protocol. If the state of $A_G$ is pure at the
start, then at the end $H(A_G)$ is at most the number of qubits in the
message registers sent between $G$ and $[n] \setminus G$, in either
direction.
\end{lemma}
\begin{proof}
Initially $H(A_G) = 0$. The operations of the parties in $G$, including
messages inside $G$, are isometries on $A_G$, and the operations of the other
parties do not act on $A_G$, so neither changes $H(A_G)$ by
Lemma~\ref{lem:local_ops}(\ref{item:loc_iso}). A message register $Q$ sent
between $G$ and $[n] \setminus G$ is added to or removed from $A_G$, which
changes $H(A_G)$ by at most $\log \dim Q$ by
Lemma~\ref{lem:local_ops}(\ref{item:loc_move}).
\end{proof}

\end{document}